\documentclass{aa}
\usepackage{natbib}
\usepackage{placeins}
\bibpunct{(}{)}{;}{a}{}{,}

\usepackage{graphicx}
\graphicspath{{./}{figure/}}
\usepackage{txfonts}
\usepackage{hyperref}

\begin{document}

   \title{J-PLUS: Support vector regression to measure stellar parameters}

   \author{   Cunshi Wang \inst{\ref{NAOC},\ref{UCAS}}
   \and Yu Bai \inst{\ref{NAOC}}
   \and Haibo Yuan \inst{\ref{BNU}} 
   \and Jifeng Liu\inst{\ref{NAOC},\ref{UCAS}}
   \and J.A. Fernández-Ontiveros\inst{\ref{CEFCA}}
   \and Paula R. T. Coelho \inst{\ref{USP}}
   \and F. Jim\'enez-Esteban \inst{\ref{CSIC}}
   \and Carlos Andrés Galarza\inst{\ref{ON}}
   \and R.~E.~Angulo\inst{\ref{DIPC},\ref{ikerbasque}}
   \and A.~J.~Cenarro\inst{\ref{CEFCA}}
   \and D.~Crist\'obal-Hornillos\inst{\ref{CEFCA}}
   \and R.~A.~Dupke\inst{\ref{ON},\ref{MU},\ref{Alabama}}
   \and A.~Ederoclite\inst{\ref{CEFCA}}
   \and C.~Hern\'andez-Monteagudo\inst{\ref{IAC},\ref{ULL}}
   \and C.~L\'opez-Sanjuan\inst{\ref{CEFCA}}
   \and A.~Mar\'{\i}n-Franch\inst{\ref{CEFCA}}
   \and M.~Moles\inst{\ref{CEFCA}}
   \and L.~Sodr\'e Jr.\inst{\ref{USP}}
   \and H.~V\'azquez Rami\'o\inst{\ref{CEFCA}}
   \and J.~Varela\inst{\ref{CEFCA}}
   }

   \institute{
    Key Laboratory of Optical Astronomy, National Astronomical Observatories, Chinese Academy of Sciences, \\
    20A Datun Road, Chaoyang District, Beijing 100012,People's Republic of China \label{NAOC}
    \and
    College of Astronomy and Space Sciences, University of Chinese Academy of Sciences, Beijing 100049, China \label{UCAS}
    \and 
    Department of Astronomy, Beijing Normal University, Beijing 100875, People's Republic of China\label{BNU} 
    \and
    Centro de Estudios de F\'{\i}sica del Cosmos de Arag\'on (CEFCA), Unidad Asociada al CSIC, Plaza San Juan 1, 44001 Teruel, Spain\label{CEFCA}
    \and
    Instituto de Astronomia, Geof\'{\i}sica e Ci\^encias Atmosf\'ericas, Universidade de S\~ao Paulo, 05508-090 S\~ao Paulo, Brazil\label{USP}
    \and 
    Departamento de Astrof\'{\i}sica, Centro de Astrobiolog\'{\i}a (CSIC-INTA), ESAC Campus, Camino Bajo del Castillo s/n, E-28692 Villanueva de la Ca\~nada, Madrid, Spain\label{CSIC}
    \and
    Observat\'orio Nacional - MCTI (ON), Rua Gal. Jos\'e Cristino 77, S\~ao Crist\'ov\~ao, 20921-400 Rio de Janeiro, Brazil\label{ON}
    \and
    Donostia International Physics Centre (DIPC), Paseo Manuel de Lardizabal 4, 20018 Donostia-San Sebastián, Spain\label{DIPC}
    \and
    IKERBASQUE, Basque Foundation for Science, 48013, Bilbao, Spain\label{ikerbasque}
    \and
    University of Michigan, Department of Astronomy, 1085 South University Ave., Ann Arbor, MI 48109, USA\label{MU}
    \and
    University of Alabama, Department of Physics and Astronomy, Gallalee Hall, Tuscaloosa, AL 35401, USA\label{Alabama}
    \and
    Instituto de Astrof\'{\i}sica de Canarias, La Laguna, 38205, Tenerife, Spain\label{IAC}
    \and
    Departamento de Astrof\'{\i}sica, Universidad de La Laguna, 38206, Tenerife, Spain\label{ULL}
             }
   \date{Received April 29, 2022; accepted April 29, 2022}

 
  \abstract
   {Stellar parameters are among the most important characteristics in studies of stars which, in traditional methods, are based on atmosphere models. However, time, cost, and brightness limits restrain the efficiency of spectral observations. The Javalambre Photometric Local Universe Survey (J-PLUS) is an observational campaign that aims to obtain photometry in 12 bands. Owing to its characteristics, J-PLUS data have become a valuable resource for studies of stars. Machine learning provides powerful tools for efficiently analyzing large data sets, such as the one from J-PLUS, and enables us to expand the research domain to stellar parameters.}
   {The main goal of this study is to construct a support vector regression (SVR) algorithm to estimate stellar parameters of the stars in the first data release of the J-PLUS observational campaign.}
   {The training data for the parameters regressions are featured with 12-waveband photometry from J-PLUS and are cross-identified with spectrum-based catalogs. These catalogs are from the Large Sky Area Multi-Object Fiber Spectroscopic Telescope, the Apache Point Observatory Galactic Evolution Experiment, and the Sloan  Extension for Galactic Understanding and Exploration.{ }We then label them with the stellar effective temperature, the surface gravity, and the metallicity. Ten percent of the sample is held out to apply a blind test. We develop a new method, a multi-model approach, in order to fully take into account the uncertainties of both the magnitudes and the stellar parameters. The method utilizes more than two hundred models to apply the uncertainty analysis. }
   {We present a catalog of 2,493,424 stars with the root mean square error of 160K in the effective temperature regression, 0.35 in the surface gravity regression, and 0.25 in the metallicity regression. We also discuss the advantages of this multi-model approach and compare it to other machine-learning methods.}
   {}


   \maketitle
%

\section{Introduction} \label{intro}
In modern astronomy, the large amount of raw data produced by the newest surveys is far beyond our traditional processing capacity, and this insufficient computing power has become a bottleneck to the rapid development of astrophysics. Fortunately, the development of computer science has provided us with new ways to understand and gain knowledge from these raw data.  Machine learning in particular has achieved remarkable success in offering novel solutions to complex problems based on applications of loss functions, optimization methods \citep{ruder17}, and statistical models \citep{mackay03}. Owing to its nonfunctional algorithms \citep{cortes95,svm1992,crist00,knn67,knn77,quin86,rf01}, machine learning can reveal potential patterns and important parameters that are indistinguishable using traditional scientific or statistical methods.

Stars are the cornerstone of astronomy, and stellar parameters are among the most crucial characteristics of understanding and characterizing stars. Among the most powerful and accurate methods of determining stellar parameters is spectral analysis \citep{wu11, boe18, ang18}. 

However, spectral observations can be costly. The most powerful spectroscopy survey --- the Large Sky Area Multi-Object Fiber Spectroscopy Telescope (LAMOST) --- has taken spectra of about ten million stars in low resolution and about three million in medium resolution with a four-meter telescope in the past 8 years. In contrast to spectral observations, photometric observations have much higher observational efficiency. Compared to LAMOST, the Javalambre Photometric Local Universe Survey (J-PLUS, \citealt{jplus2019}) observed 223.6 million objects in its five-year observational campaign  \footnote{\url{http://J-PLUS.es/datareleases/data_release_dr2}}. Therefore, many studies have been working on estimating stellar parameters from photometry-based data by constructing models of photometric observation results \citep{bailer11, sic12, sic14}. Their modeling requires highly precise photometric data from a few surveys, while another mighty tool --- machine learning --- can reveal the distribution or pattern hidden inside the photometry with rougher data. 

Machine learning, a cross-disciplinary subject of statistics, optimization, and computer science, creates algorithms that can process data based on a well-chosen sample set with features. Different algorithms assign different models and reveal potential patterns in the sample \citep{mackay03, shai14}. The algorithms select features from samples and optimize the parameters based on a given loss function. The loss function evaluates the difference between the predicted result and the true value. Machine-learning algorithms have been applied in many different disciplines, including finance, medical science, and computer vision.

Machine-learning technology has shown the ability to obtain valuable information from multi-band photometric data. Several studies have made some efforts to determine stellar parameters from photometric surveys. \citet{bai19r} have derived stellar effective temperatures from $Gaia$ second data release using a random forest (RF) algorithm. \citet{bai19} have presented RF models to categorize objects such as stars, galaxies, and quasi-stellar objects (QSOs) and classified and obtained the effective temperature of stars. \citet{lu15} developed a scheme to achieve stellar parameters from the Least Absolute Shrinkage and Selection Operator (LASSO) algorithm and support vector regression (SVR) models. \citet{yang21} designed a cost-sensitive artificial neural network and achieved stellar parameters from two million stars from the J-PLUS first data release (DR1). \citet{gala22} developed the stellar parameters estimation based on the ensemble methods (SPEEM) pipeline, which is a stack of feature searching, normalization, and multi-output regressor. The SPEEM is based on RF and extreme gradient boosting (XGB). These works are all based on machine-learning algorithms which indicates their powerful ability to reveal potential patterns within data.

J-PLUS has also been used to gain knowledge of objects ranging from our Solar System to the deep universe, \footnote{\url{http://J-PLUS.es/survey/science}} such as the coma cluster \citep{jim19}, low metallicity stars \citep{whit19, gala22},  and galaxy formation \citep{nog19}. J-PLUS has 12-waveband photometry, which makes the survey ideal for the application of machine learning.
 
In this study, we adopt the SVR algorithm to obtain the stellar parameters, which include the effective temperature ($T_{\rm eff}$), the surface gravity (log $g$), and the metallicity ([Fe/H]) of the stars in the J-PLUS DR1 catalog. To construct the training sample, we use data from LAMOST, the Apache Point Observatory Galactic Evolution Experiment (APOGEE, \citealt{zas13}), and the Sloan Extension for Galactic Understanding and Exploration (SEGUE, \citealt{segue09}; Sect. \ref{data}).

We adopt the SVR \citep{svr15,svr97,cortes95} algorithm and determine the kernel scales (Sect. \ref{svr}). We construct 80 training sets based on the uncertainties of the stellar parameters. Each parameter of any object in the 80 sets obeys a Gaussian distribution  (Sect. \ref{dataenhance}). In all, we construct more than two hundred different models. The blind test is presented in Sect. \ref{BT}. 

The result of applying our method to the entire J-PLUS DR1 catalog is presented in Sect. \ref{result}.  In Sect. \ref{d1} and \ref{d2}, we compare different methods for constructing the training sample and the inconsistency of stellar parameters among different pipelines (Sect. \ref{solpipe}). We discuss the distribution of the regressed stellar parameters in Sect. \ref{dis5.4}, and compare our result with \citet{yang21} in Sect. \ref{dis5.5}. We present a conclusion in Sect. \ref{conclusion}.

\section{Data} \label{data}

\subsection{J-PLUS}
J-PLUS is conducted by the Observatorio Astrof\'{\i}sico de Javalambre (OAJ, Teruel, Spain; \citealt{oaj}). It uses the 83\,cm Javalambre Auxiliary Survey Telescope (JAST80) and T80Cam, which is a panoramic camera of 9.2k $\times$ 9.2k pixels that provides a $2\deg^2$ field of view (FoV) with a pixel scale of 0.55 arcsec pix$^{-1}$ \citep{t80cam}. The J-PLUS contains a 12-passband filter system that comprises five broad ($u, g, r, i, z$) and seven medium bands from 3000 to 9000 \AA. \citet{jplus2019} illustrate the scientific goals and the observational and image reduction strategy of J-PLUS.

J-PLUS DR1 covers an area of 1,022 $ {\rm deg}^2$ on the sky with a magnitude limit of $\sim 21.5$ for a S/N of $\sim$ 3. These 12 bands provide a large sample for us to characterize the spectral energy distribution of the detected sources \citep{jplus2019}. The 12-band magnitudes are adopted as our training features, which are $u$, $J0378$, $J0395$, $J0410$, $J0430$, $g$, $J0515$, $r$, $J0660$, $i$, $J0861$, and $z$. We name them mag1 to mag12 for simplicity.

\citet{yuanp}  recalibrated the J-PLUS DR1 catalog using stellar color regression. The method is described in detail in \citet{yuan15}. The catalog in \citet{yuanp} contains 13,265,168 objects, including 4,126,928 objects with all 12 magnitudes. In \citealt{wang21}, the objects in the recalibrated J-PLUS catalog are classified using the support vector machine (SVM) algorithm, which distinguishes them among the classes STAR, GALAXY, and QSO. Here they chose the 12 J-PLUS magnitudes as features and used their corresponding uncertainties as weight for the SVM. They provide two catalogs based on the 12 density contours on the 12 magnitudes of the training sample. The objects that fall into all the 12 contours are assigned as interpolations, and the others are assigned as extrapolation. Interpolations have better classification accuracy than extrapolations. In this study, we use all objects classified as STAR in these two catalogs for the stellar parameter regressions.

\subsection{LAMOST spectra}
LAMOST is a northern spectroscopic survey situated at Xinglong Observatory, China. LAMOST is able to observe 4,000 objects simultaneously with a 20 $ {\rm deg}^2$ FoV. The main scientific project of LAMOST aims to understand the structure of the Milky Way \citep{deng12} and external galaxies. We used the A, F, G, and K catalogs in LAMOST Data Release 7 (DR7), low resolution spectra (LRS), and medium resolution spectra (MRS) \footnote{\url{http://dr7.lamost.org/catalogue}}.

LAMOST LRS have a limiting magnitude of about $20$ in  the $g$ band and its S/N  is higher than 6 on dark nights or 15 on bright nights. In MRS, the S/N is always larger than 10. The parameters are given by the LAMOST Stellar Parameter pipeline (LASP), and their uncertainties are mainly from the stellar S/N and chi-square of the best-matched theoretical spectrum \citep{wu14,liu20}. The parameter differences of the LRS data and MRS data are very small, on average 17.6 K for $T_{\rm eff}$, 0.028 dex for log $g$  ($g$ in ${\rm (cms^{-2})}$), and 0.084 dex for [Fe/H], according to our test. The internal uncertainties of LASP are estimated to be $\Delta T_{\rm eff} \sim 80 \rm K $, $\Delta {\rm log} g \sim 0.15 \rm dex$, and $\Delta {\rm [Fe/H]} \sim 0.09 \rm dex$ \citep{wanglm20}. We cross-matched these catalogs with J-PLUS DR1 to within one arcsec using the Tool for OPerations on Catalogues And Tables (TOPCAT, \citealt{topcat}). There are 216,114 and 25,170 cross-matched stars in the LRS and MRS catalogs, respectively.

\subsection{APOGEE}
APOGEE \citep{zas13} has observed about 150,000 stars in the Milky Way and obtained precise stellar information, including stellar atmospheric parameters and radial velocities. The APOGEE Stellar Parameter and Chemical Abundance Pipeline (ASPCAP, \citealt{aspcap15}) uses a chi-square minimization to determine the stellar parameters. ASPCAP has high precision in  stellar parameters ($\Delta T_{\rm eff} \sim 2\% $, $\Delta {\rm log} g \sim 0.1 \rm dex$, and $\Delta {\rm [Fe/H]} \sim 0.05 \rm dex$). \citet{aspcap18} compared the spectra and determined the uncertainties of the stellar parameters ($\Delta T_{\rm eff} \sim 100 \rm K $, $\Delta {\rm log} g \sim 0.2 dex$, and $\Delta {\rm [Fe/H]} \sim 0.1 dex$).

We adopted the APOGEE catalog to enlarge our training sample. Using the Sloan Digital Sky Survey (SDSS) Catalog Archive Server Jobs\footnote{\url{http://skyserver.sdss.org/casjobs/}}, we extracted table aspcapStar and find 12,931 stars that satisfy our one arcsec cross-match tolerance.

\subsection{SEGUE}
SEGUE \citep{segue09} is designed to obtain images in the $u,g,r,i$, and $z$ wavebands for 3,500 square degrees of the sky located primarily at low galactic latitudes ($|b|<35^{\circ}$). It delivers observations of about 240,000 stars with a $g$ band magnitude between 14.0 and 20.3 mag and moderate-resolution spectra from 3,900 to 9,000 \AA.  The stellar parameters are presented in SDSS Data Release 7 for these Milky Way stars with S/N greater than ten. The SEGUE Stellar Parameter Pipeline (SSPP; \citealt{sspp1,sspp2,sspp3}) developed a multi-method technology to calculate the stellar parameters which includes non-linear regression models, a minimal distance of observed spectra and grids of synthetic spectra, and the correlations between spectral lines and color relations. The average uncertainties of SSPP are $\Delta T_{\rm eff} \sim 130 \rm K$, $\Delta {\rm log} g \sim 0.21 \rm dex$, and $\Delta {\rm [Fe/H]} \sim 0.11 \rm dex$. One arcsec tolerance cross-match yields 25,487 stars.

\subsection{RAVE}
\citet{bai19}  adopted the RAdial Velocity Experiment (RAVE, \citealt{rave2020}) catalog for their training sample. Our one arcsec tolerance cross-matching returns only 70 stars.  Moreover, the RAVE catalog includes [m/Fe] not [Fe/H]. The conversion from [m/Fe] to [Fe/H] could import bias or deviation when using an empirical formula. Therefore, the RAVE catalog is not included in our training sample.

\subsection{Normalization}
The training sample was constructed with J-PLUS DR1 and three spectral surveys, 12,931 stars from APOGEE, 216,114 from LAMOST LRS, 25,170 from LAMOST MRS, and 25,487 from SSPP. After removing the stars with missing photometric observations in one or more bands, there are 279,702 stars left. 

Prior to training our model, we centered all parameters $x_i$ in our training data (both the targets and input magnitudes) by subtracting their mean: $x_{i,t}=x_i - {\rm E}(x_i)$ \footnote{E stands for mathematical expectation.}. This normalization reduces the upper and lower bound of the parameter distribution and makes them all zero-mean distributed, which can accelerate the training speed \citep{shai14}.

During the prediction procedure, the query input magnitudes are centered in the same fashion, using the respective means from the training data, before being fed to the model. The model output $x_{i,p}^*$ is then returned to the true target parameter space following: $x_{i,p} = x_{i,p}^* + {\rm E}(x_i)$ to retrieve the prediction $x_{i,p}$.

We kept the parameters of the same stars from different spectroscopy surveys. For example, star A appears in both LAMOST and APOGEE, and the corresponding effective temperatures are $\rm Teff_1$ and $\rm Teff_2$. In our sample set, there would be a star $\rm A_1$ with effective temperature $\rm Teff_1$ and another star $\rm A_2$ with $\rm Teff_2$ . The stellar parameters are only merged when all parameters are the same. 

\subsection{Contours}
The prior distributions of the parameters in the training sample set are shown in Appendix \ref{appc}. The effective temperature is mostly distributed from $5000 \sim 6000K$, with log $g$ about $4 \sim 4.8$ and a metallicity similar to that of our Sun, which means that most stars are in their main sequence stage. There are also some giants with lower surface gravity ($2 \sim 3$), lower metallicity (<-1), and lower effective temperature ($4000 \sim 5000K$). 

However, the LAMOST stellar parameters are extracted from the spectra whose S/N is higher than 10. This criterion makes our LAMOST sample brighter than the limit magnitude of J-PLUS stars. Such disagreement would decrease the accuracy of prediction. To solve this problem, we applied the method to control the interpolation and extrapolation in \citet{wang21}. This method could increase the precision in prediction. To quantify the density of the sample space, we considered 12 sub-space combinations, which are (mag1, mag2, mag3), ... , (mag10, mag11, mag12), (mag11, mag12, mag1), and (mag12, mag1, mag2), and calculated the 95\% density contours to make clear how they cycle through the parameter space. The sample-dense space can be approached by the intersection space inside all these twelve density contours. Stars situated in all contours are assigned as interpolations.


\section{Methodology} \label{method}

\subsection{Support vector regression}
\label{svr}
The SVR \citep{svr15,svr97,cortes95} algorithm is a regression method based on the SVM (SVM, \citealt{cortes95, svm1992, crist00, shai14}). The data located inside the margin given by SVR algorithm are not involved in the calculation. These data gave us the flexibility to define the tolerance in regression. 

For a nonlinear regression problem, we first embedded the sample space into a new feature space with a higher dimension and transformed it into a linear regression problem. To transform the problem to the higher dimensional feature space, the SVR algorithm makes use of the kernel trick, which represents the inner product of the image of objects mapped from the sample space to the feature space with a kernel function.  This process can accelerate the calculation by doing calculations in a lower dimension. In the feature space, the algorithm then fits the data linearly.

Similar to the SVM algorithm, the SVR also has a strip area (called a tube). The difference between the two algorithms is that SVM maximizes the minimal distance between each sample to the strip, while the SVR considers how to minimize the width of the strip to put all samples in. The width of the strip area is called the margin. The vectors from the final line to the samples that finally determined the strip area are called the support vector. Details on the algorithm are given in \citet{svr04}.

The root mean square error (RMSE) is one of the most useful standards for evaluating the efficiency of regression. The RMSE of $\theta$ is given by $\sqrt{{\bf E}((\hat{\theta}-\theta)^2)}$, where $\hat{\theta}$ is the estimate of $\theta$. In regression, the RMSE is given by 
$${\rm RMSE}=\sqrt{\frac{1}{n}\sum_{t=1}^n (\hat{y_t}-y_t)^2 , } $$ where y is the stellar parameter.

\begin{figure}
 \includegraphics[width=0.45\textwidth]{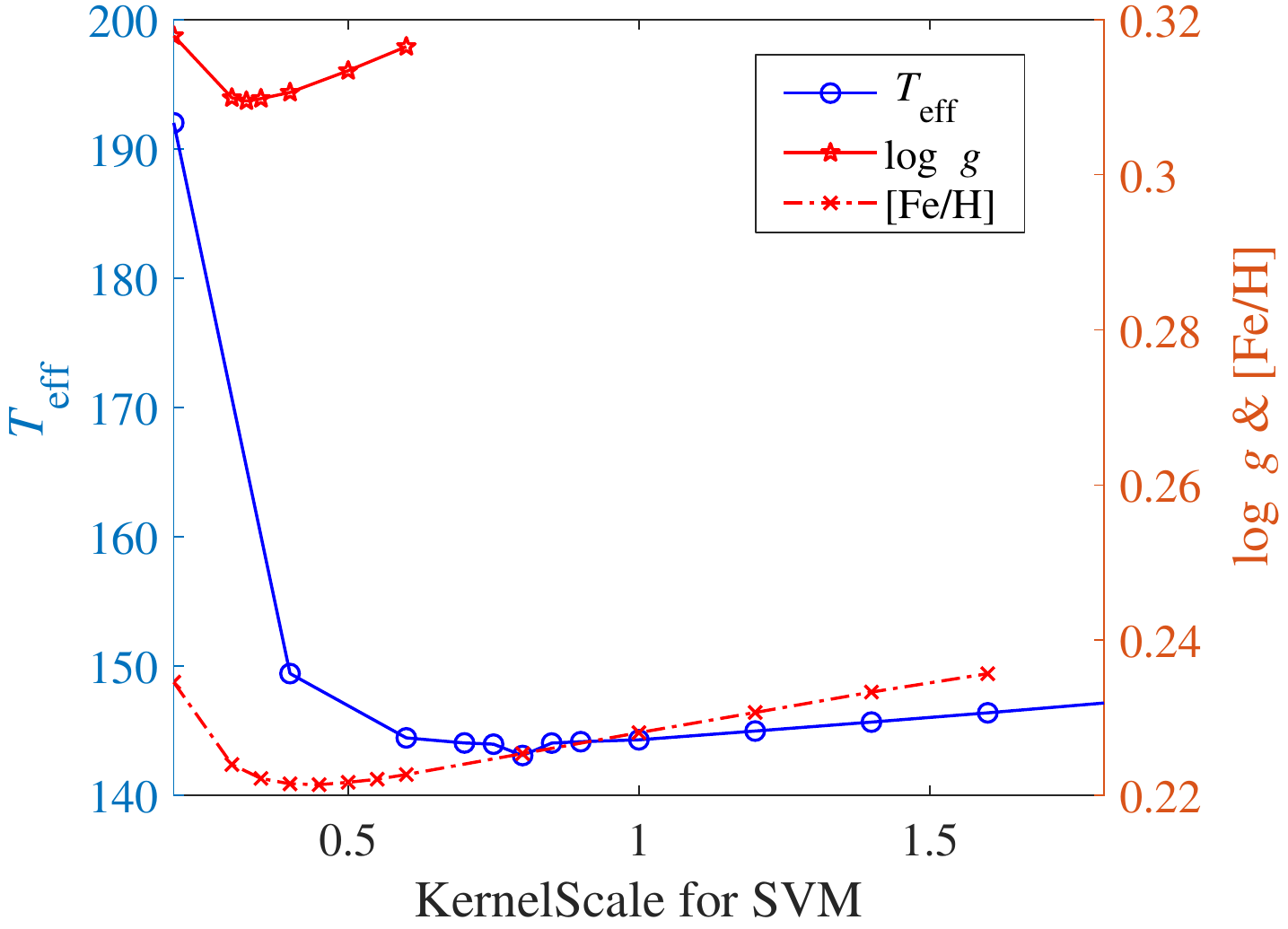}
\caption{RMSEs of the stellar parameters vs. the kernel scale. The blue line is the effective temperature. The surface gravity is the red solid line with the star markers, and the metallicity is the dotted line with the cross markers. \label{RMSEfig}}
\end{figure}

In the pre-training, we tested both magnitudes and colors as the input features with the same preprocessing. As a result, the RMSE of the magnitudes was lower than that of the mag(n-1)$-$mag(n) color under the default setting of SVR training (with Gaussian kernel 0.83). Then, we constructed independent models with different Gaussian kernel scales for each stellar parameter. We present the RMSE as a function of different kernel scales in Fig. \ref{RMSEfig}. We used the kernel scales with the lowest RMSEs for the training. We also tested the mag(n-1)$-$mag(n) and mag1$-$mag(n) color\citep{yang21}, and optimized the kernel scale with the Bayesian optimizer. Table \ref{rmses} presents the adopted kernel scales and the corresponding RMSEs, which are based on ten-fold validation. We adopted 0.8 as the kernel scale of $T_{\rm eff}$, 0.325 as the kernel scale of log $g$, and 0.45 as the kernel scale of [Fe/H].

\begin{table*}
\caption{Adopted kernel scales and RMSEs. \label{rmses}}
\centering
\begin{tabular}{l|cccccc}
\hline \hline
\noalign{\smallskip}
 & Kernel scale & Magnitude & Kernel Scale C1 & Color1 & Kernel Scale C2 & Color2 \\
\noalign{\smallskip}
\hline
\noalign{\smallskip}
$T_{\rm eff}$   & 0.8 & 144.059 & 1.01 & 165.71 & 5.77 & 195.77 \\
log $g$ & 0.325 & 0.310 & 1.646 & 0.312 & 0.415 & 0.316 \\
$\rm [Fe/H]$ & 0.45 & 0.221 & 1.944 & 0.233 & 0.440 & 0.233 \\
\hline
\end{tabular}
  \tablefoot{The  "Magnitude" column lists the RMSE of using magnitudes as features, while the column of "Color1" is the mag(n-1)$-$mag(n) color and "Color2", presents the optimized mag1$-$mag(n) color.} The 'Kernel Scale' columns match the magnitudes, color1, and color2, respectively. 
\end{table*}

\subsection{Data enhancement}
\label{dataenhance}
We duplicated the sample into 80 sets to enhance the cardinality of data for each stellar parameter. In image recognition, the method is usually carried out by rotating or mirroring the image and producing more samples for the training. We applied this method here by generating random training sets to fully use the uncertainties of both the magnitudes and the stellar parameters. 

For each star in our training sample, after the centering process, we generated 80 stars with its 12 magnitudes, 3 stellar parameters, and 15 corresponding uncertainties. These 80 stars have Gaussian distributions of all the magnitudes and the stellar parameters following $\widetilde{x_{i,t}} = x_{i,t} + \sigma(x_{i,t})e_i$, where $\sigma(x_{i,t})$ is the uncertainty of $x_{i,t}$, and $\sigma_i \sim {\rm N}(0,1)$.  For example, the mean temperature is 6,000K and the $\sigma$ is 100K for the star with $T_{\rm eff}=6000 \pm 100$K. The simulation process did not change the original distribution or introduce new errors (Appendix \ref{appc}).

Each of the 80 constructed samples has 251,732 stars. We trained 80 models using the SVR algorithm with the same kernel scale in Sect. \ref{svr} for each stellar parameter, which resulted in a total of 240 different models (same scheme but different sample set) for $T_{\rm eff}$, log $g,$ and [Fe/H]. We then used the corresponding 80 regression models to retrieve a distribution of 80 predictions for a given query observation. Then, we used Gaussian  functions to fit these predicted distributions to obtain the centers and standard deviations. 

The data enhancement method takes the uncertainties of stellar parameters into account, which are not considered in other studies. The absence of these uncertainties could cause unexpected errors in the prediction since the spectral precision is not involved in the model construction.


\subsection{Model validation}
\label{BT}
We applied model validation to illustrate the effectiveness and to avoid potential overfitting. A widespread method for model validation is blind tests, which can reveal potential overfitting and quantify a model's ability to generalize to new, previously unseen data. We reserved 10\%, 27,970 stars as our blind test sample, producing 80 sets of tests for each stellar parameter according to our data enhancement procedure. Similar to the construction of the sample set, we did not use any criterion to preselected data in order to avoid the selection effect.

We provide the RMSE and normalized RMSE (NRMSE) of our blind test for both interpolation and extrapolation in Table \ref{btrmse}, where the ${\rm NRMSE} = \frac{\sqrt{\sum_{i=1}^n(\hat{x_i}-x_i)^2}}{max(x_i)-min(x_i)}$. The prediction has the best result in effective temperature and the least in surface gravity. The extrapolation contains 445 stars, which suffered from a large deviation in the blind test. The result shows that the training model is better restricted in the space in which the samples are densely distributed. 

\begin{table*} 
\centering
\caption{Blind test RMSEs \label{btrmse}}
\begin{tabular}{c|cc|cc|cc}
\hline \hline
\noalign{\smallskip}
  & BT RMSE & BT NRMSE & Inter RMSE & Inter NRMSE & Extra RMSE & Extra NRMSE \\
\noalign{\smallskip}
\hline
\noalign{\smallskip}
$T_{\rm eff}$ & 159.6 & 0.0283 & 150.2 &  0.0266 & 454.0 & 0.0805 \\
log $g$ & 0.3453 & 0.0677 &  0.3380 &  0.0663 & 0.6543 & 0.1283 \\
$\rm [Fe/H]$ & 0.2503 & 0.0484 &  0.2428 &  0.0470 & 0.5402 & 0.1045 \\
\hline  
    \end{tabular}
    \tablefoot{Inter stands for the interpolation and Extra stands for the extrapolation.}
\end{table*}

\begin{figure}
\includegraphics[width=0.55\textwidth]{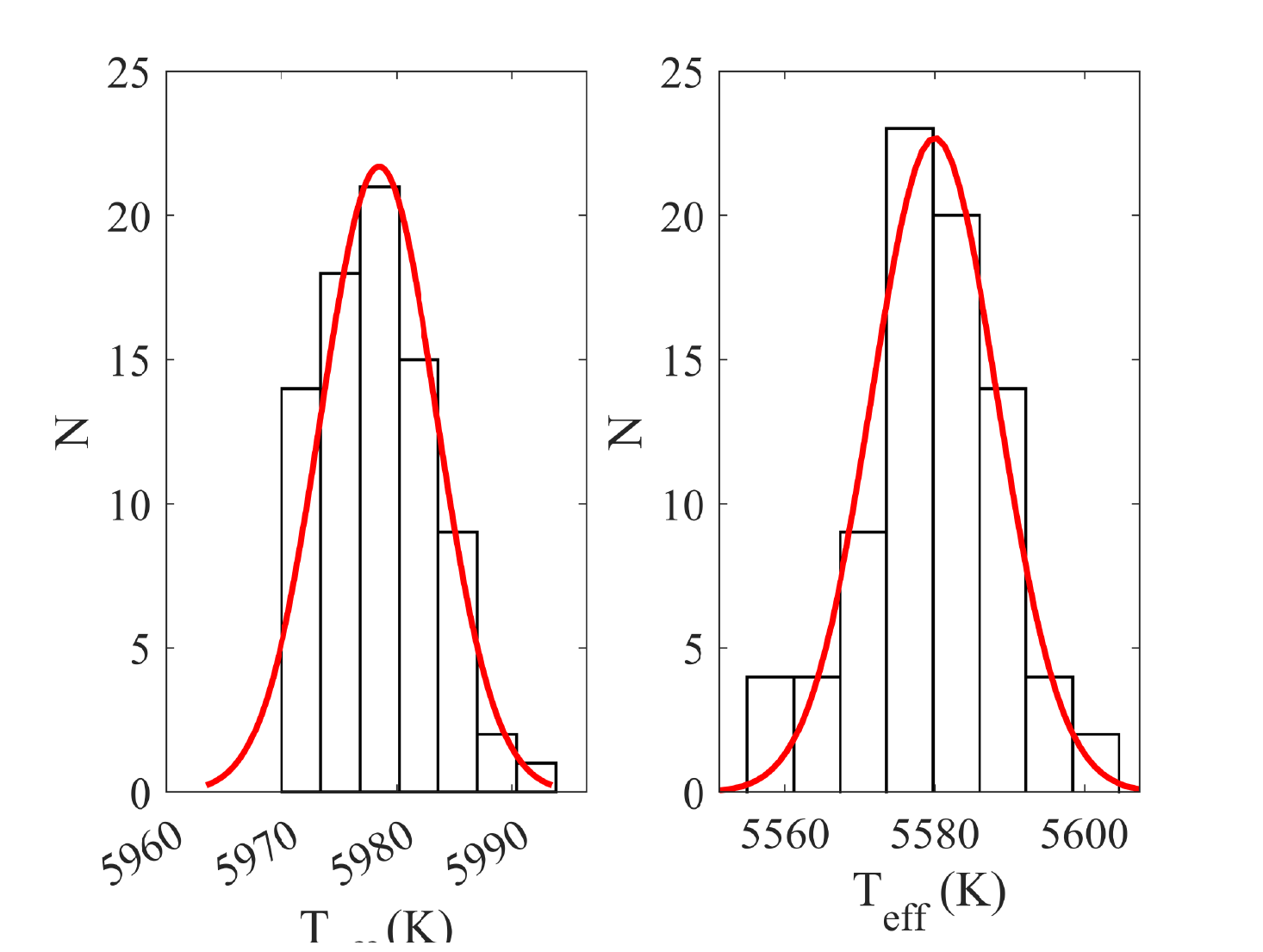}
\caption{Two examples of the $T_{\rm eff}$ distributions. The left panel is the distribution of $T_{\rm eff}$ of 26109-16130 in the blind test. The right panel is the distribution of  26016-295 in the prediction. The red lines are Gaussian fits. \label{Fig45}}
\end{figure}

Figure \ref{Fig45} shows the $T_{\rm eff}$ distribution of one star (26109-16130) in our blind tests. After applying the Lilliefors test \citep{Lill67}, the $p$-value of the distribution is 0.375, greater than 0.05, implying that it may follow a Gaussian distribution. However, this may not generally be the case (Section \ref{dis5.4}). The results of the blind tests for log $g$ and [Fe/H] are presented in Appendix \ref{appb}.

\section{Result} \label{result}
We  then applied 240 stellar parameter models to the interpolation and extrapolation catalogs from \citet{wang21}, separating our predictions into two catalogs. There are 2,493,424 objects in the interpolation catalog and 233,924 objects in the extrapolation catalog. In the classification, the authors restrained the 12-band magnitudes and constructed 12 contours based on the sample set distribution. All classified objects were assigned to the area inside or outside the contours and categorized as either interpolation or extrapolation. The distribution of inside objects is similar to the samples and therefore they have higher accuracy. In short, the stars in the interpolation catalog have a higher classification confidence than those in the extrapolation catalog. 

We also applied this method in the prediction. There are 1,898,154 stars situated in our training set contours, and 595,270 stars are categorized as extrapolation in the classification interpolation catalog. In the classification extrapolation catalog, there are 13,274 interpolations and 220,650 extrapolations.  We then input the 12-band magnitudes of each star into the $3 \times 80$ models and obtained the stellar parameters of the interpolation and extrapolation catalogs. An example of the Gaussian fit is given in the right panel of Fig. \ref{Fig45}. Table \ref{interp} is a stellar parameter catalog of the interpolation catalog.

\begin{table*}
\centering
\caption{Parameter catalog of the interpolation \label{interp}}
\begin{tabular}{lcccccccc}
\hline \hline
\noalign{\smallskip}
ID & R.A. (deg) & Dec. (deg) & Teff  (K) & $\Delta$Teff (K) & log $g$ (dex) & $\Delta$log $g$ (dex) &  [Fe/H] (dex) & $\Delta$[Fe/H] (dex) \\
\noalign{\smallskip}
\hline
\noalign{\smallskip}
26016-5* & 255.4683 & 22.7600 & 6831 & 101 & 3.71 & 0.14 & $-$1.20 & 0.10 \\
26016-15 & 255.3051 & 22.7608 & 5950 & 15 & 4.15 & 0.07 & $-$0.36 & 0.04 \\
26016-16* & 255.3675 & 22.7611 & 5659 & 127 & 3.89 & 0.08 & $-$0.79 & 0.10 \\
26016-50* & 255.1639 & 22.7644 & 6034 & 80 & 3.71 & 0.13 & $-$1.19 & 0.17 \\
26016-55* & 255.6631 & 22.7617 & 4061 & 120 & 3.93 & 0.06 & $-$0.67 & 0.09 \\
26016-64* & 254.7562 & 22.7650 & 6175 & 58 & 3.73 & 0.13 & $-$1.18 & 0.15 \\
26016-68 & 255.7539 & 22.7623 & 6467 & 13 & 3.59 &  0.07 & $-$1.63 & 0.06 \\
26016-70 & 255.3743 & 22.7649 & 6137 & 41 & 3.54 & 0.14 & $-$1.62 & 0.10 \\
26016-74 & 255.5894 & 22.7630 & 6162 & 14 & 3.87 & 0.06 & $-$1.04 & 0.05 \\
26016-84* & 254.8371 & 22.7663 & 5532 & 151 & 3.84 & 0.04 & $-$0.76 & 0.09 \\
\hline
\end{tabular}
\tablefoot{ID stands for the J-PLUS ID. RA and Dec. are the right ascension and the declination in degree (J2000). The extrapolation stars have an asterisk on their ID, which indicates that their stellar parameters are not very reliable. The $\Delta$ shows the uncertainty. }
\end{table*}



Both median and mean values of the uncertainties are presented in Table \ref{respipcmp}. The difference in mean versus median shows the existence of a few stars with a large residual that raise the mean in both the interpolation and extrapolation catalogs. These uncertainties were decreased by about $3\% \sim 5\%$ by applying the contours to constrain the prediction sample. The uncertainties of the prediction are smaller than the residuals among different pipelines, implying that our regressions have fairly good precisions (see Sect. \ref{solpipe}).  This also implies that the uncertainties are mainly generated from the pipeline difference and observation uncertainties.

\begin{table*} 
 \centering
\caption{Uncertainties of the classification interpolation and the extrapolation catalogs \label{respipcmp}}
    \begin{tabular}{c|cccc}
\hline \hline
\noalign{\smallskip}
  & Inter median & Inter mean & Extra median & Extra mean \\
\noalign{\smallskip}
\hline
\noalign{\smallskip}
$T_{\rm eff}$  (K) & 14.0 & 34.7 & 59.5 & 71.6  \\
log $g$ (dex)& 0.0689 & 0.0780 & 0.0268 & 0.0515 \\
$\rm [Fe/H]$ (dex) & 0.0566 & 0.0677 & 0.0187 & 0.0477\\
\hline  
    \end{tabular}
    \tablefoot{Inter stands for the classification interpolation catalog and Extra stands for the classification extrapolation catalog. We present both median and mean values in this table.}
\end{table*}

\section{Discussion} \label{discuss}
\subsection{Training and blind test}
\label{d1}

We tested three methods to regress the parameters. In the first method, we trained a single SVR model using the magnitude uncertainties as weights, but excluding the stellar parameter uncertainties. This method aims at making a comparison with our multi-model simulation. The weight is given by $w_i=lg\sum_{j=1}^12(\frac{1}{e_{ij}^2})$ for an object $x_i,$ where the $e_{ij}$ shows the magnitude uncertainty of mag$j$. We adopted a logarithmic scale to reduce potential steep cliffs in the feature space \citep{wang21}. In the second and third tests, we randomly constructed 80 training sets for each stellar parameter. The difference between the two tests is whether LAMOST MRS were included in the training sample. We constructed the second test to show the generalization ability of extrapolating data. The first and second tests use LAMOST MRS as a blind test, while the third one includes all the stars in the training sample and holds $10\%$ for the blind test. The third method is the final model we applied. It has lower RMSEs than the single model. We present the results of these three tests in Table \ref{bt3}. 

\begin{table}
 \centering
\caption{RMSEs of the blind tests in the three tests \label{bt3}}
\begin{tabular}{lccc}
\hline \hline
\noalign{\smallskip}
Method & $T_{\rm eff}$ (K) & log $g$ (dex) & [Fe/H] (dex) \\
\noalign{\smallskip}
\hline
\noalign{\smallskip}
  Single Model  & 239.9 & 0.373 & 0.296 \\
  Generalization Ability & 239.6 & 0.400 & 0.313 \\
  Final Model & 159.6 & 0.345 & 0.250  \\
  \hline
\end{tabular}

\end{table}

\subsection{Multi-model approach}
\label{d2}

\begin{table*}
\centering
\caption{Gaussian fits of the residuals for our blind tests and other works. \label{btp} }
\begin{tabular}{lccccccc}
\hline \hline
\noalign{\smallskip}
Parameter & $\mu$(BT) & $\Delta$(BT) & $\Delta$(Train) & $\Delta$(Bai) & $\Delta$(Yang) & $\mu$(Galarza) & $\Delta$(Galarza) \\
\noalign{\smallskip}
\hline
\noalign{\smallskip}
$T_{\rm eff}$ (K) & 2.3125 & 159.5216 & 84.6750 & 100 $\sim$ 200 & $\sim$ 55 & 41 & 61 \\
log $g$ (dex) & 0.0520 & 0.3413 & 0.1200 & & 0.15 & 0.11 & 0.17\\
$\rm [Fe/H]$ (dex) & 0.0366 & 0.2476 & 0.0660 & & 0.07 & 0.09 & 0.14\\
\hline
\end{tabular}
\tablefoot{BT stands for the blind test. Bai, Yang, and Galarza present the uncertainties from \citealt{bai19}, \citealt{yang21}, and \citealt{gala22}, respectively. \citealt{bai19} only contains effective temperature regression. $\mu$ is the center of the Gaussian fit and $\Delta$ is the standard deviation. We present the median values of the training uncertainties.}
\end{table*}

\citet{bai19} constructed a single model to train the stellar parameters and enlarge the sample to constrain the discrepancy caused by different pipelines. The uncertainty of the stellar parameter is not included in their calculation. \citet{yang21} also constructed one deep learning artificial neural network (ANN) --- $CSNet$ --- and gained stellar parameters with a precision of $\sim 55 \rm K$ in effective temperature, 0.15 dex in log $g,$ and 0.07 dex in [Fe/H]. \citet{gala22} presented a multi-regressior pipeline, SPEEM, based on three single XGB algorithms and validated the result with SSPP. The errors for each stellar parameter ($T_{\rm eff}$, log $g$, [Fe/H]) are $41 \pm 61$ K, $0.11 \pm 0.17$ dex, and $0.09 \pm 0.14$ dex, respectively. More information is presented in Table \ref{btp}. The precision of all these single models is similar to our results.

In our study, the RMSE of a single model is similar to the RMSEs of 80 models, indicating that the multi-model approach performs similarly to the weight-controlled model. We can incorporate the uncertainties of both features and target parameters, which is the main advantage of our approach. Furthermore, if a certain photometric measurement suffers a higher uncertainty, its contribution to the regression would decrease significantly, independently of the uncertainties in the remaining 11 measurements. This could introduce biases into the models and result in inaccurate predictions. Such bias is avoided in our results, since all the uncertainties of the 12-band magnitudes are included in the model construction.

We also find that the uncertainties of the stellar parameters are larger than the blind test's RMSEs. If these uncertainties are not considered in the calculation, the error from the spectral fitting pipeline would not propagate to the final regression. The uncertainties of the prediction would be highly underestimated in this case.

\subsection{Pipelines}\label{solpipe}

Different pipelines result in different stellar parameters (Fig. \ref{Fig56}, see more in Appendix \ref{appa}), and we adopted all parameters from all pipelines. We selected the stars that have parameters from multiple spectroscopy surveys and show the differences and variances in Table \ref{pipelinetab}. In Table \ref{btp} and Table \ref{pipelinetab}, the differences in the blind test are smaller than those caused by pipelines. When the stellar parameters from different pipelines are included in our training, the regression models are unbiased among these different pipelines and  can thus be used to provide general predictions from different surveys. Including a larger variety of surveys in the training data will improve the generalization ability. 

\begin{figure}
  \flushright
  \includegraphics[width=0.5\textwidth]{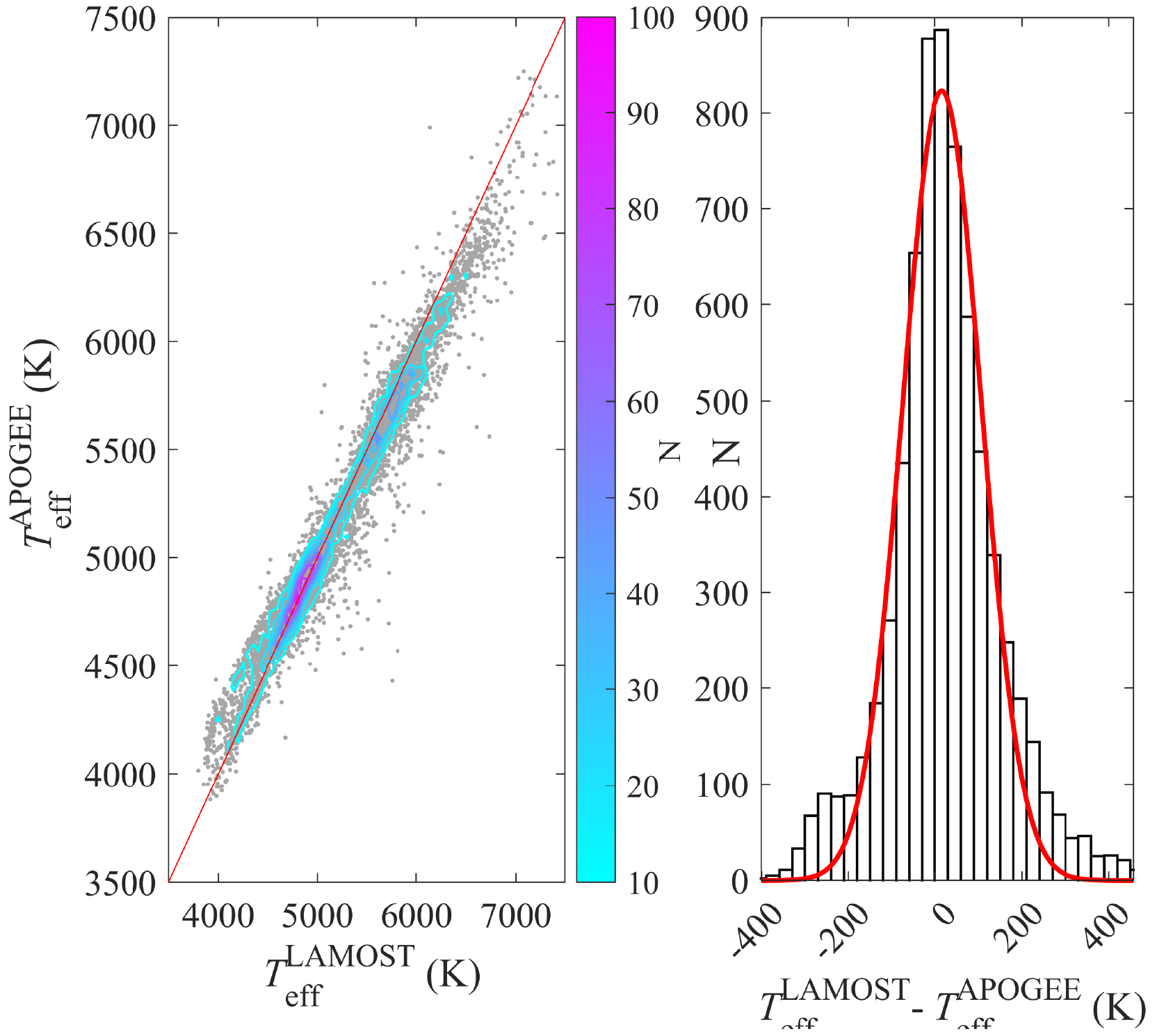}
  \caption{One-to-one correlation of $T_{\rm eff}$ between APOGEE and LAMOST. Data points are in gray. The color bar is the density contours. The Gaussian fit of the residual is presented in the right panel. The mean of this normal distribution is -27.7873, and its standard deviation is 143.2126.\label{Fig56}}
\end{figure}



\begin{table}
 \centering
\caption{Differences among pipelines \label{pipelinetab}}
\begin{tabular}{c|cccc}
\hline\hline
\noalign{\smallskip}
Stellar parameter & pipelines & overlap & $\mu$ & $\sigma$ \\
\noalign{\smallskip}
\hline
\noalign{\smallskip}
        & L$-$A & 6,956 & 27.8 & 143.2 \\
  $T_{\rm eff}$ & L$-$S & 3,318 & $-$88.4 & 143.1 \\
        & S$-$A & 64 & $-$112.3 & 149.7  \\
        \noalign{\smallskip}
  \hline
  \noalign{\smallskip}
    & L$-$A &  & 0.048 & 0.191 \\
  log $g$ & L$-$S &  & 0.206 & 0.353 \\
    & S$-$A &  & $-$0.150 & 0.228  \\
        \noalign{\smallskip}
  \hline
  \noalign{\smallskip}
    & L$-$A &  & 0.008 & 0.099 \\
  ${\rm [Fe/H]}$ & L$-$S &  & 0.068 & 0.184 \\
    & S$-$A &  & $-$0.111 & 0.144  \\
    \noalign{\smallskip}
    \hline
\end{tabular}
  \tablefoot{The mean and standard variance of the difference of pipelines. In this table, the "L" stands for LAMOST, while "S" for SSPP and "A" for APOGEE. Overlap stands for the number of cross-match overlapping between these catalogs. The unit of effective temperature is in K, and dex for log $g$ and [Fe/H].}
\end{table}


Other studies usually adopt data from one survey to avoid biases among different pipelines \citep{bu20,lu15}. Another method is to set a priority for each catalog. Some studies have concluded that a larger sample size can decrease the bias caused by pipelines, for example \citet{bai19}. 

Although a single catalog and priority-based methods do not suffer the systematic error caused by pipelines, the models built from them probably do not perform well for other surveys. Our training sample contains different surveys, and their biases propagate to the final regression models. Therefore, our models would have a wider application.

One catalog-based method may be good for regression of the same catalog, but J-PLUS is a more general catalog with a large sky coverage and many photometric wavebands. J-PLUS requires models from hybrid samples for general applications. 

There may not exist a regressor that can properly fit the data for all the surveys; this is analogous to the no-free-lunch theorem, which says that a machine-learning algorithm that can solve every problem does not exist (\citealt{shai14}). This implies that the prediction would be biased if we use single catalog or priority-based methods to regress the data in another catalog. In machine learning, a larger amount of data does not always work, but higher diversity does \citep{wang21}.

\subsection{Distribution of stellar parameters} 
\label{dis5.4}

While our data augmentation process employs Gaussian-based uncertainty sampling, the predicted stellar parameters may not necessarily follow a Gaussian distribution. The SVR embeds the samples to a higher-dimensional feature space by using a nonlinear embeding mapping. The nonlinear mapping may not hold the distribution of sample to the feature space.

To quantify whether our approach preserves the shape of the distribution, we applied the Lilliefors test \citep{Lill67} to the blind test samples. About $19.01\%$ of the $T_{\rm eff}$, $19.49\%$ for log $g,$ and $19.39\%$ for [Fe/H] pass the  test with a significance level of 0.05. When the level decreases to 0.01, the number increases to $39.94\%$ of the $T_{\rm eff}$, $41.40\%$ for log $g$ and $40.54\%$ for [Fe/H]. Given the low number statistics in our predicted distributions of only 80 samples, it is possible that this test underestimates the true fraction of Gaussian distributed predictions. To estimate the influence of the number of samples, we constructed 27,970 sets (which was the same size as the blind test size) of random numbers based on $N(100,25)$ and find that about $94.87\%$ passed the test with a significance level of 0.05. When we change the number of sets to 800, the result increases to $95.11\%$. The difference is less than $1\%$. Therefore,\ there is no evidence that the resulted stellar parameters follow a Gaussian distribution.

\subsection{Comparison with \citet{yang21}}
\label{dis5.5}
To determine the robustness of our predictions, we performed a comparison with the stellar parameter catalog from \citet{yang21} \footnote{available at \url{http://www.j-plus.es/ancillarydata/dr1_stellar_parameters_elemental_abundances}}. We selected the reliable stars in both catalogs, which were stars in the interpolation catalog and the stars with stellar parameter \texttt{FLAG}=0 in \citet{yang21}. The cross-match yielded 2,008,654 stars. The average stellar parameter differences are -247.9764 K for $T_{\rm eff}$, -0.1984 dex for log $g$, and -0.0998 dex for [Fe/H]. There are some extreme values in \citet{yang21}, and some of them are unreliable. For example, the maximum and minimum value of surface gravity are 882.44 dex and -237.24 dex and 460.39 dex and -567.16 dex for [Fe/H]. These stars may be situated at the edge of the feature space and, thus, are subject to overfitting by the ANN model.

\begin{figure*}
    \centering
    \includegraphics[width=0.45\textwidth]{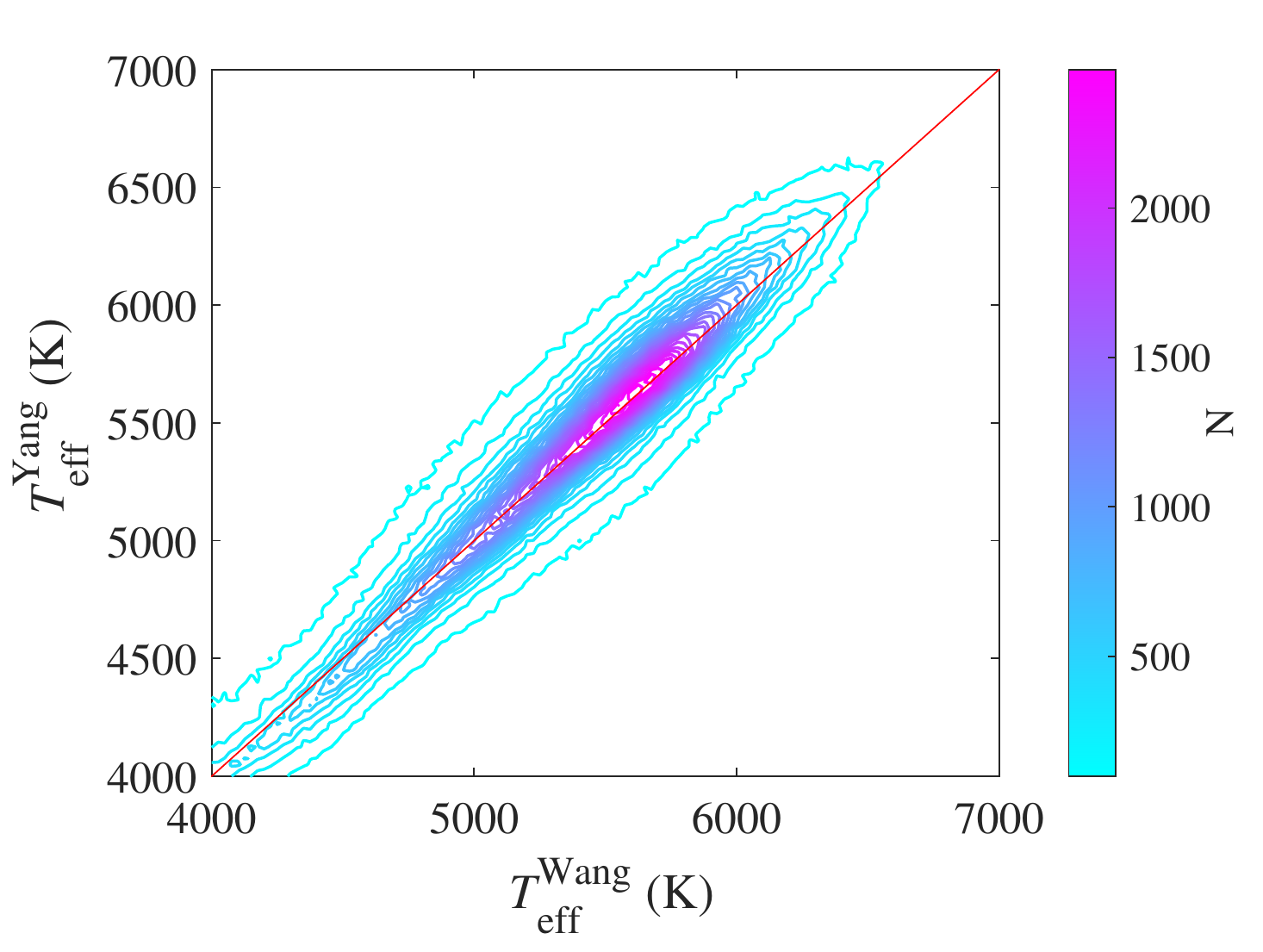}
    \includegraphics[width=0.4\textwidth]{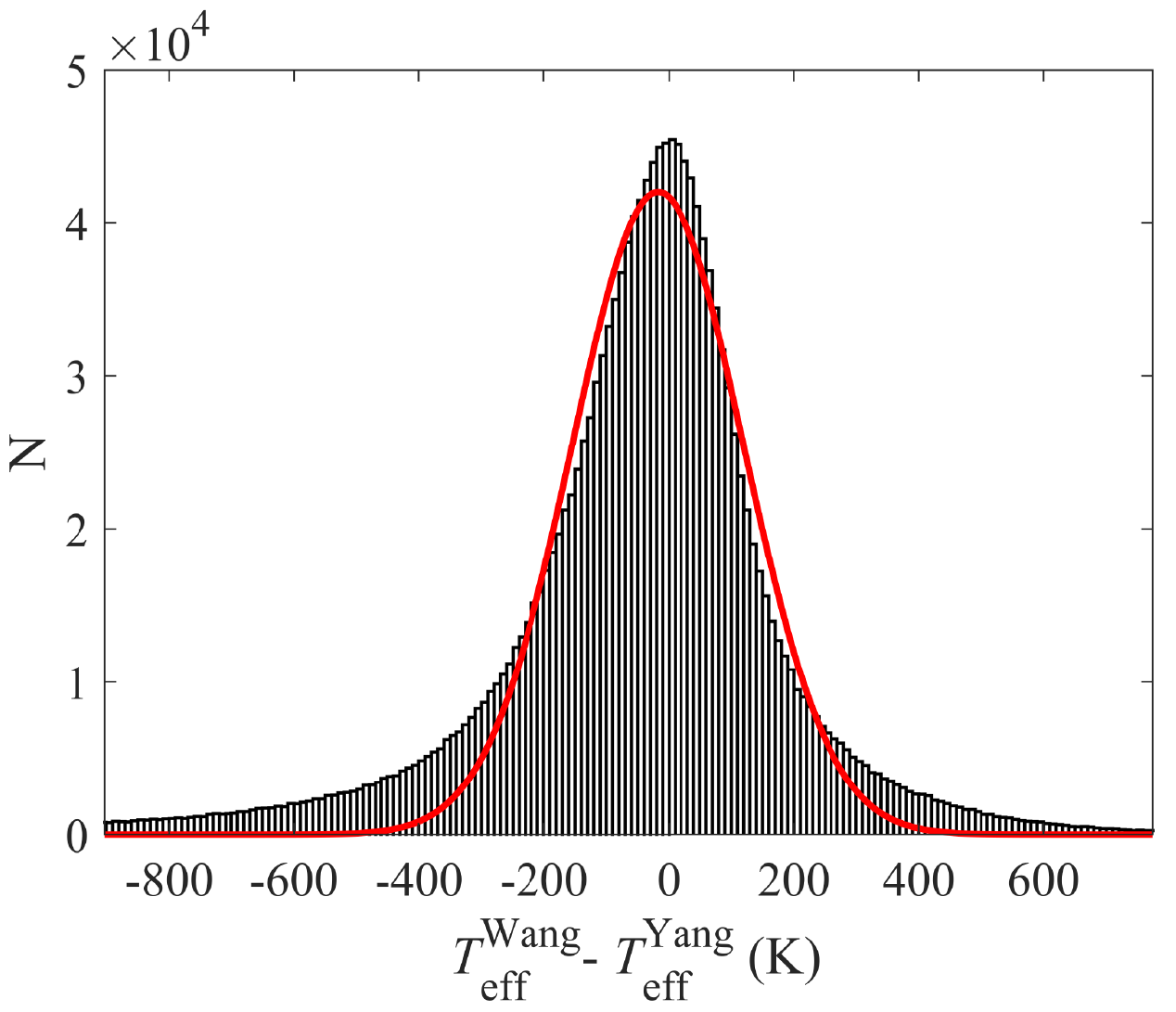}
    \includegraphics[width=0.45\textwidth]{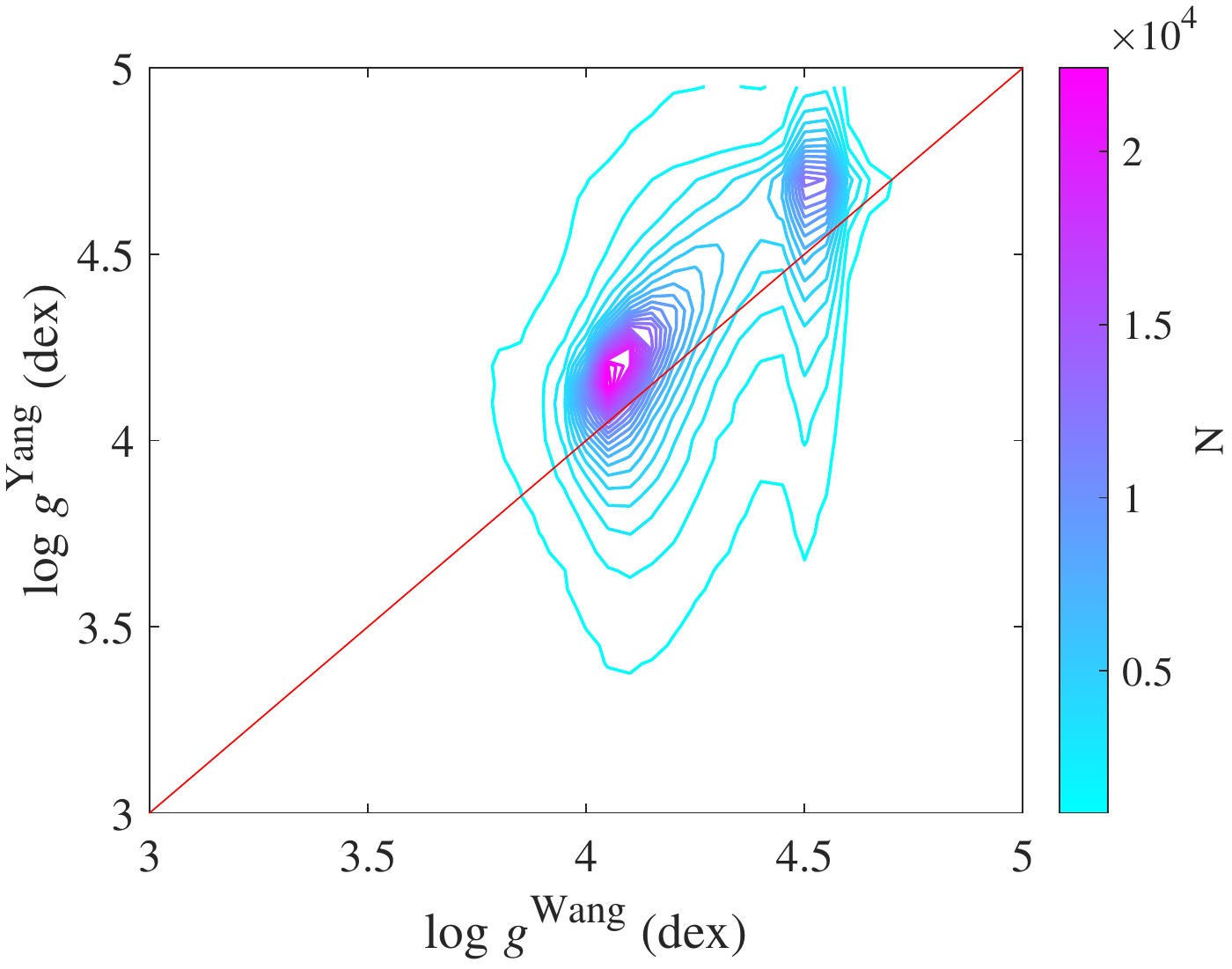}
    \includegraphics[width=0.4\textwidth]{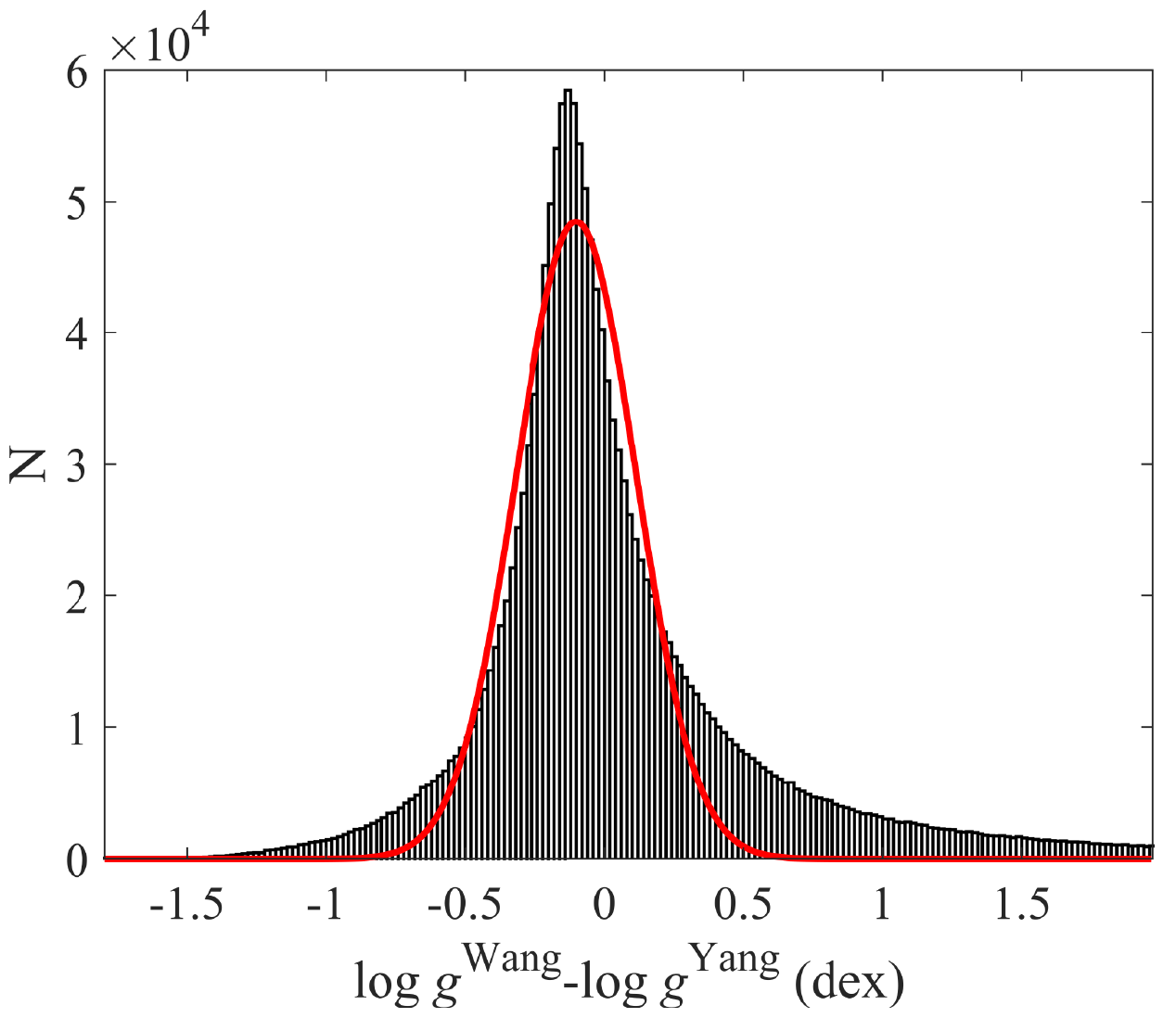}
    \includegraphics[width=0.45\textwidth]{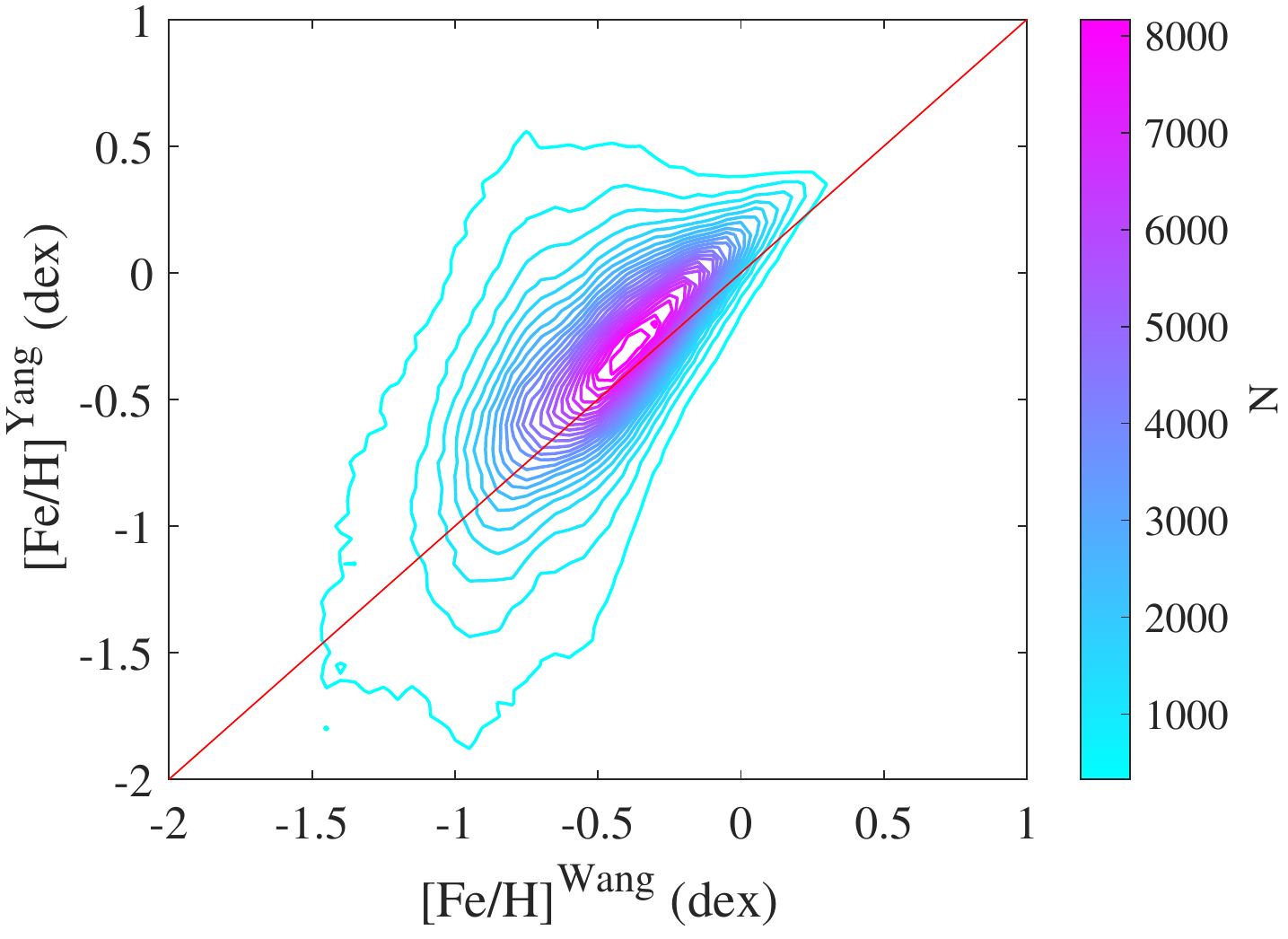}
    \includegraphics[width=0.45\textwidth]{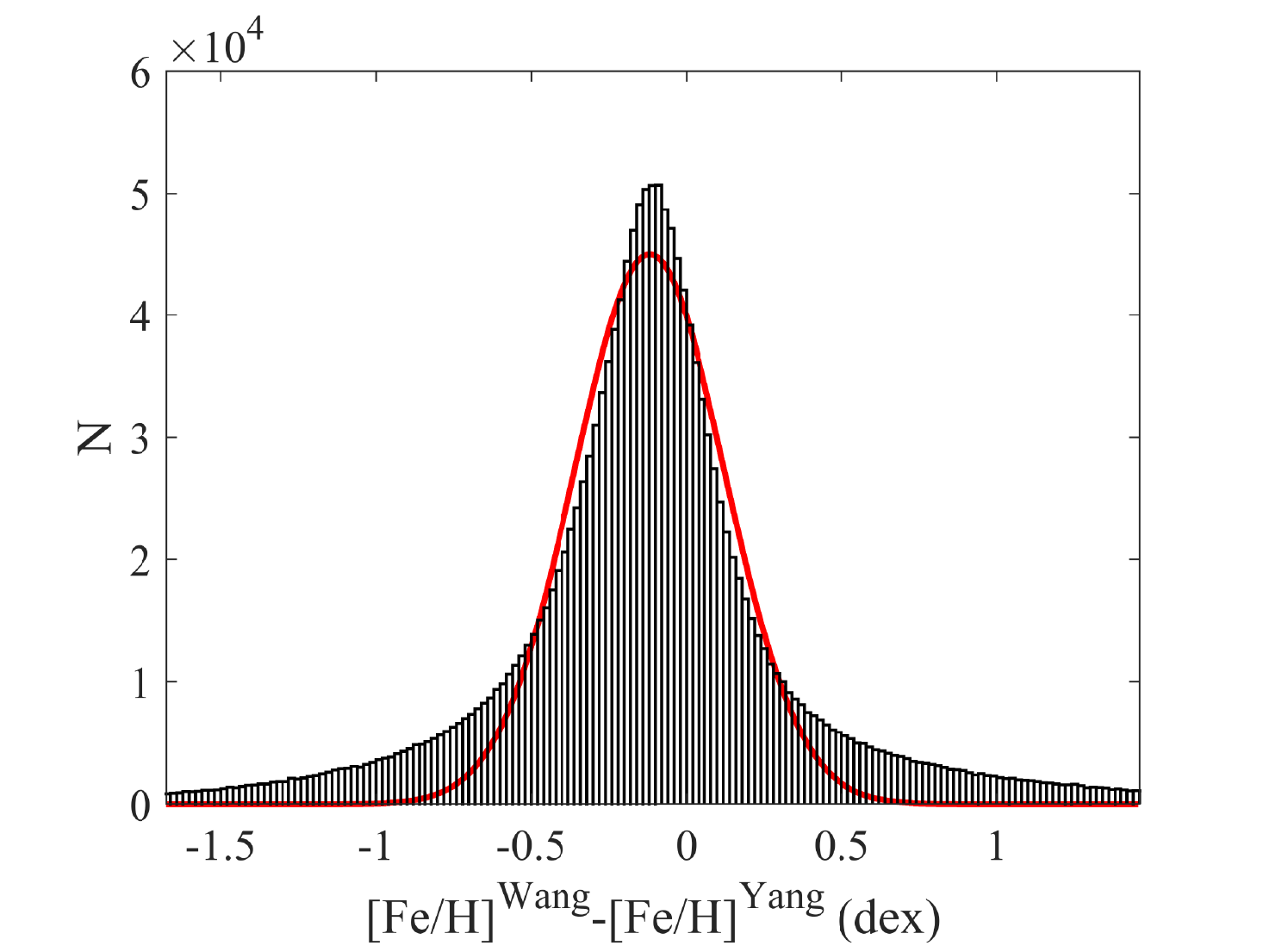}
    \caption{Comparison with \citet{yang21}. The three panels are $T_{\rm eff}$, log $g$, and [Fe/H], from top to bottom. Only the density contours are shown in these figures. \label{yangetal}}
\end{figure*}

The training sample ranges of \citet{yang21} are about 4000 $\sim$ 7500 K for $T_{\rm eff}$, 0 $\sim$ 5 dex for log $g$, and -3 $\sim$ 1 dex for [Fe/H]. We restricted their results to these ranges and obtain 1,663,053 stars. The average differences decrease to -63.6817 K for $T_{\rm eff}$, 0.0886 dex for log $g$, and -0.1069 dex for [Fe/H] (Fig. \ref{yangetal}). These differences are similar to our blind test uncertainties. The main difference between our work and \citet{yang21} is the control of the extrapolations. These extreme values might fall into the extrapolation category and suffer from overfitting. The training sample of \citet{yang21} is not an official LAMOST release, and the training sample difference may also be a minor reason for such a stellar parameter difference.

\section{Conclusion} \label{conclusion}
In this work, we predicted stellar parameters, the effective temperature $T_{\rm eff}$, the surface gravity log $g,$ and the metallicity [Fe/H] for J-PLUS DR1 by using $3 \times 80 $ SVR models. We chose stars from spectra-based surveys (LAMOST, APOGEE, and SEGUE) to construct our training sample in order to improve the reliability of the sample and gained 279,702 stars from the cross-match. We normalized the features of these stars by subtracting their average in order to accelerate the calculation. We held out ten percent of the set  for the blind test and the remaining 251,732 stars were used for training. We used 12 three-dimensional density contours to distinguish the prediction reliability for a given star. We presented two catalogs that correspond to the classification catalogs of \citet{wang21}. For the classification interpolation catalog, we present 1,898,154 stars with interpolated stellar parameters and 595,270 stars with extrapolations in our regression. For their classification extrapolation catalog, on the other hand, we present 13,274 interpolated and 220,650 extrapolated predictions with our SVR. Regarding the construction of the models, multi-model simulations give better results than the single model.  The RMSEs of our validations are 159.6 for $T_{\rm eff}$, 0.3453 for log $g,$ and 0.2502 for $[Fe/H]$. Using different catalogs as a sample set may increase the generalization ability of our model. Lastly, we compared our results to \citet{yang21} and find a decent agreement with their work, with discrepancies comparable to the RMSEs that we found in our blind test.

\FloatBarrier

\begin{acknowledgements}
This work was supported by the National Natural Science Foundation of China (NSFC) through grants NSFC-11988101/11973054/11933004 and the National Programs on Key Research and Development Project (grant No.2019YFA0405504 and 2019YFA0405000). Strategic Priority Program of the Chinese Academy of Sciences under grant number XDB41000000. 
H, B, Yuan, acknowledge funding from National Natural Science Foundation of China (NSFC) through grants NSFC-12173007.
JAFO acknowledges the financial support from the Spanish Ministry of Science and Innovation and the European Union -- NextGenerationEU through the Recovery and Resilience Facility project ICTS-MRR-2021-03-CEFCA.
F.J.E. acknowledges financial support by the Spanish grant PGC2018-101950-B-I00 and MDM-2017-0737 at Centro de Astrobiología (CSIC-INTA), Unidad de Excelencia María de Maeztu.
We acknowledge the science research grants from the China Manned Space Project with NO.CMS-CSST-2021-B07. \\

Based on observations made with the JAST80 telescope at the Observatorio Astrofísico de
Javalambre (OAJ), in Teruel, owned, managed, and operated by the Centro de Estudios de Física del Cosmos de Aragón (CEFCA). We acknowledge the OAJ Data Processing and Archiving Unit (UPAD) for reducing the OAJ data used in this work.
Funding for the J-PLUS Project has been provided by the Governments of Spain and Aragón through the Fondo de Inversiones de Teruel; the Aragón Government through the Research Groups E96, E103, and E16\_17R; the Spanish Ministry of Science, Innovation and Universities (MCIU/AEI/FEDER, UE) with grants PGC2018-097585-B-C21 and PGC2018-097585-B-C22; the Spanish Ministry of Economy and Competitiveness (MINECO) under AYA2015-66211-C2-1-P, AYA2015-66211-C2-2, AYA2012-30789, and ICTS-2009-14; and European FEDER funding (FCDD10-4E-867, FCDD13-4E-2685). The Brazilian agencies FINEP, FAPESP, and the National Observatory of Brazil have also contributed to this project.

Guoshoujing Telescope (the Large Sky Area Multi-Object Fiber Spectroscopic Telescope LAMOST) is a National Major Scientific Project built by the Chinese Academy of Sciences. Funding for the project has been provided by the National Development and Reform Commission. LAMOST is operated and managed by the National Astronomical Observatories, Chinese Academy of Sciences.

Funding for the Sloan Digital Sky Survey IV has been provided by the Alfred P. Sloan Foundation, the U.S. Department of Energy Office of Science, and the Participating Institutions. SDSS-IV acknowledges support and resources from the Center for High-Performance Computing at the University of Utah. The SDSS website is \url{http://www.sdss.org/}.
SDSS-IV is managed by the Astrophysical Research Consortium for the Participating Institutions of the SDSS Collaboration including the Brazilian Participation Group, the Carnegie Institution for Science, Carnegie Mellon University, the Chilean Participation Group, the French Participation Group, Harvard-Smithsonian Center for Astrophysics, Instituto de Astrofísica de Canarias, The Johns Hopkins University, Kavli Institute for the Physics and Mathematics of the Universe (IPMU)/University of Tokyo, Lawrence Berkeley National Laboratory, Leibniz Institut für Astrophysik Potsdam (AIP), Max-Planck-Institut für Astronomie (MPIA Heidelberg), Max- Planck-Institut für Astrophysik (MPA Garching), Max-Planck-Institut für Extraterrestrische Physik (MPE), National Astronomical Observatories of China, New Mexico State University, New York University, University of Notre Dame, Observatário Nacional/MCTI, The Ohio State University, Pennsylvania State University, Shanghai Astronomical Observatory, United Kingdom Participation Group, Universidad Nacional Autónoma de México, University of Arizona, University of Colorado Boulder, University of Oxford, University of Portsmouth, University of Utah, University of Virginia, University of Washington, University of Wisconsin, Vanderbilt University, and Yale University.

\end{acknowledgements}

\bibliographystyle{aa}
\bibliography{ref}



\begin{appendix}

\section{Sample} \label{appc}
We present the distribution of a sample set in this section. The standard sample set distributions are shown in red blocks. We then split the sample set into 80 new randomly constructed sample sets to perform our simulation. The distribution contains all constructed objects of the 80 sets and is presented in the white block. The overlapping part of these blocks will fade to pink, and we can see that these figures are nearly all in pink.

\begin{figure*}
    \centering
    \includegraphics[width=0.45\textwidth]{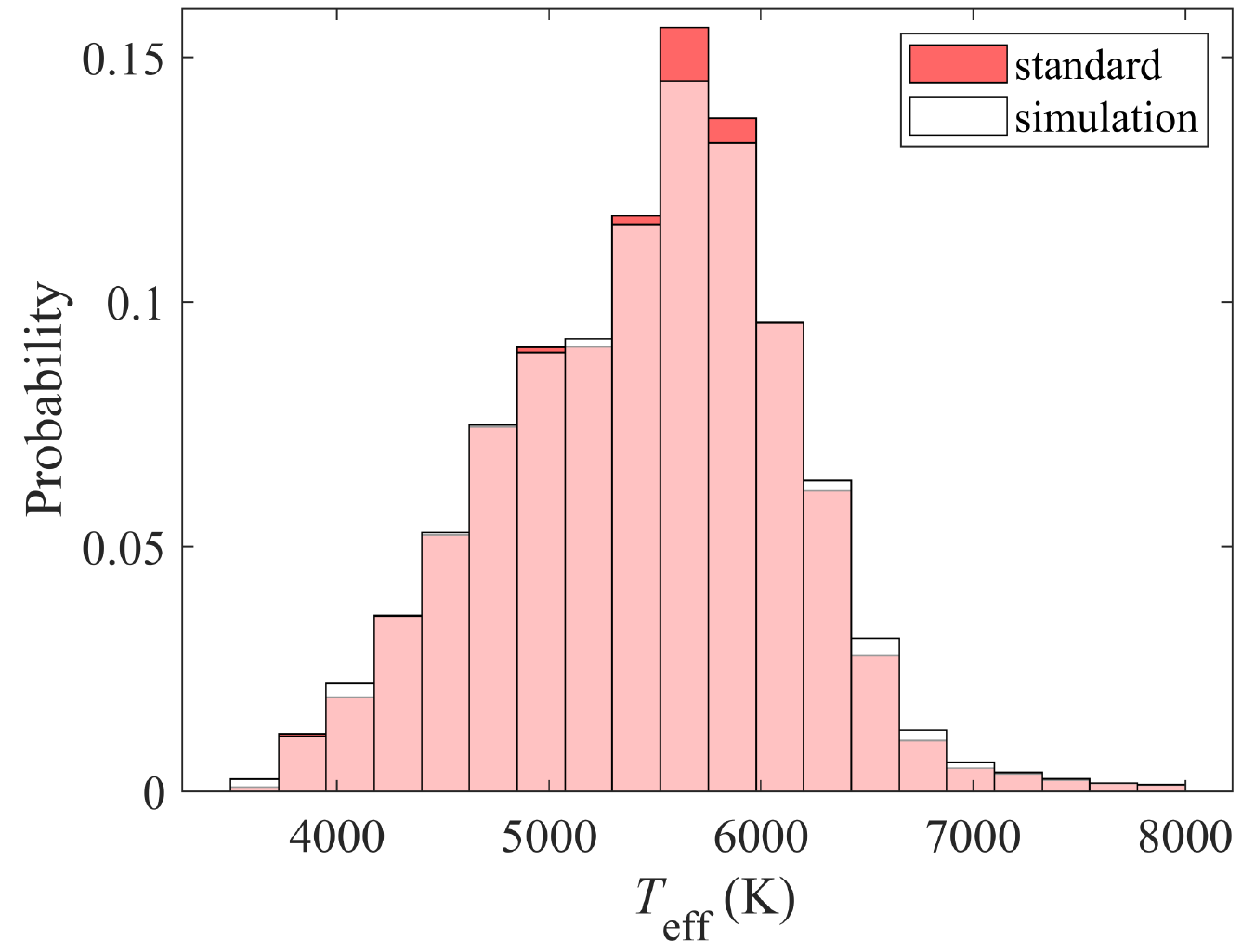}
    \includegraphics[width=0.45\textwidth]{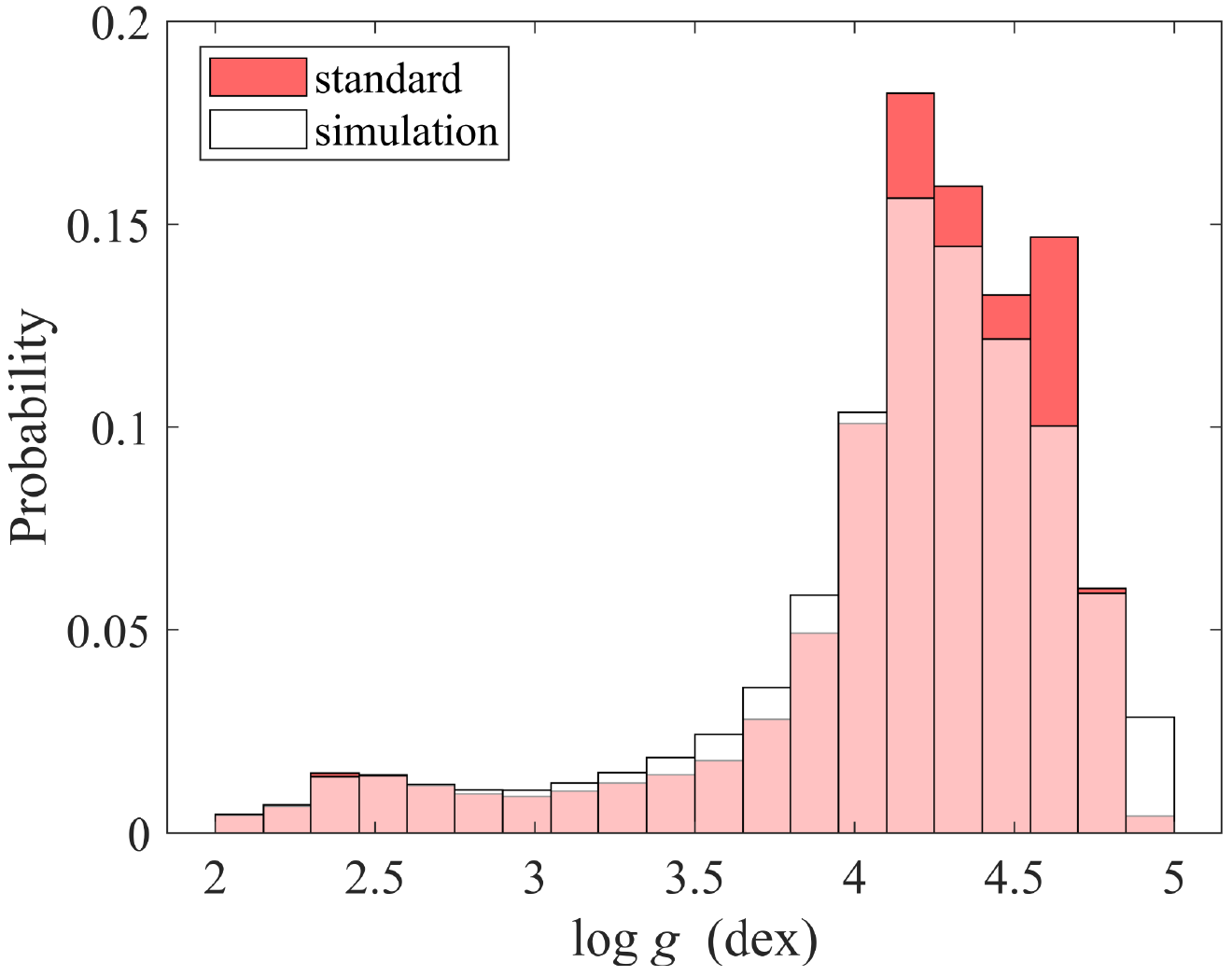}
    \includegraphics[width=0.45\textwidth]{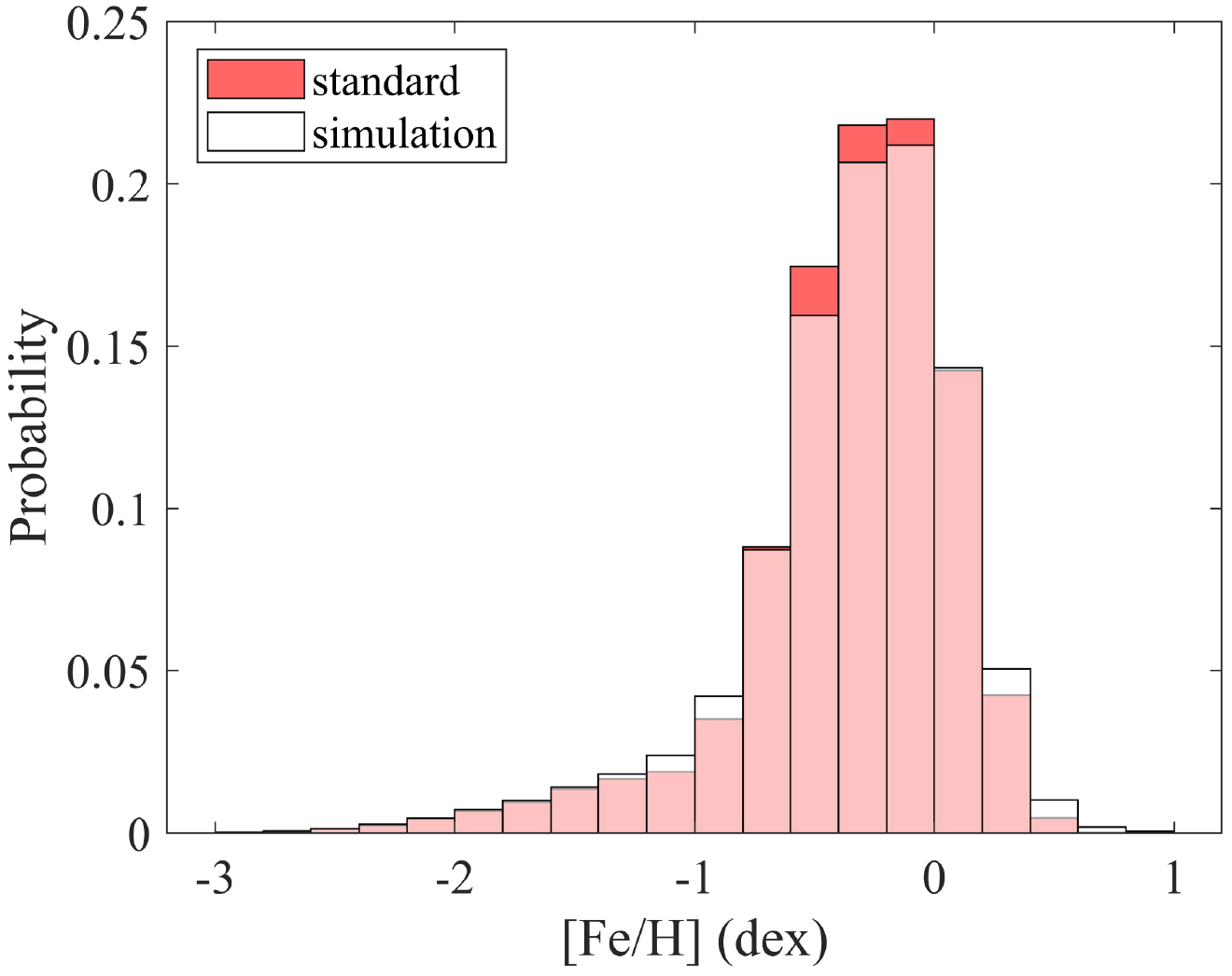}
    \caption{Stellar parameter distributions of the sample set. The red block shows the distribution of standard sample set. The white block shows the simulation distribution. The overlap is shown in pink (red+white). The y axis shows the probability.}
\end{figure*}

\begin{figure*}
    \centering
    \includegraphics[width=0.33\textwidth]{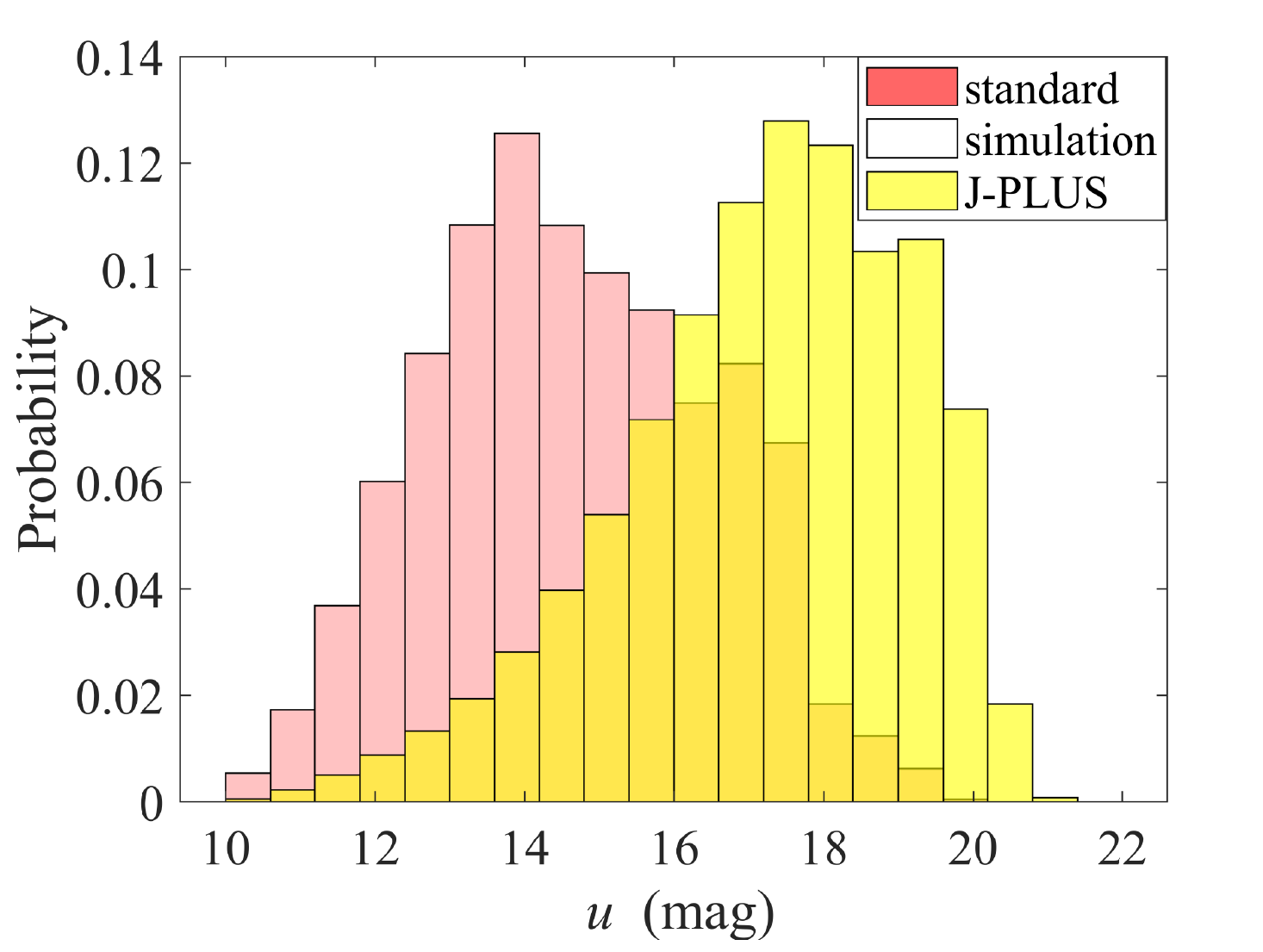}
    \includegraphics[width=0.33\textwidth]{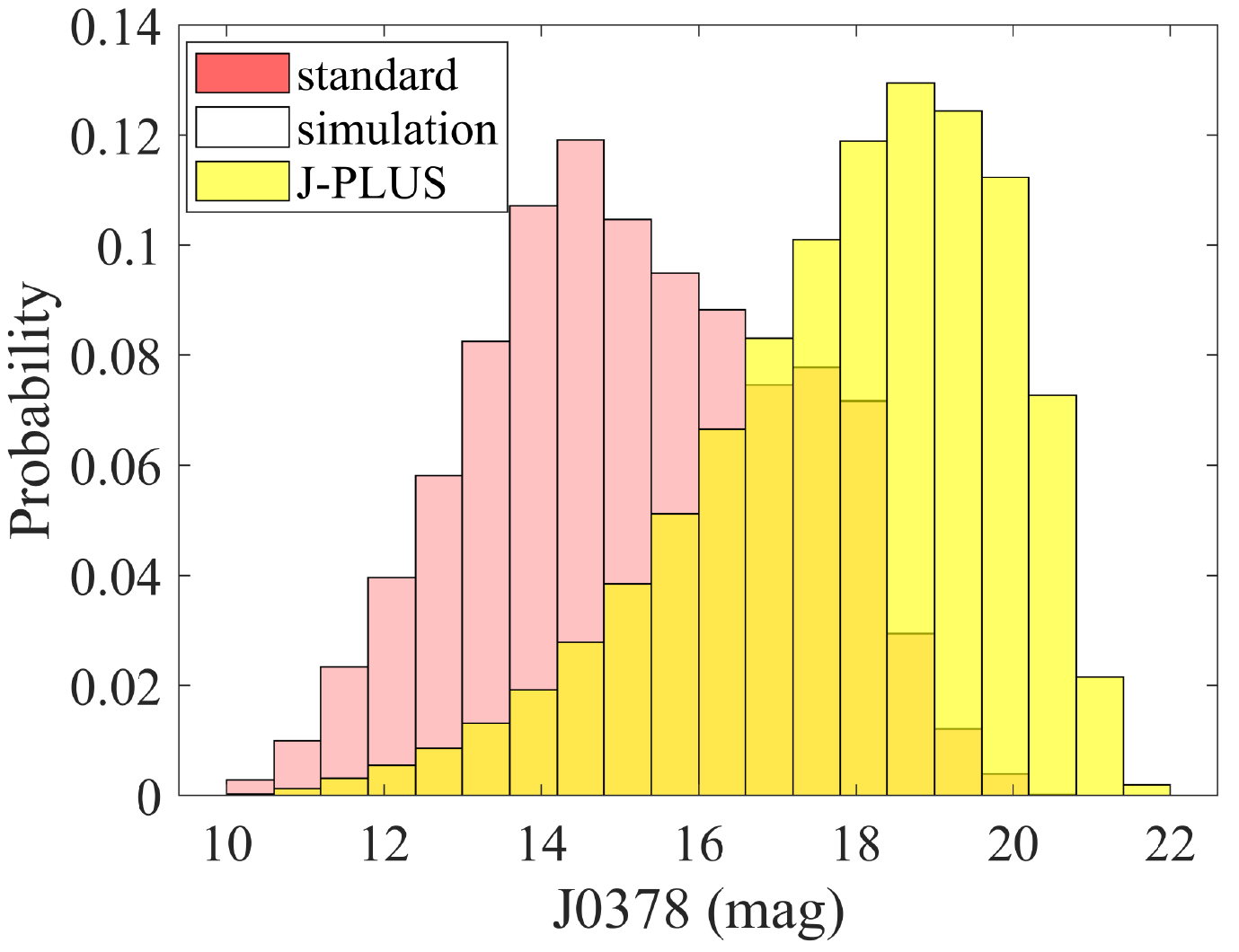}
    \includegraphics[width=0.33\textwidth]{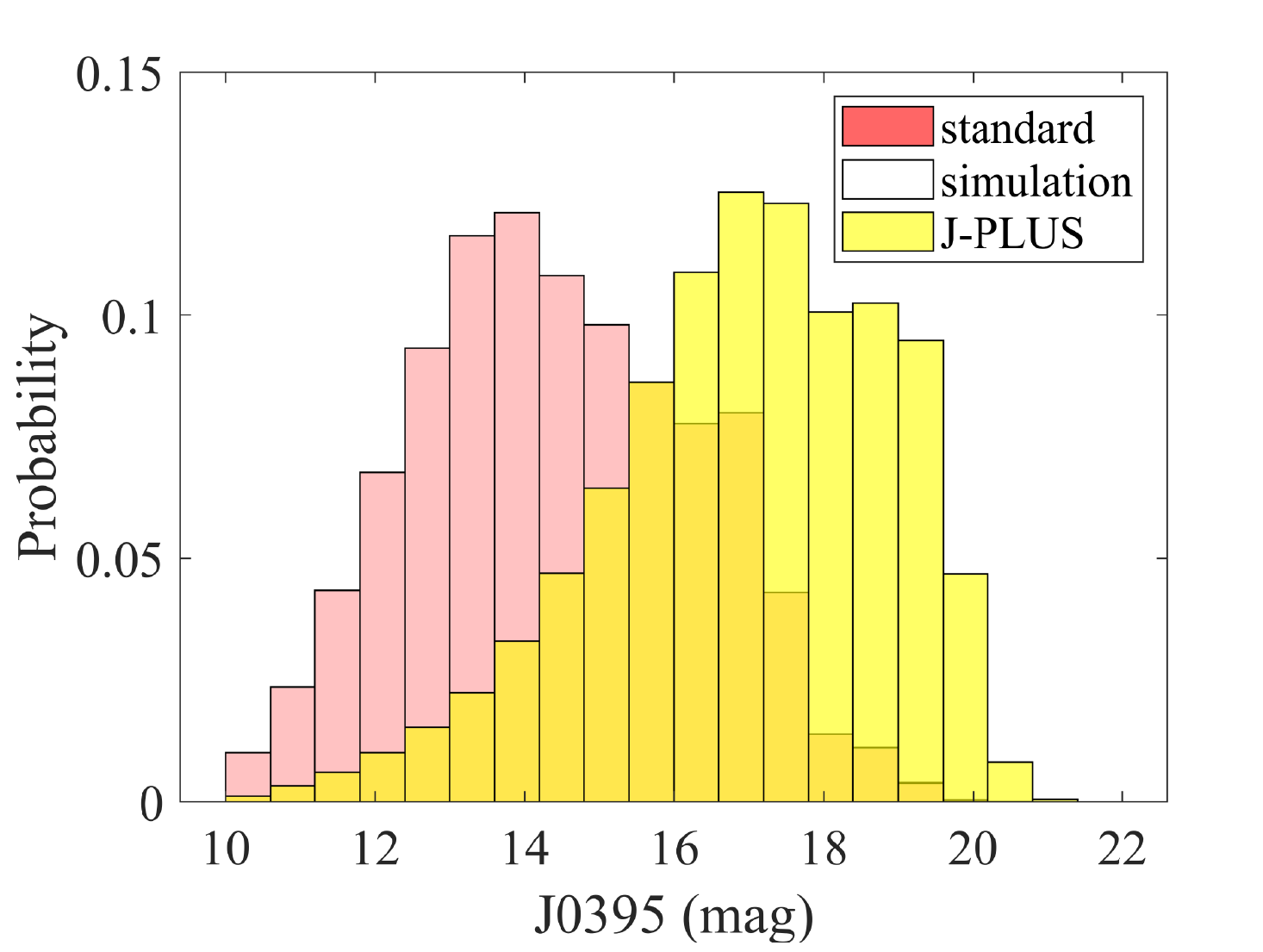}
    \includegraphics[width=0.33\textwidth]{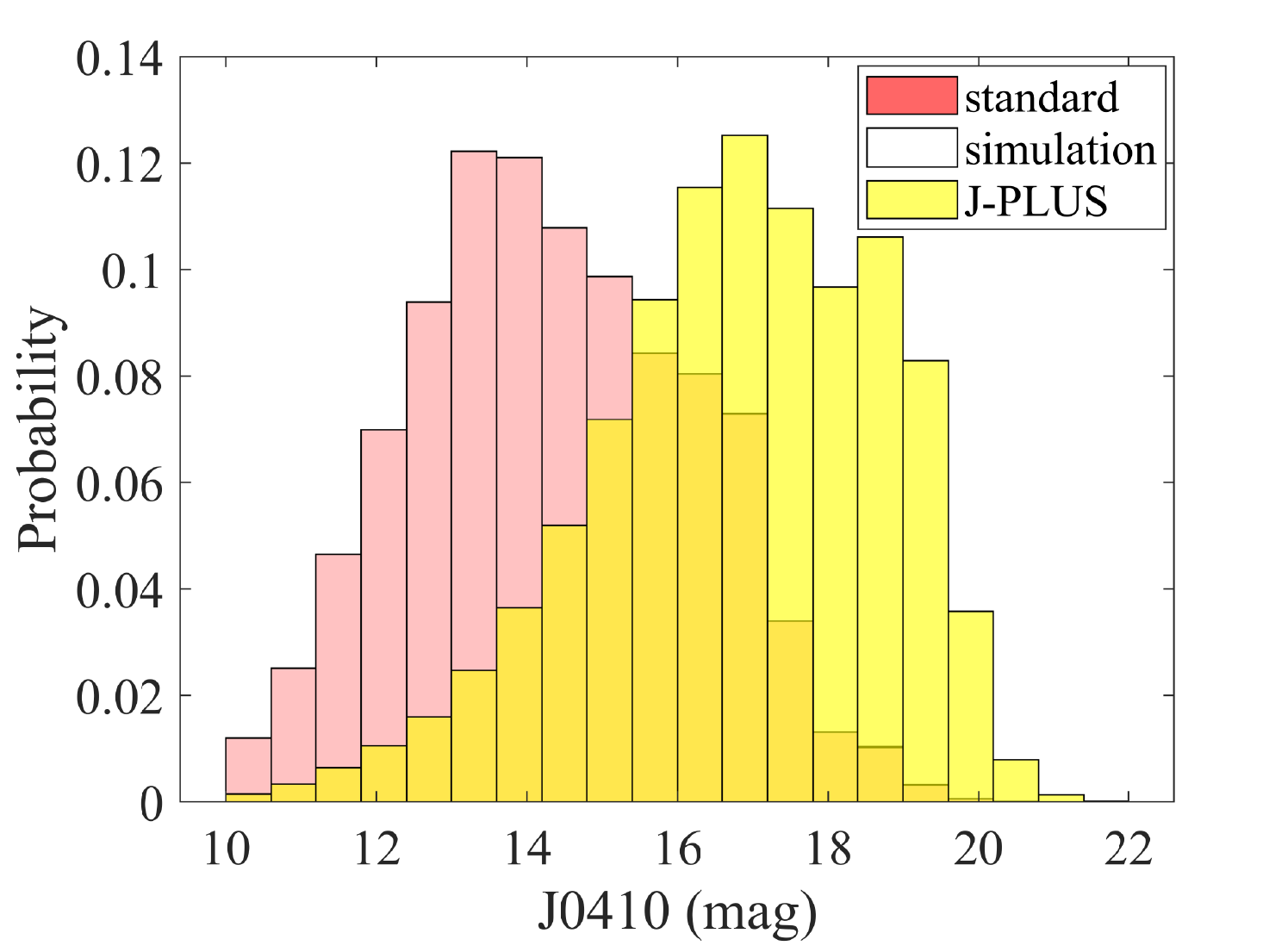}
    \includegraphics[width=0.33\textwidth]{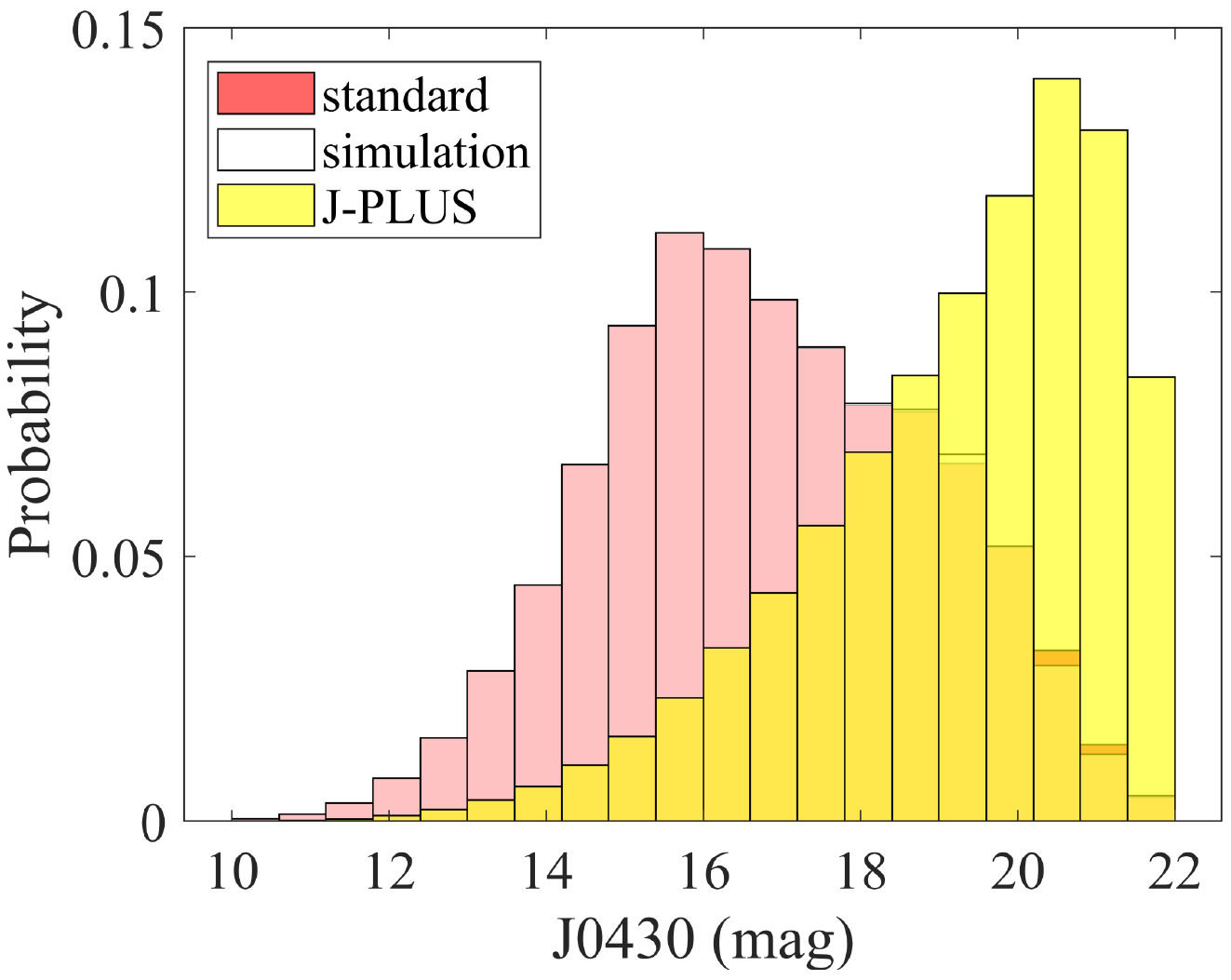}
    \includegraphics[width=0.33\textwidth]{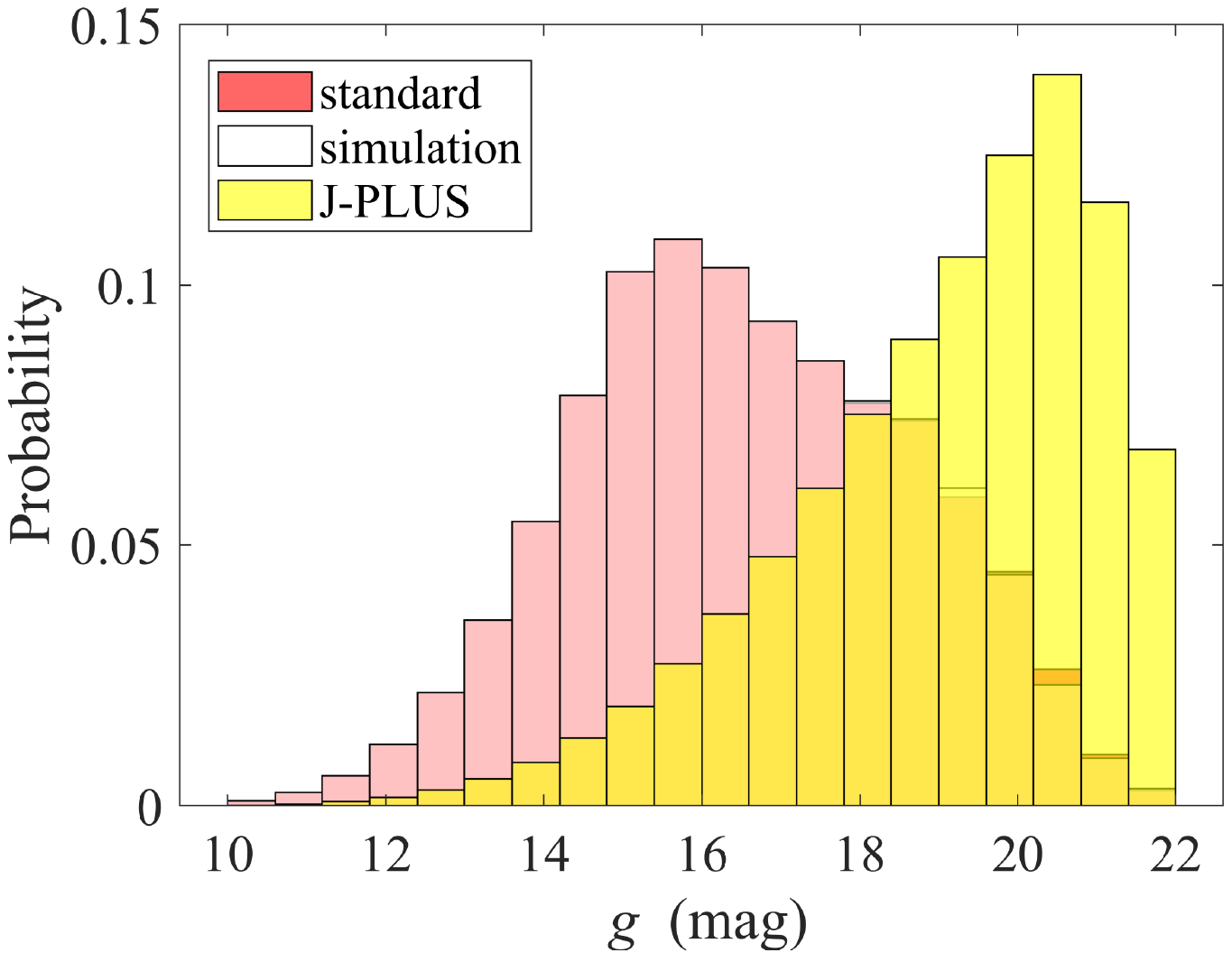}
    \includegraphics[width=0.33\textwidth]{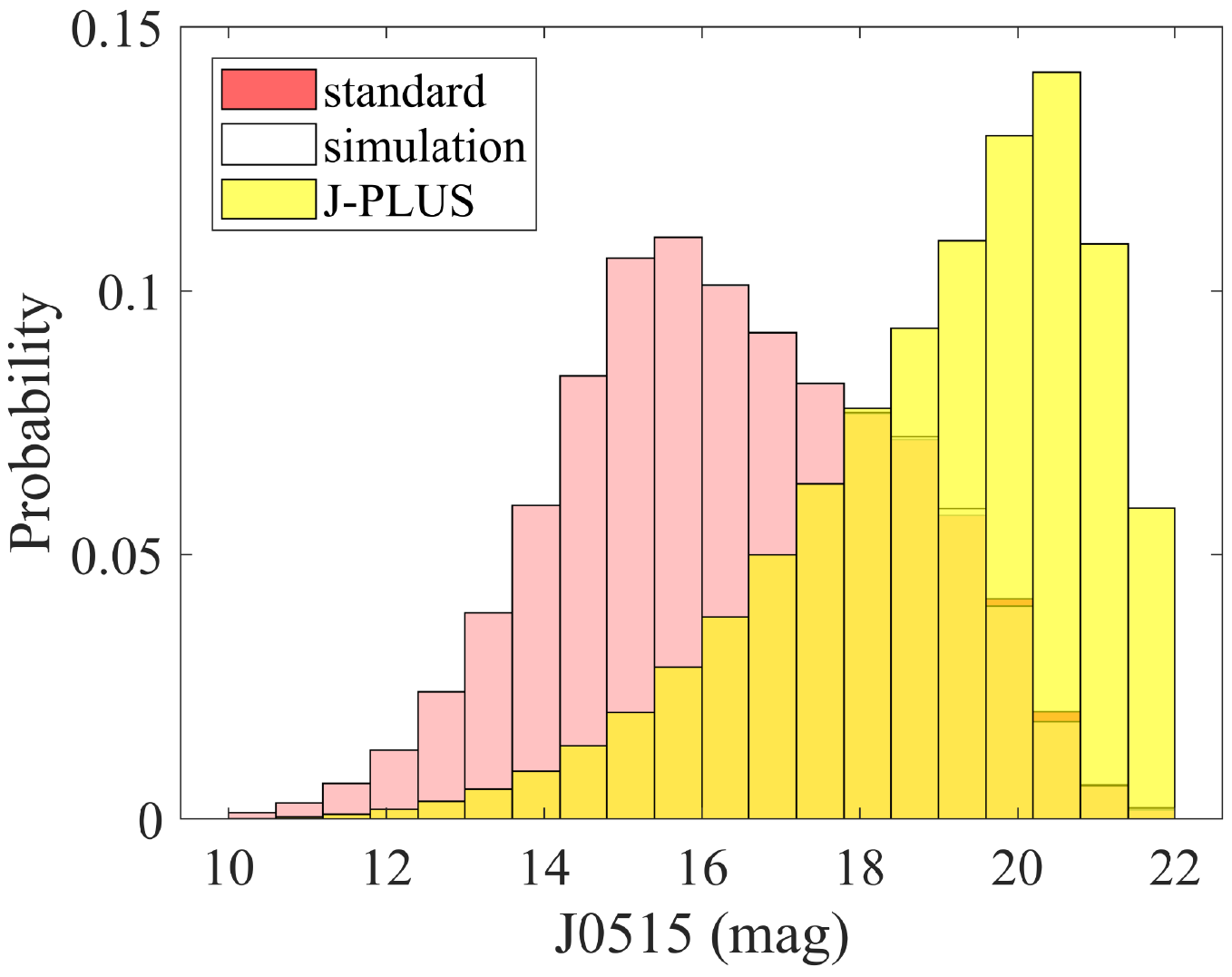}
    \includegraphics[width=0.33\textwidth]{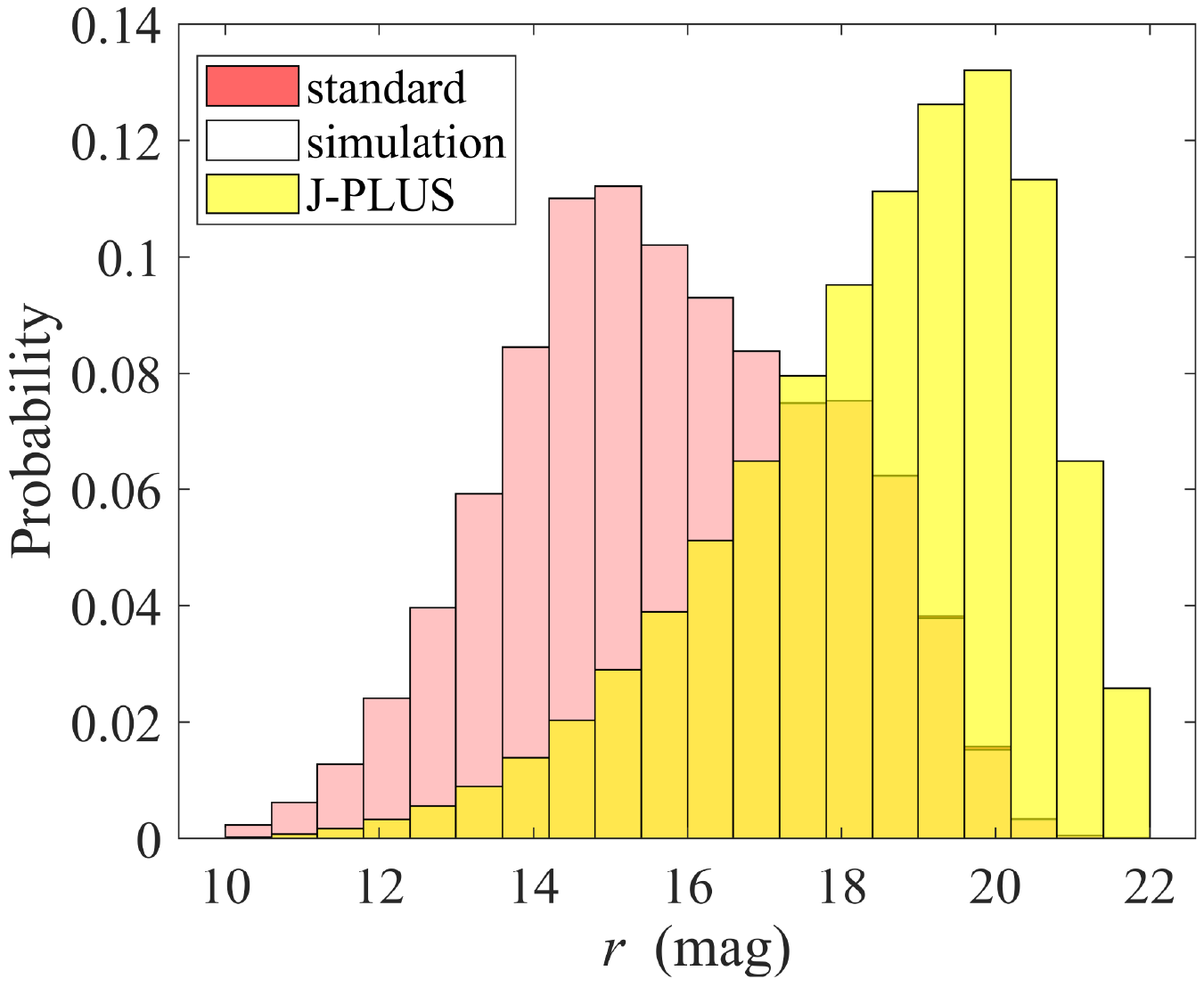}
    \includegraphics[width=0.33\textwidth]{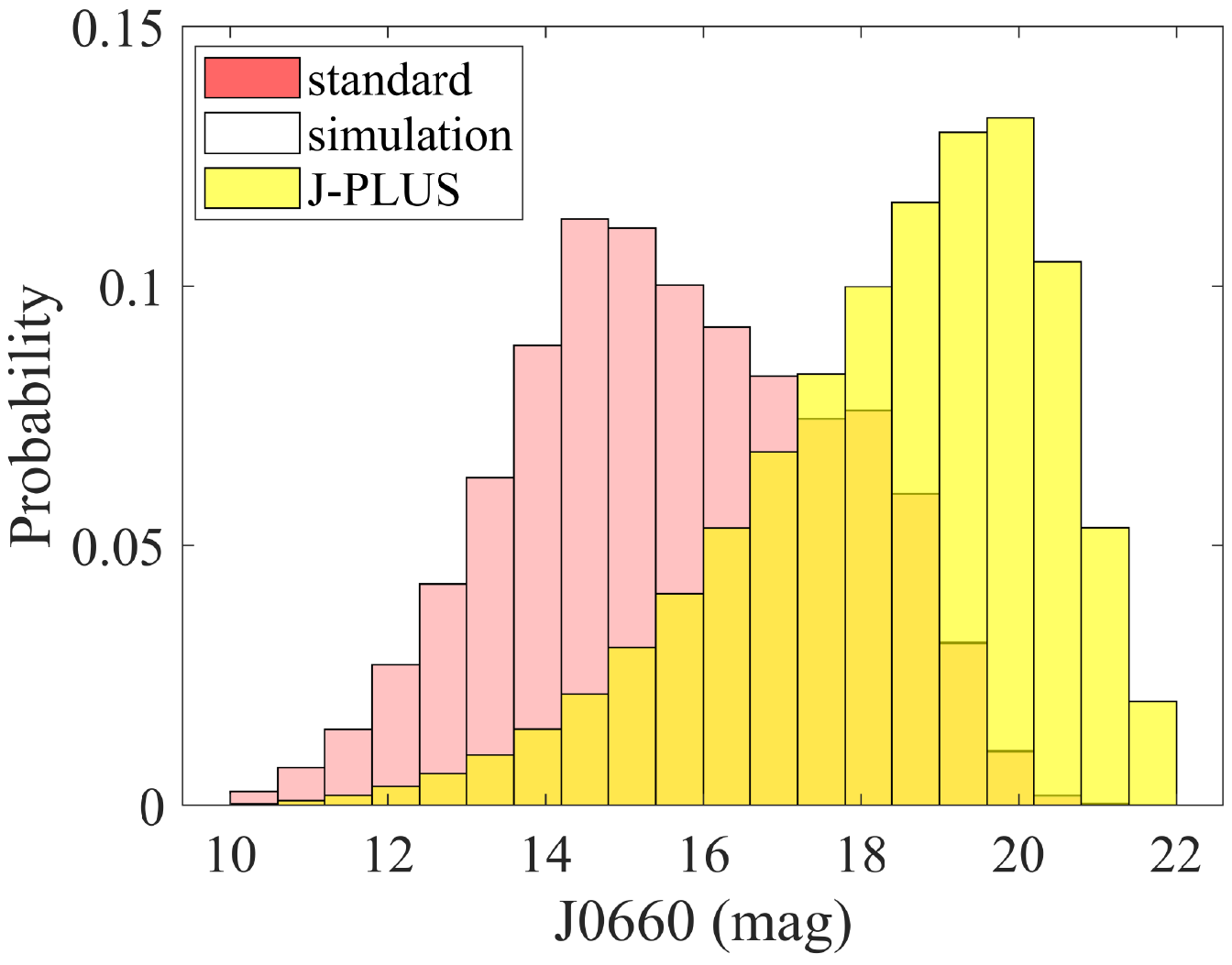}
    \includegraphics[width=0.33\textwidth]{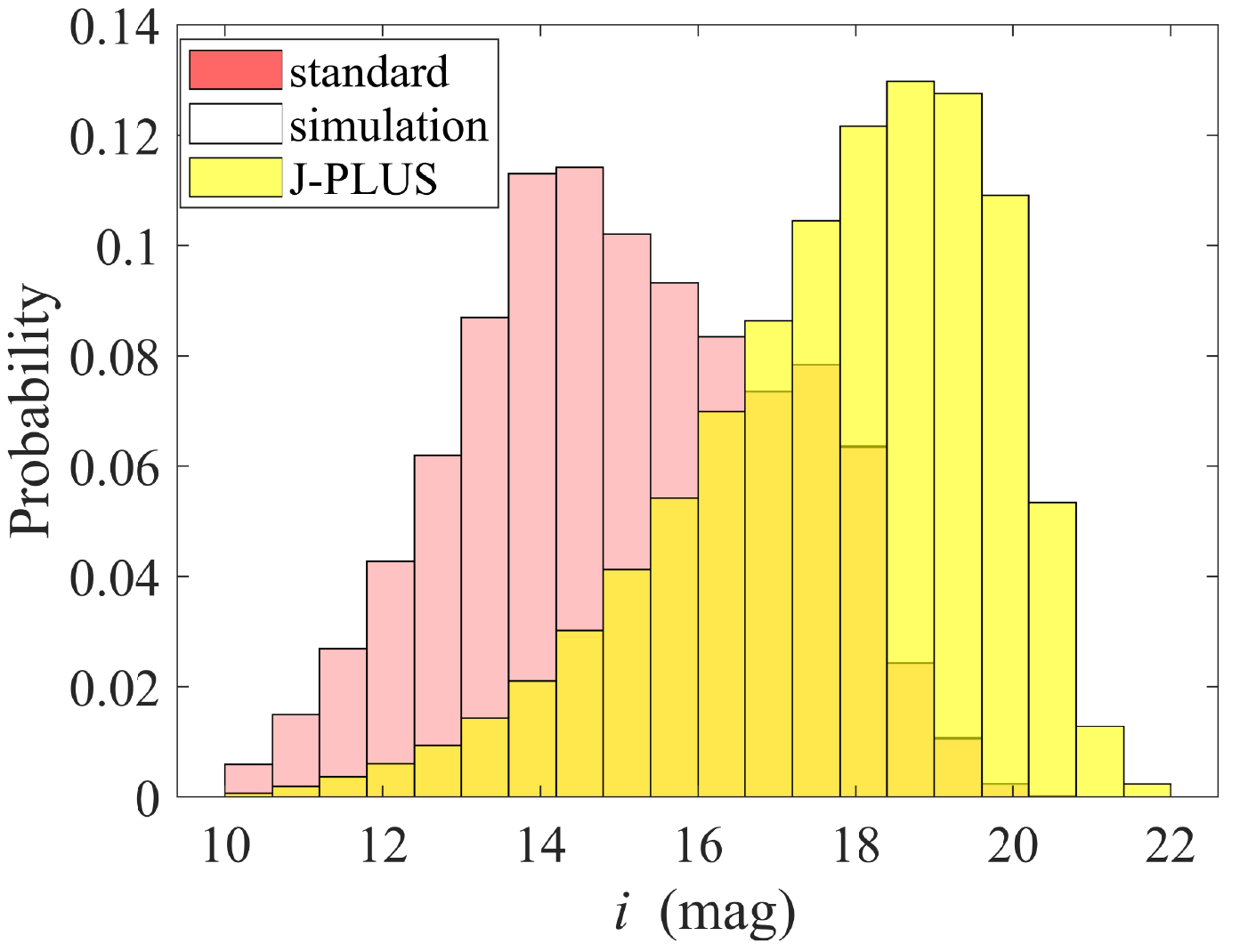} 
    \includegraphics[width=0.33\textwidth]{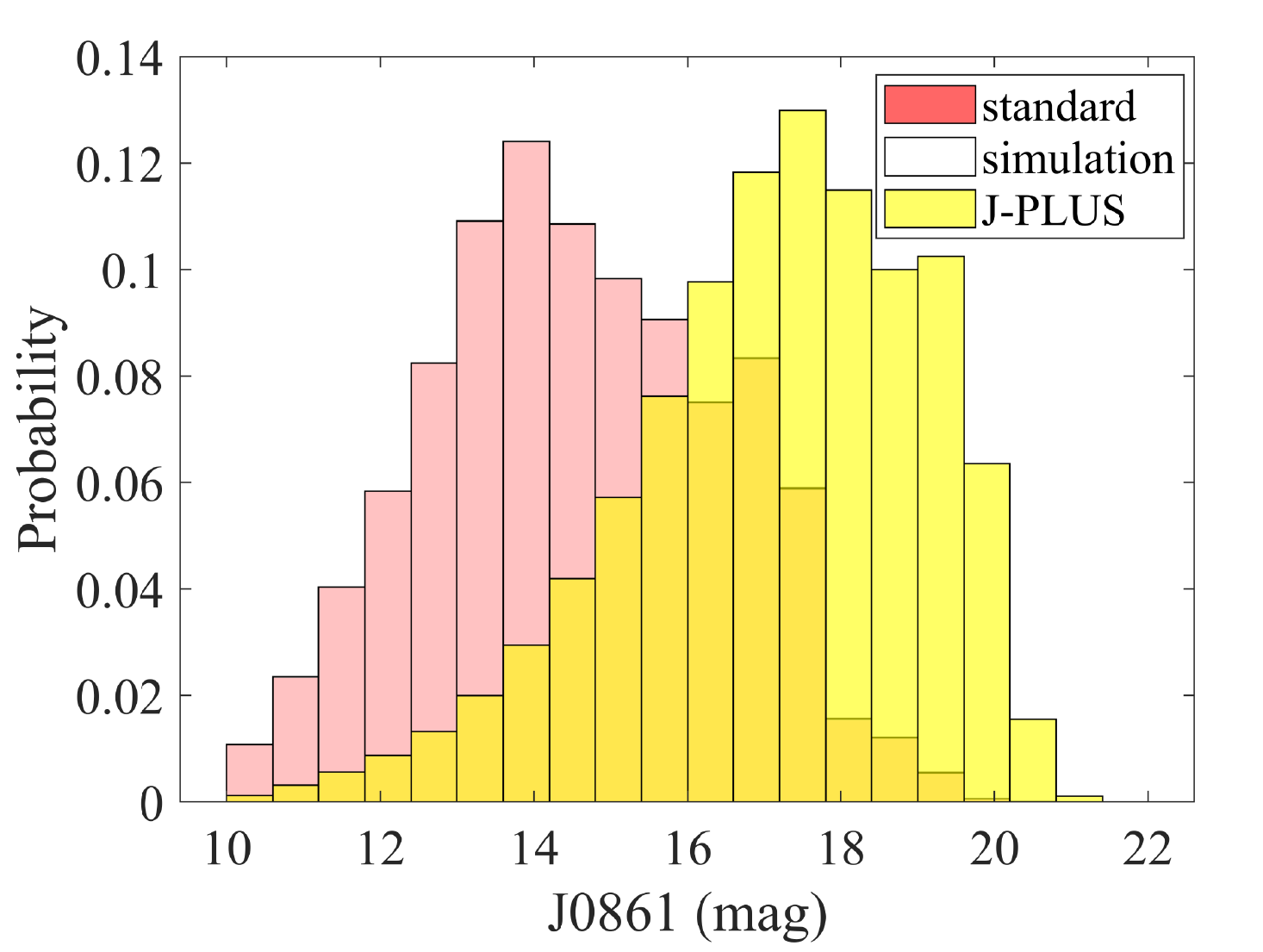}
    \includegraphics[width=0.33\textwidth]{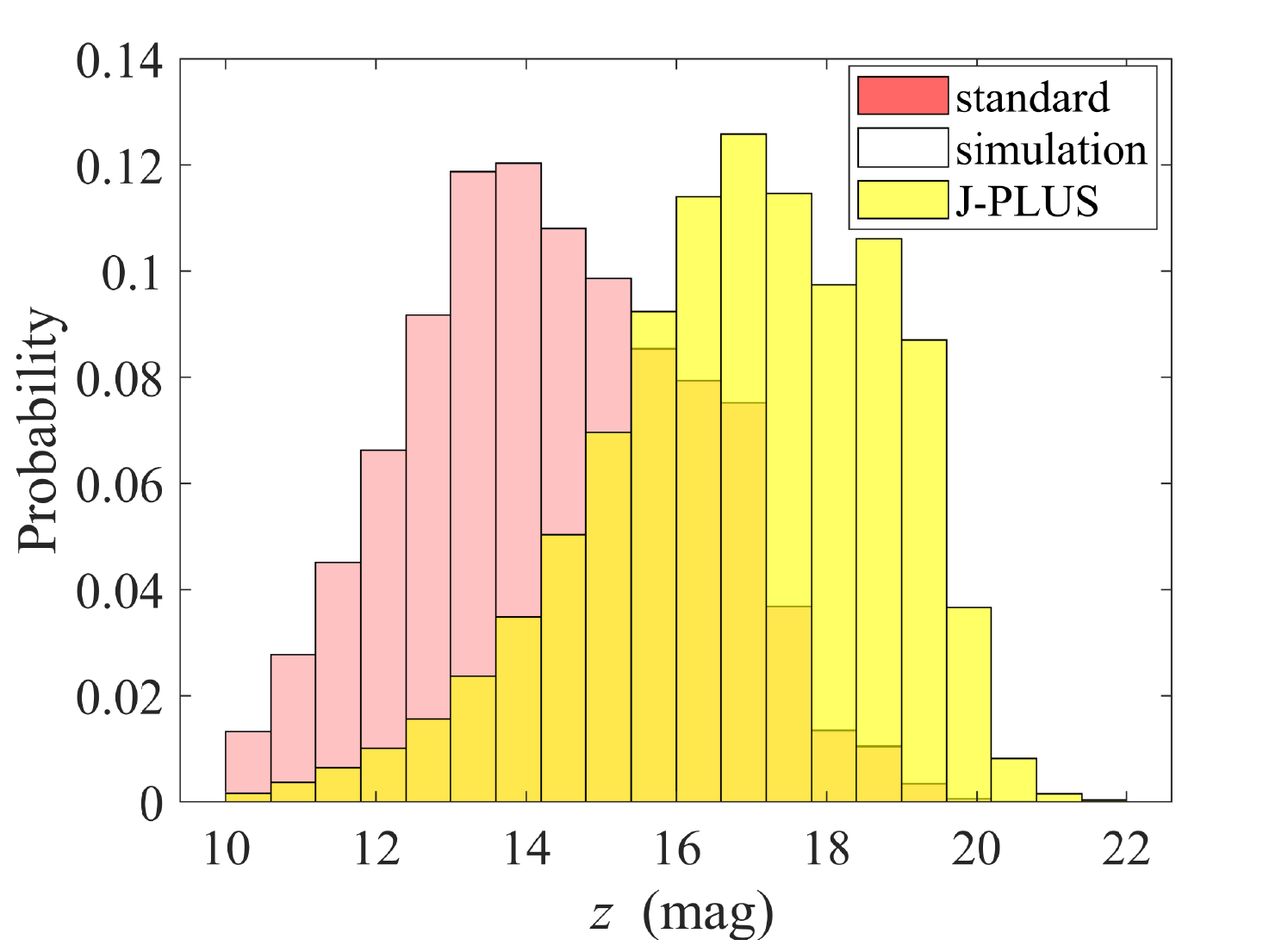}
    \caption{Magnitude distributions of the sample set. The yellow block shows the J-PLUS distribution.}
\end{figure*}

\FloatBarrier

\section{Blind test result}\label{appb}
We plot the results of the blind test, showing the differences between the predicted and the true parameters. We fit the residuals with Gaussian functions.

\begin{figure*}
    \centering
    \includegraphics[width=0.45\textwidth]{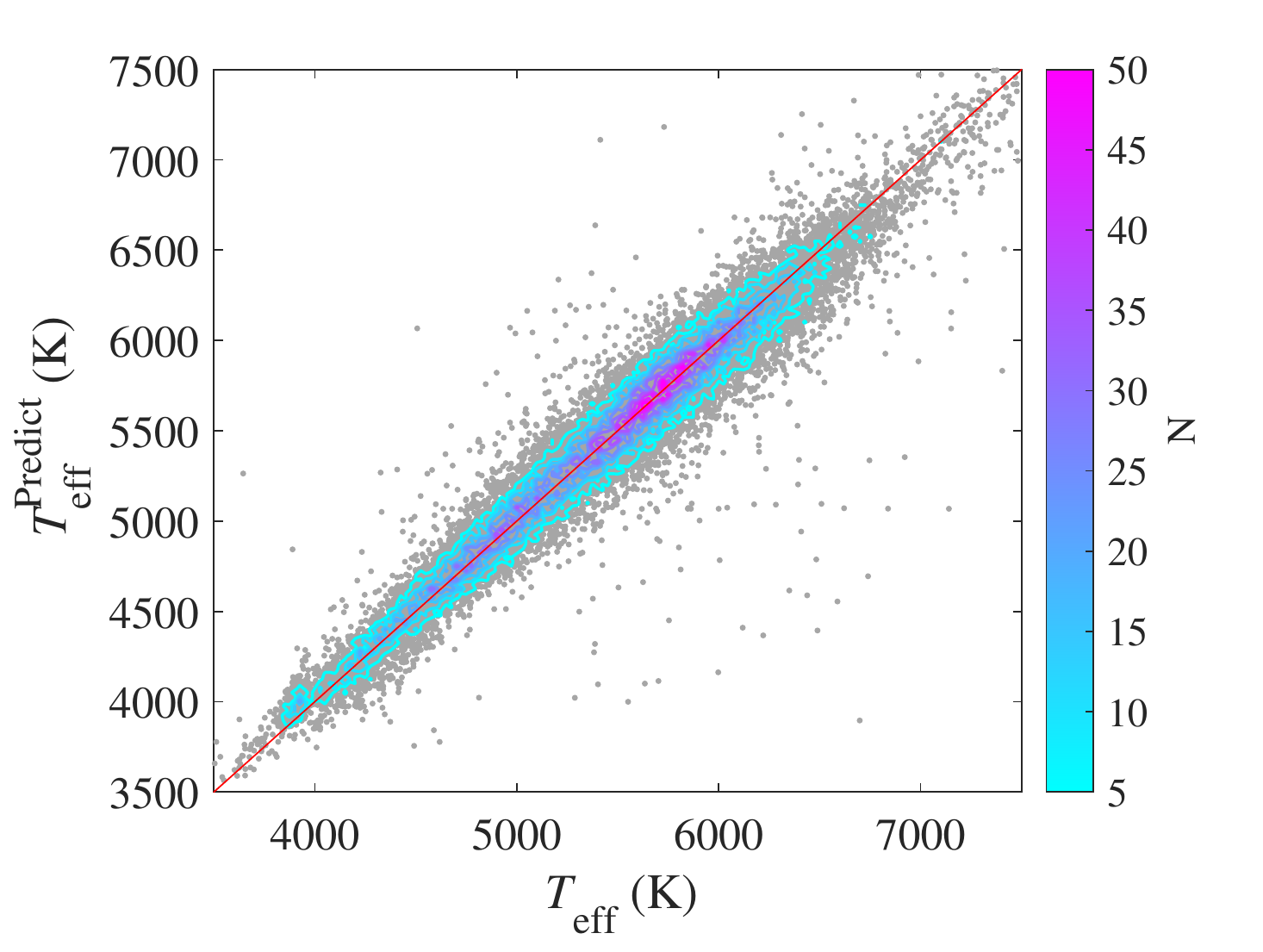}
    \includegraphics[width=0.45\textwidth]{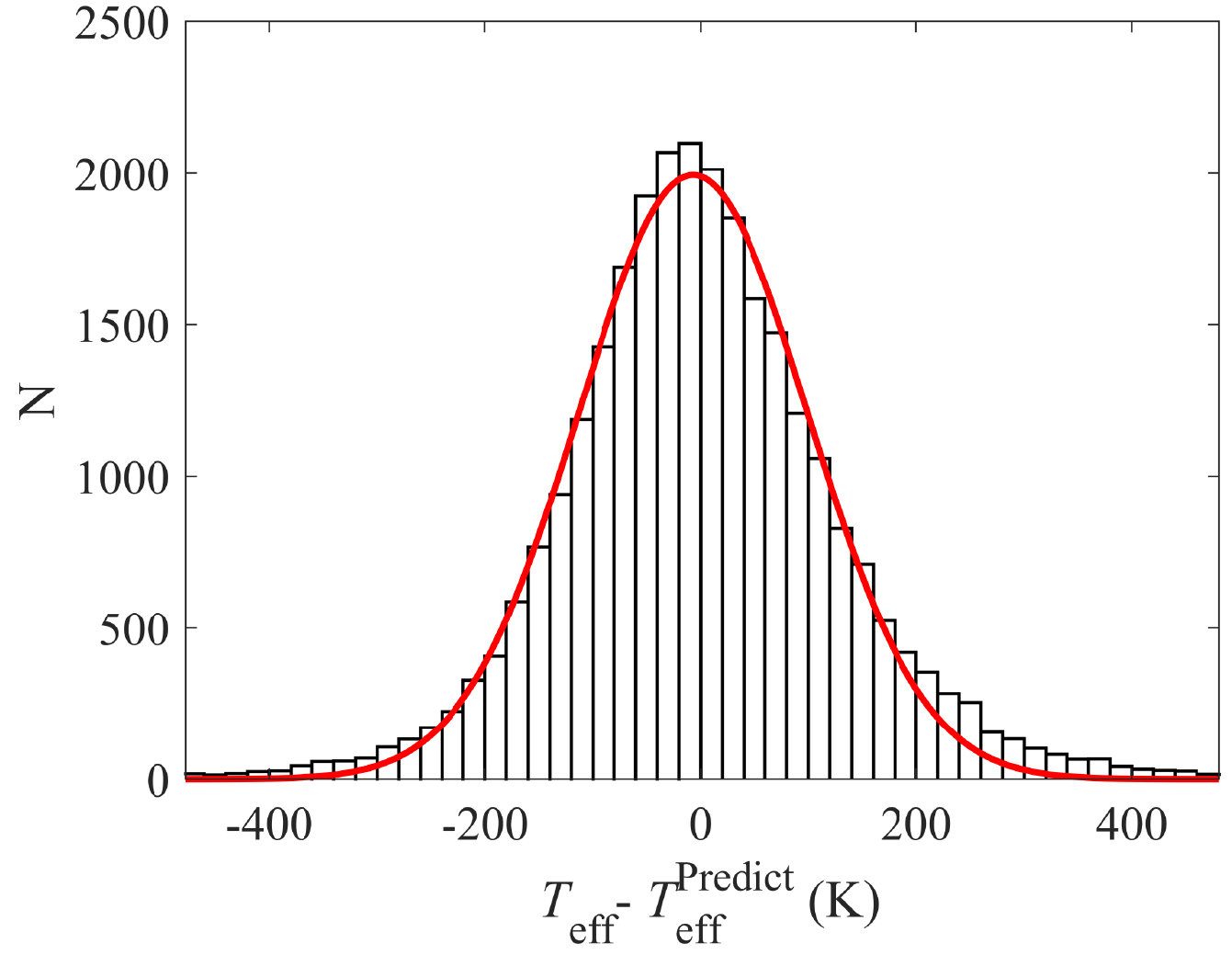}
    \includegraphics[width=0.45\textwidth]{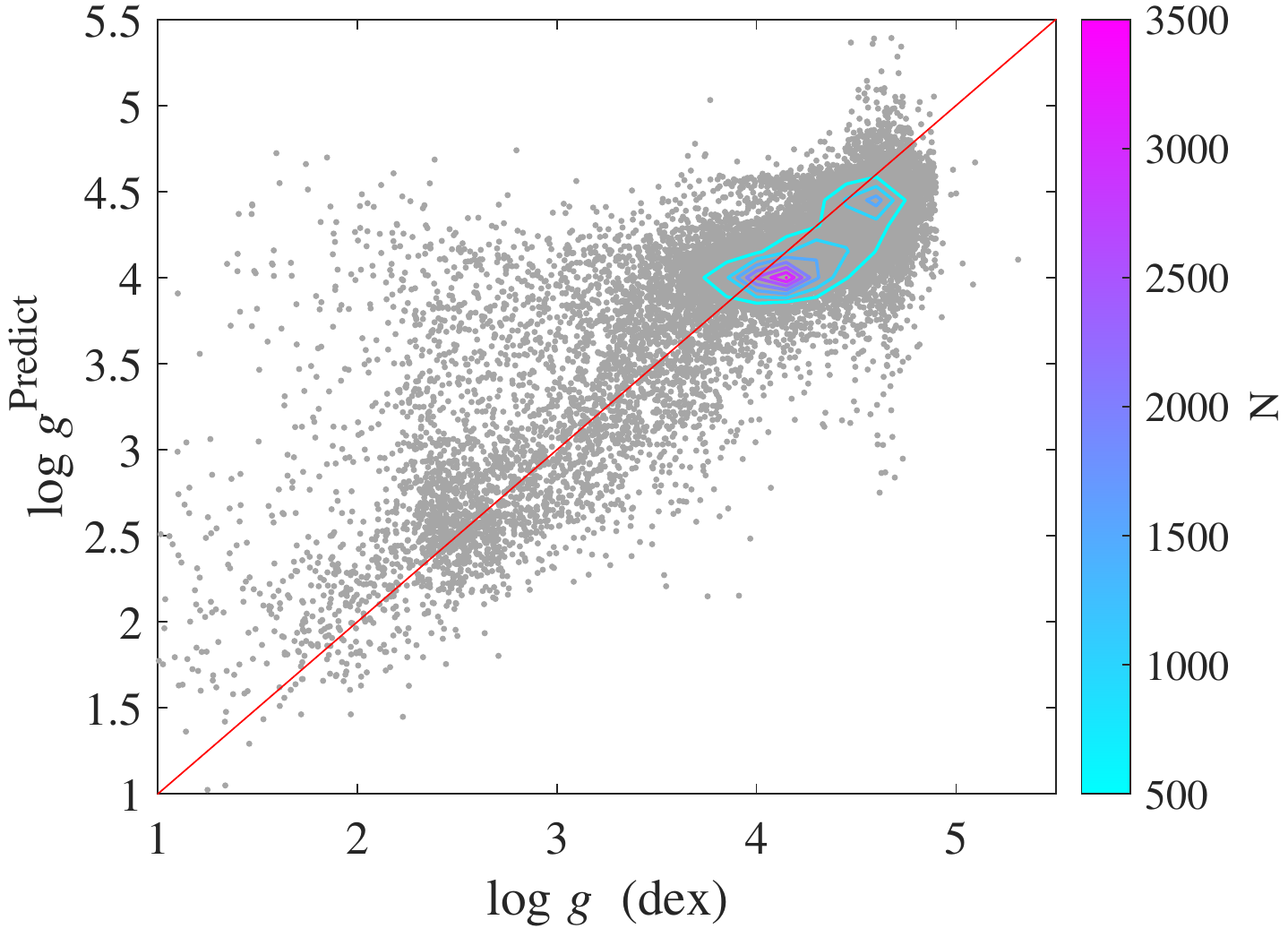}
    \includegraphics[width=0.45\textwidth]{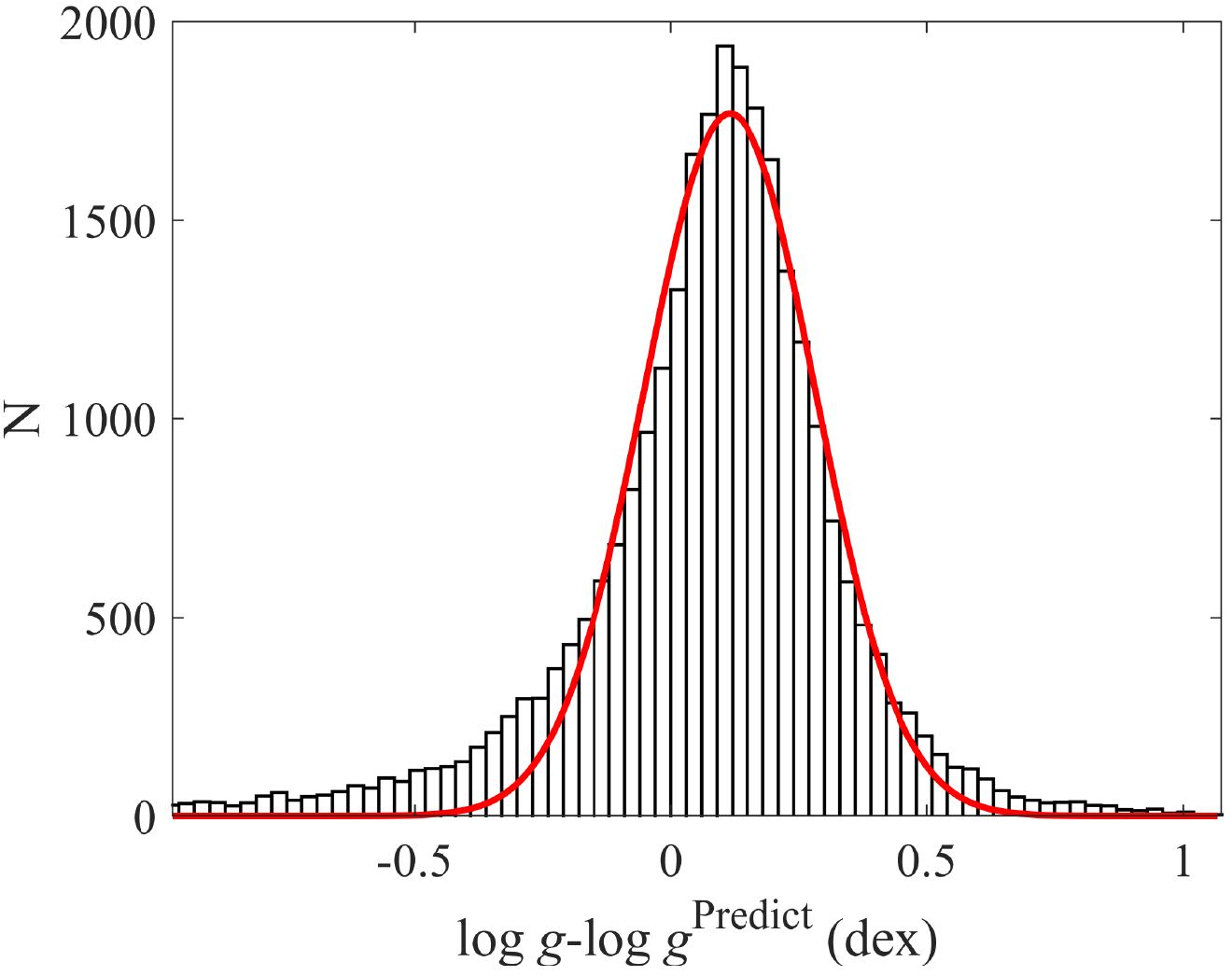}
    \includegraphics[width=0.45\textwidth]{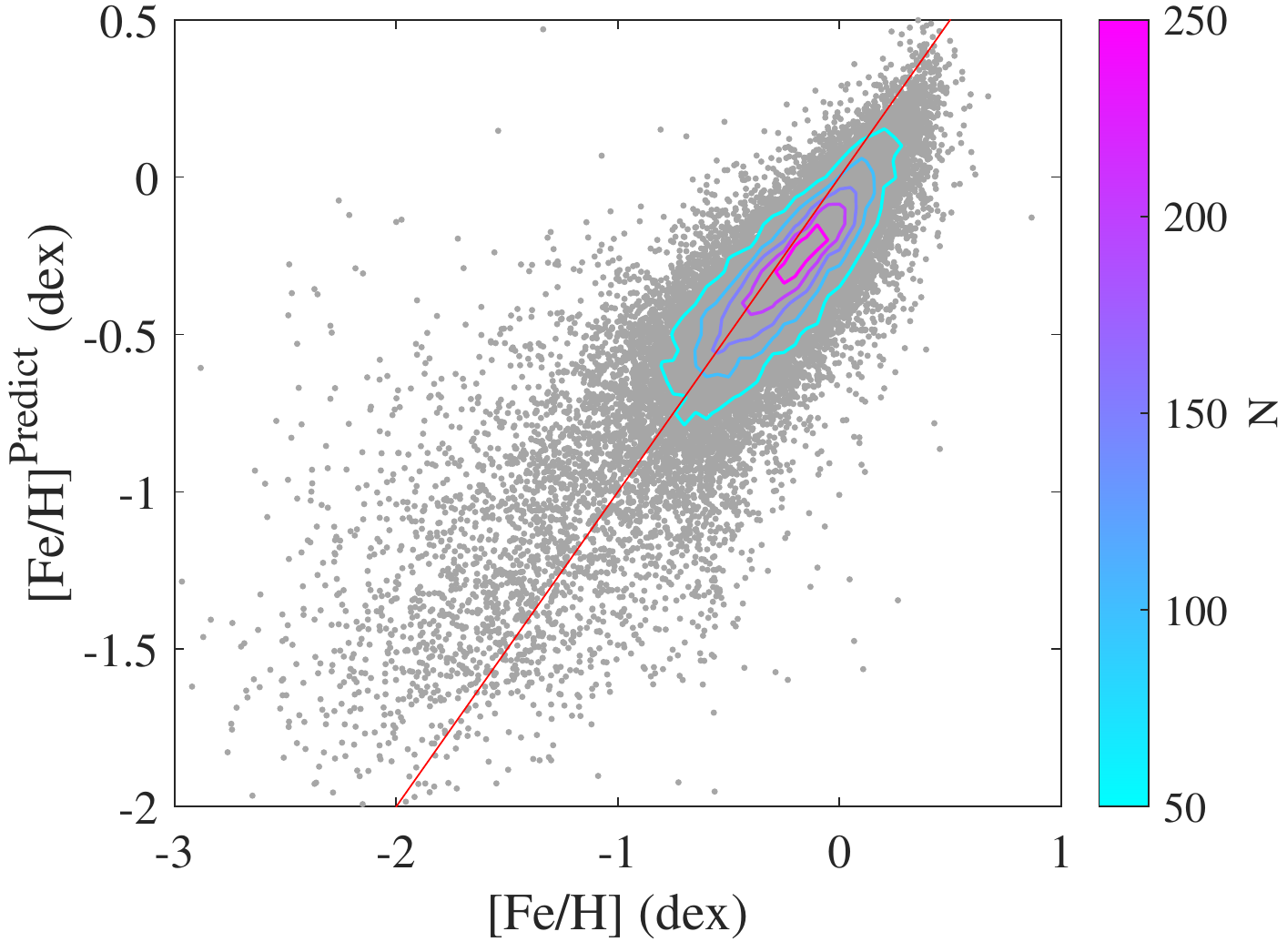}
    \includegraphics[width=0.45\textwidth]{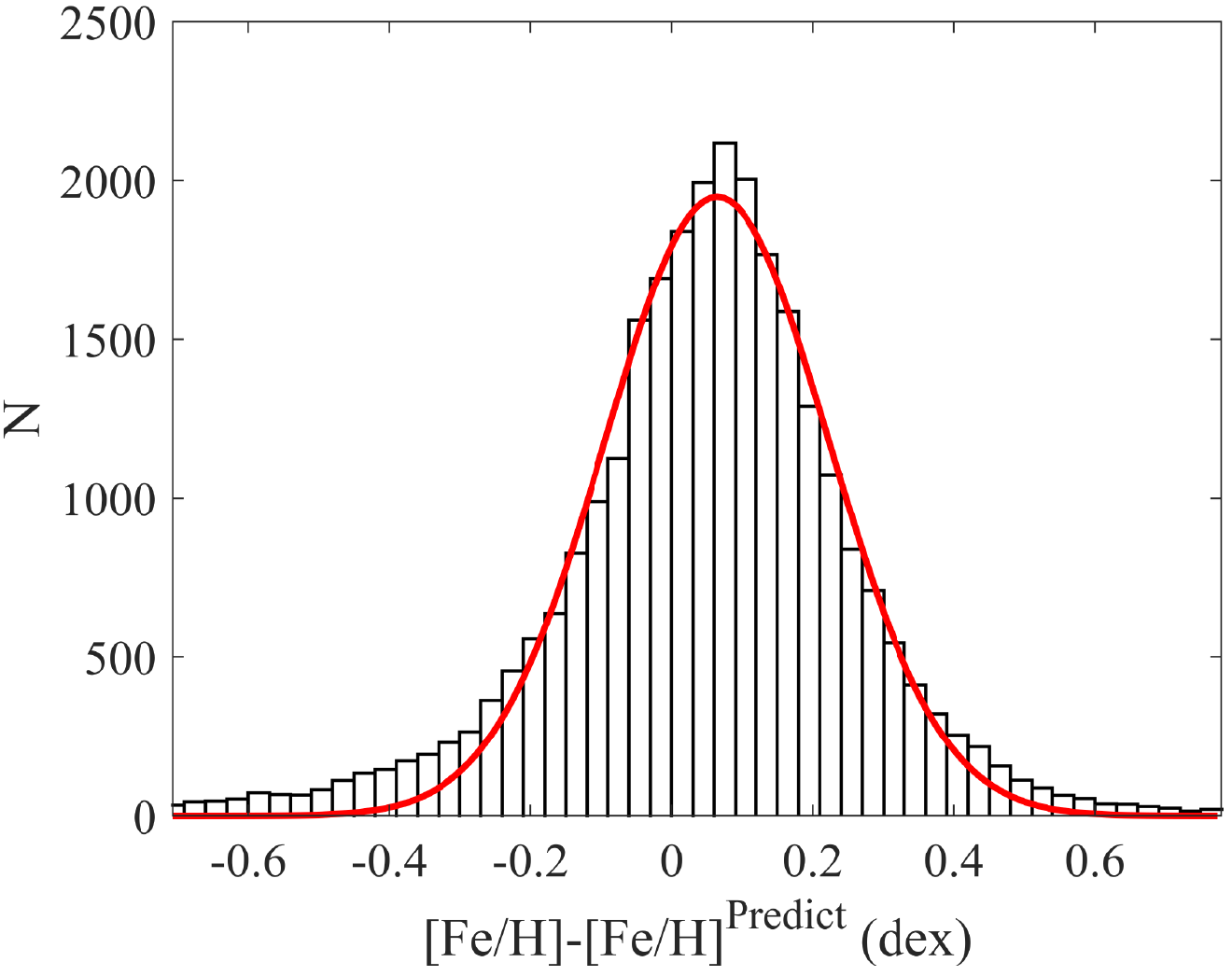}
    \caption{Results of the blind test validation. The upper two panels are $T_{\rm eff}$. The middle two panels are log $g$, and the lower two panels are [Fe/H].}
\end{figure*}

\FloatBarrier

\section{Pipeline differences}
\label{appa}
We present the differences among the pipelines. The residuals are fitted by Gaussian functions, and the means and variances are shown in Table \ref{pipelinetab}.

\begin{figure*}
    \centering
    \includegraphics[width=0.45\textwidth]{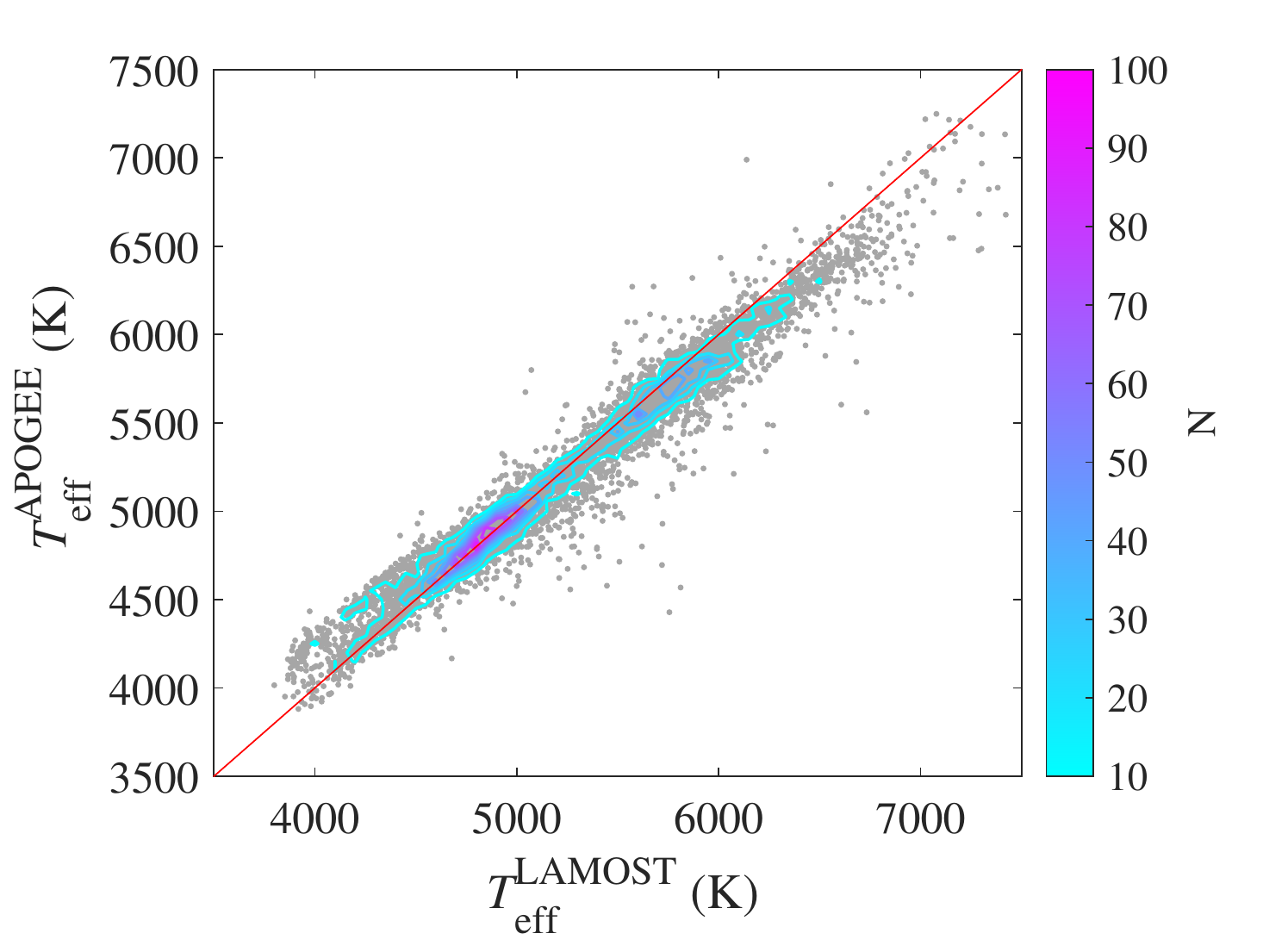}
    \includegraphics[width=0.45\textwidth]{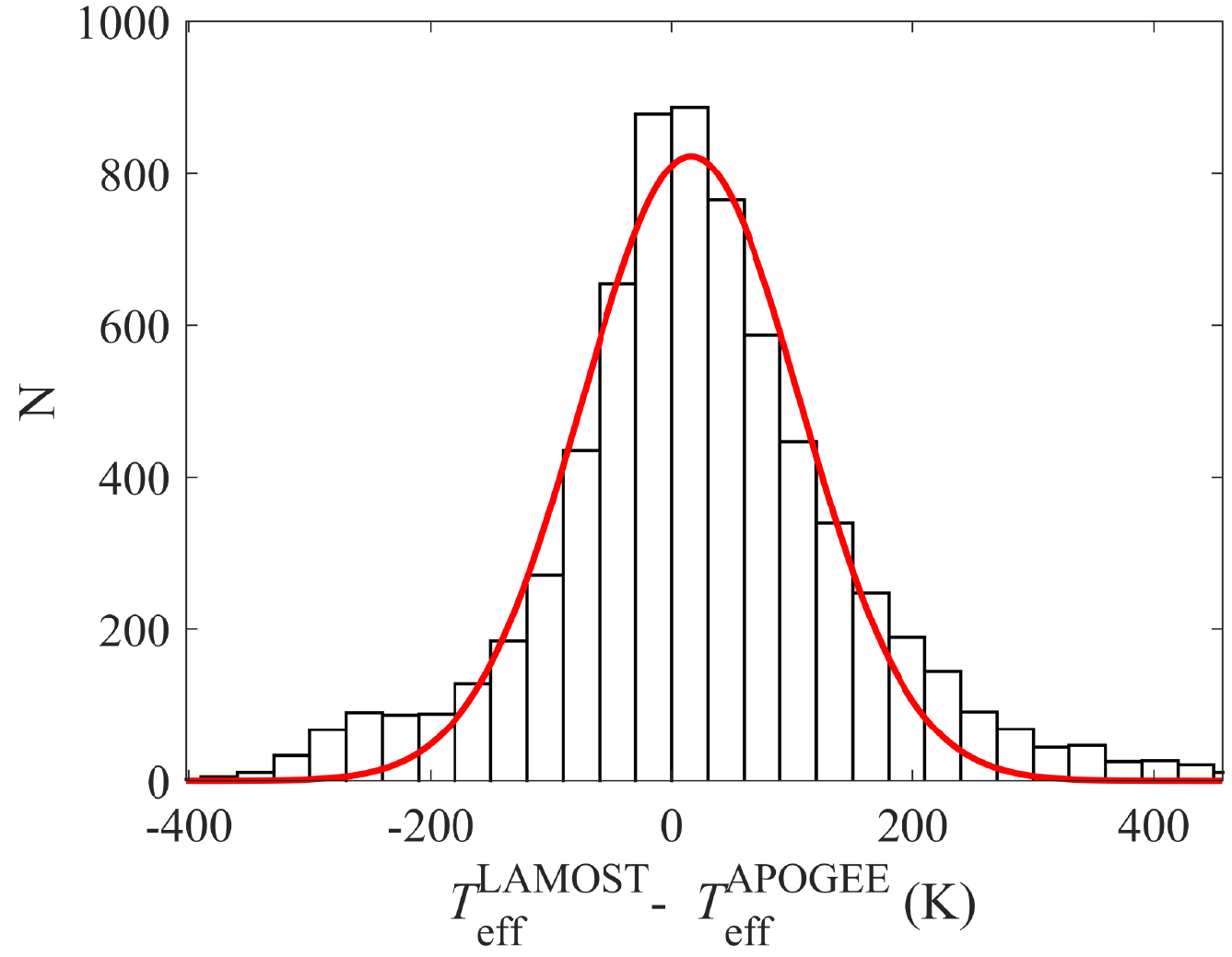}
    \includegraphics[width=0.45\textwidth]{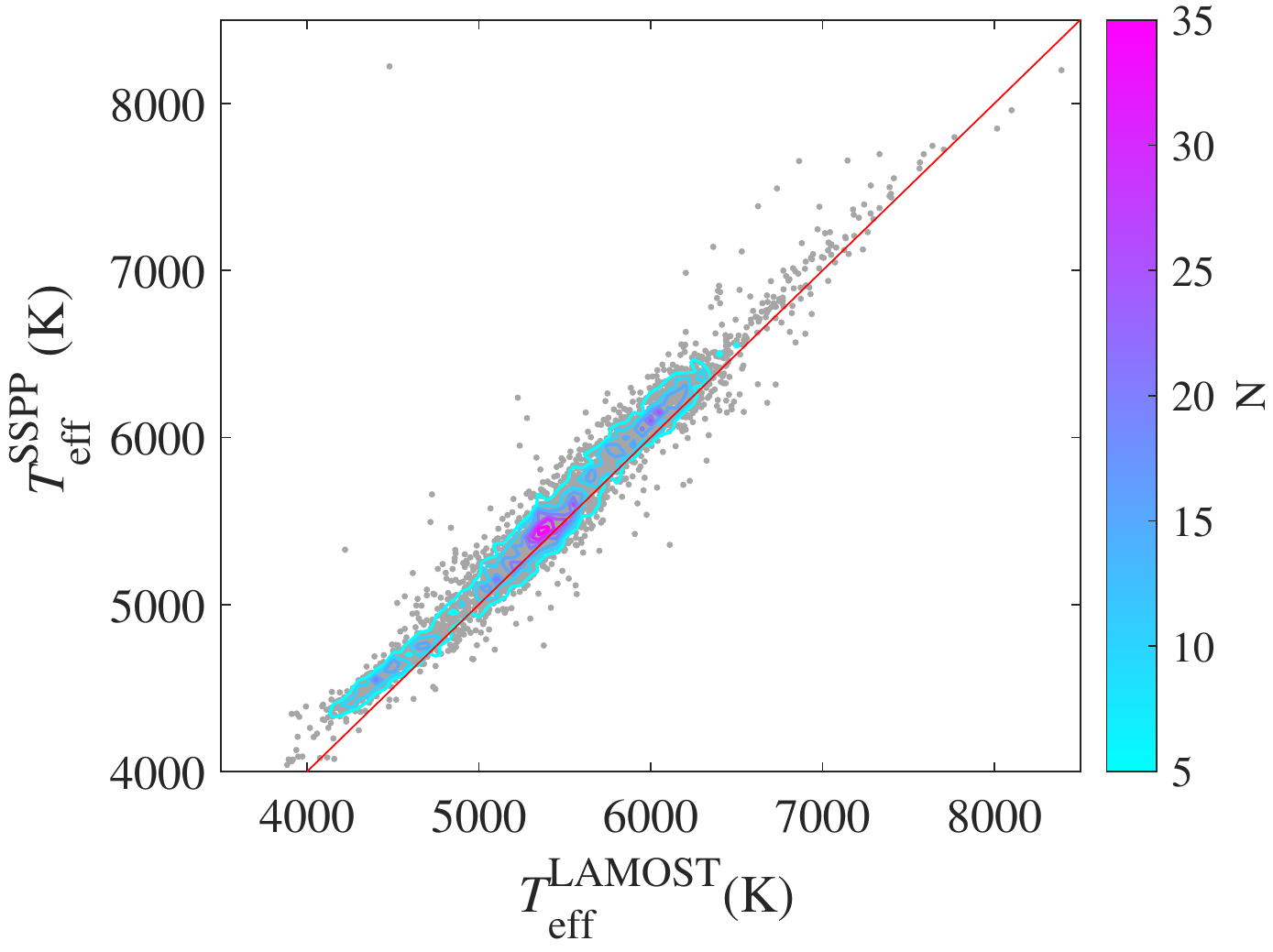}
    \includegraphics[width=0.45\textwidth]{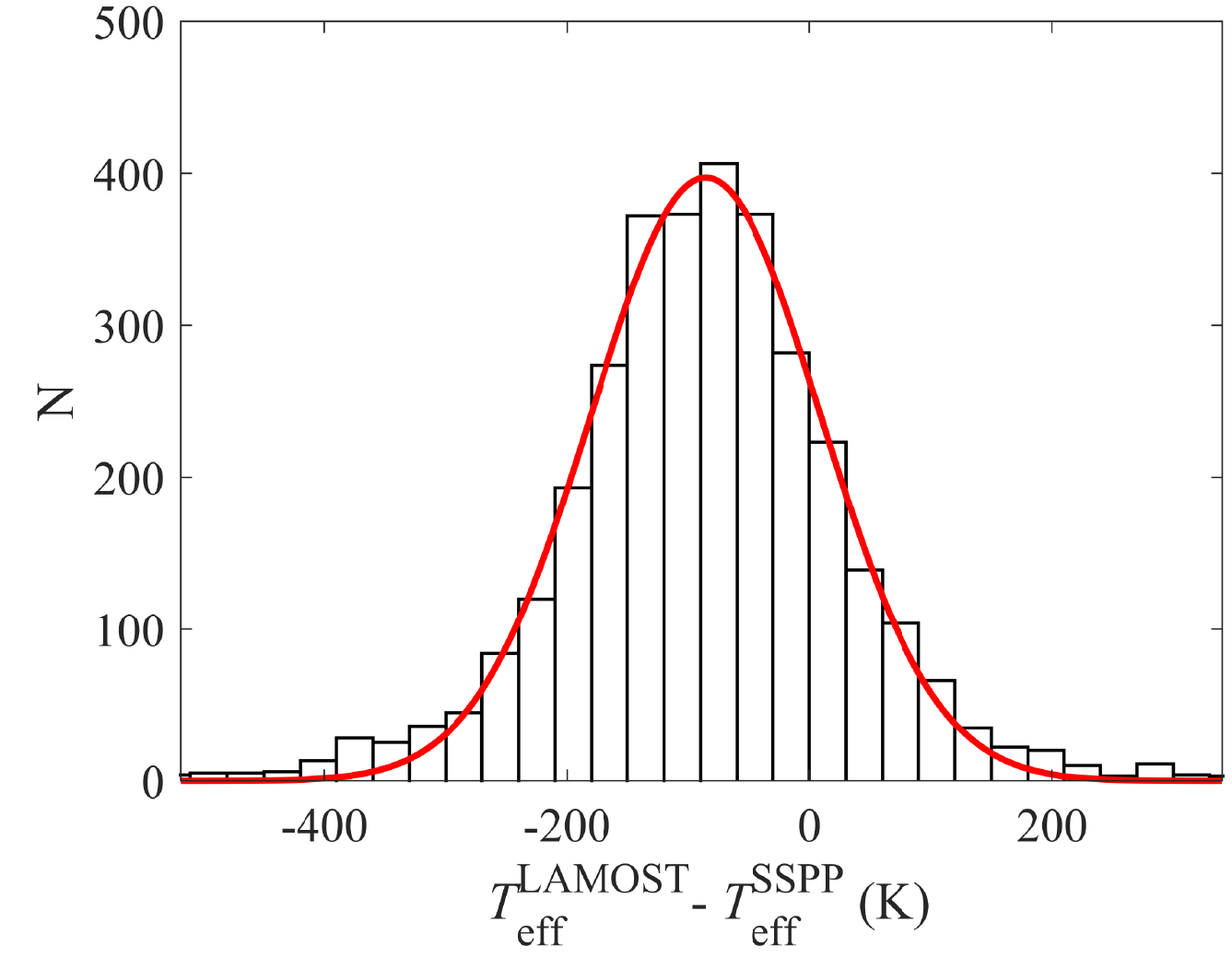}
    \includegraphics[width=0.5\textwidth]{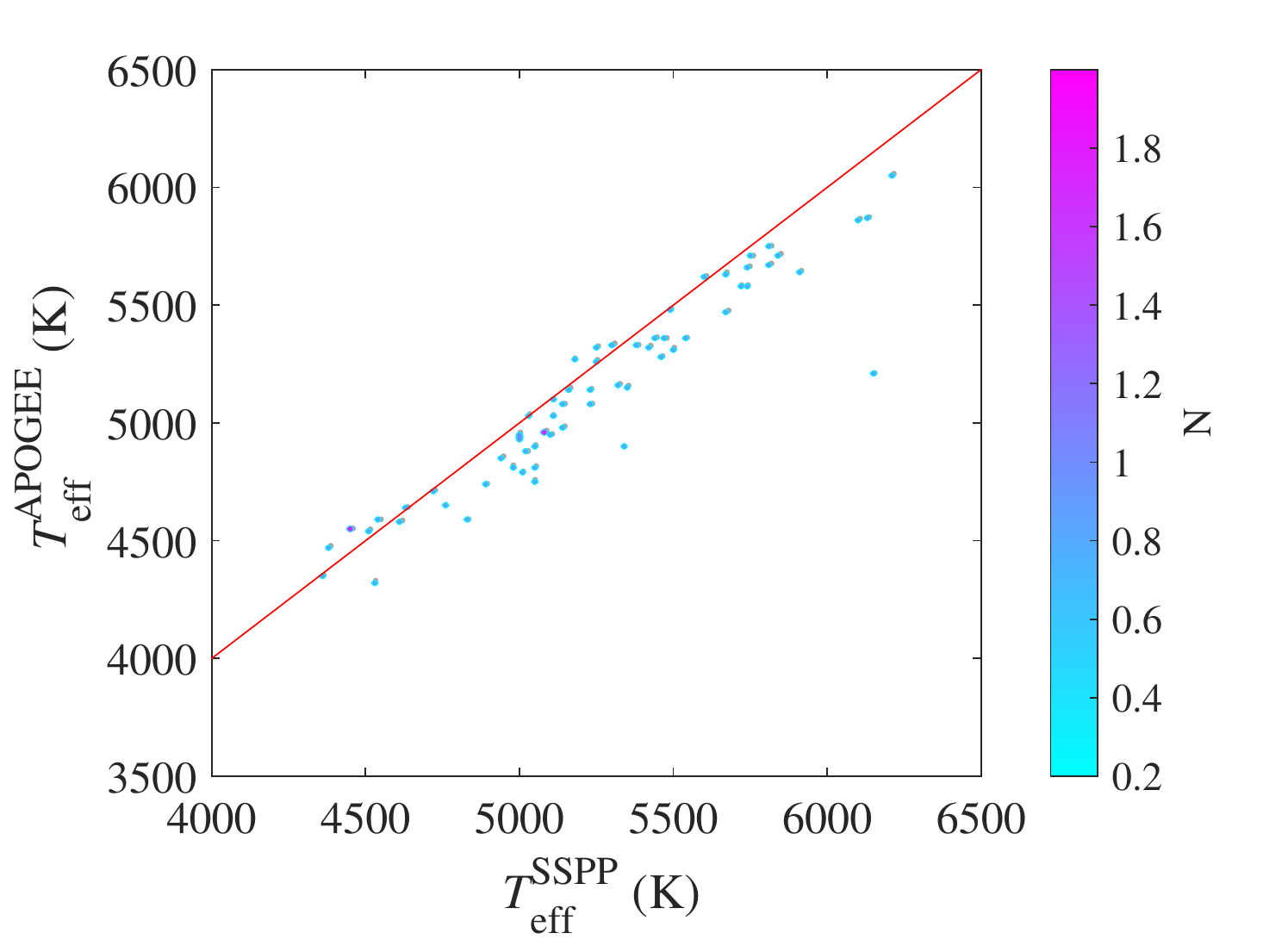}
    \includegraphics[width=0.45\textwidth]{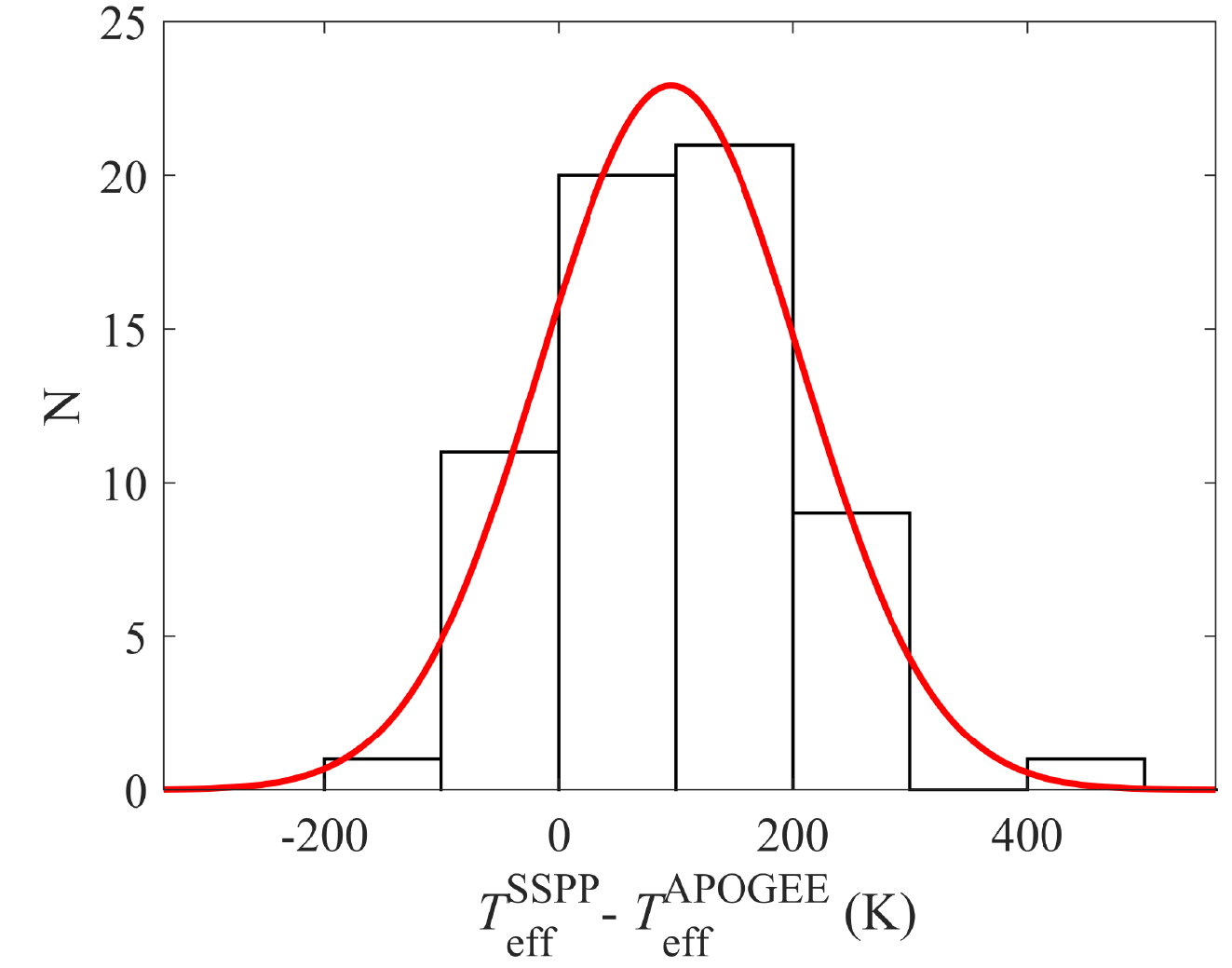}
    \caption{Pipeline differences of $T_{\rm eff}$. The upper two panels show the difference between LAMOST and APOGEE. The middle two are the difference between LAMOST and SSPP, and the lower two are the difference between SSPP and APOGEE.}
\end{figure*}

\begin{figure*}
    \centering
    \includegraphics[width=0.45\textwidth]{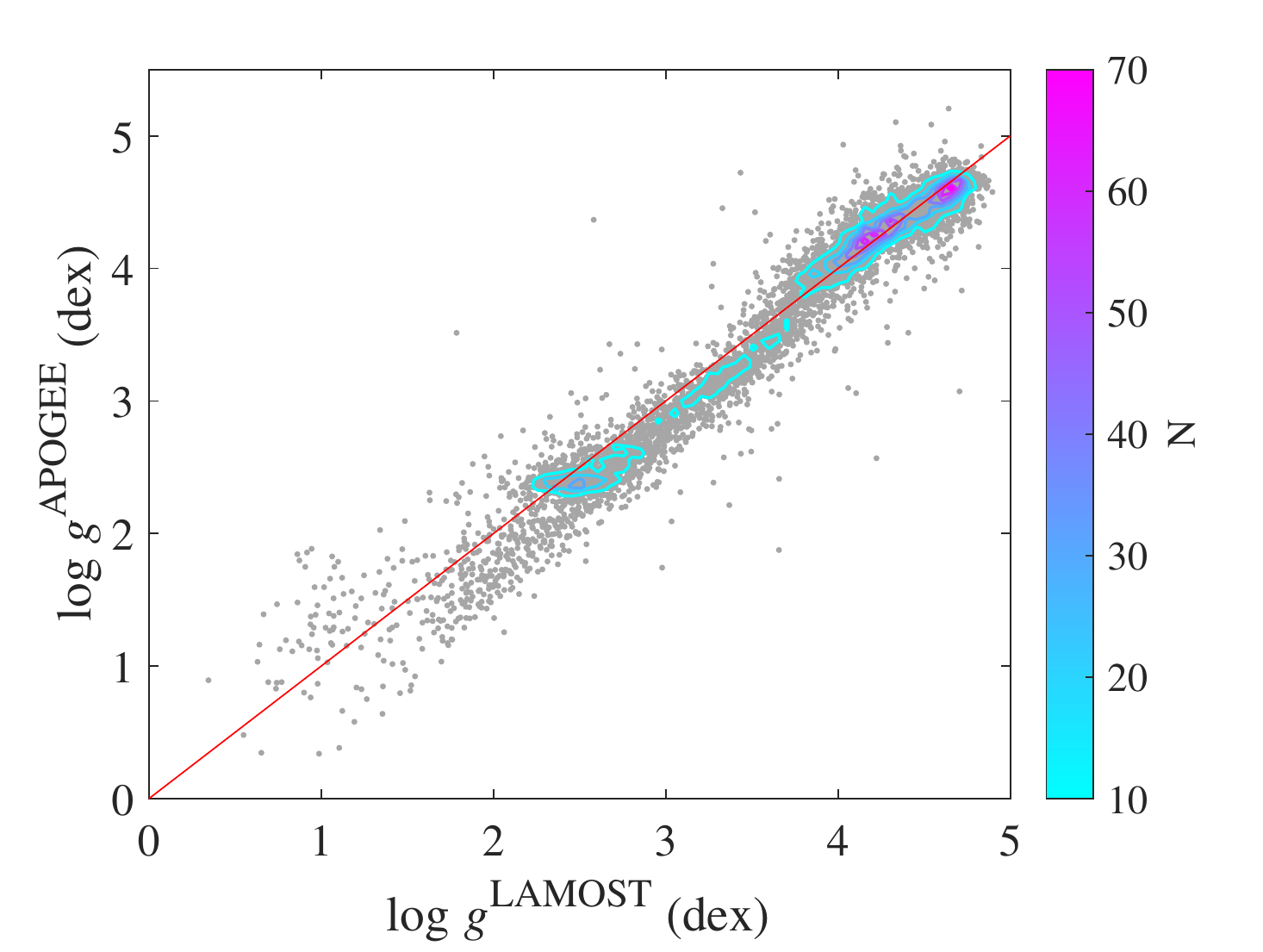}
    \includegraphics[width=0.45\textwidth]{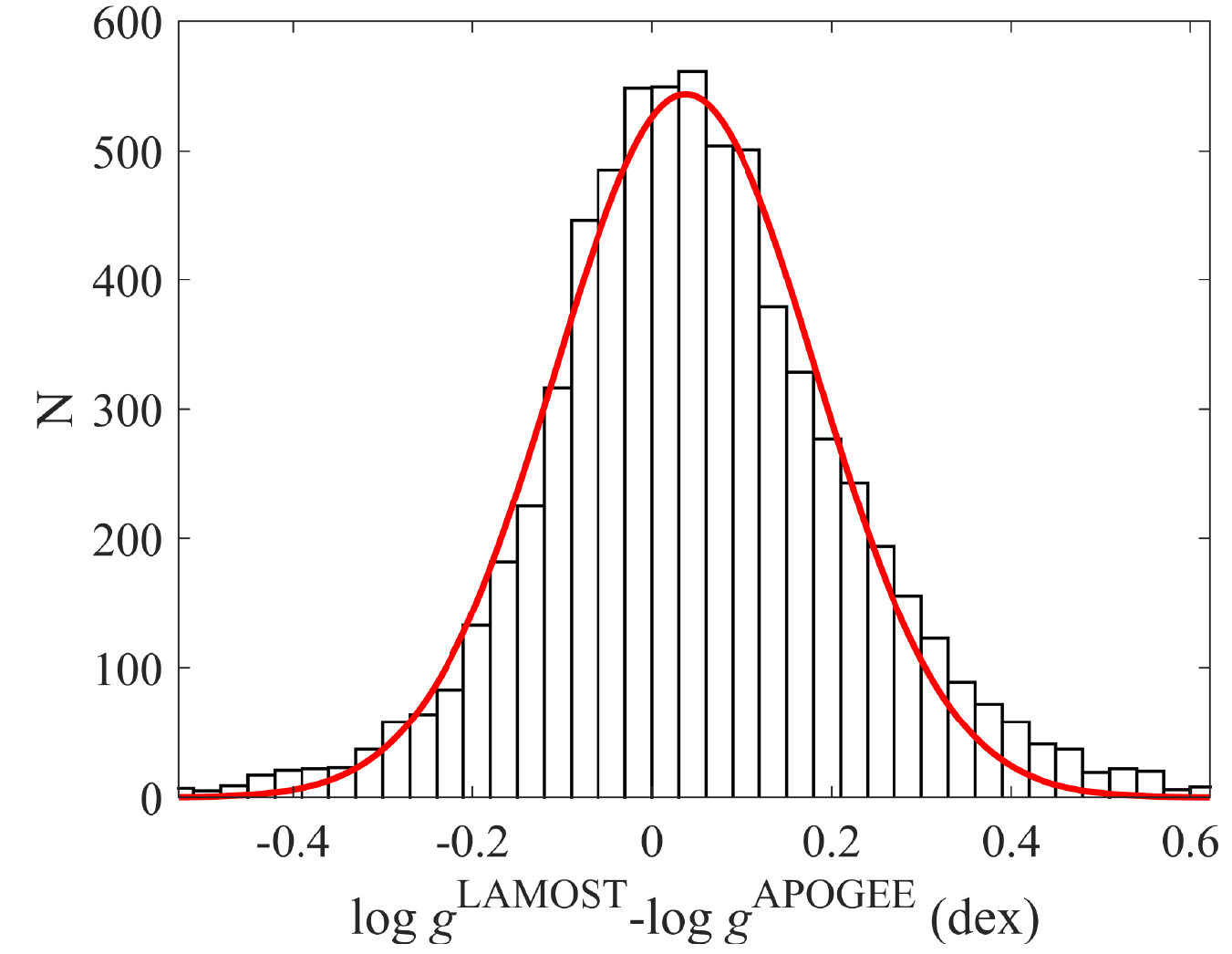}
    \includegraphics[width=0.45\textwidth]{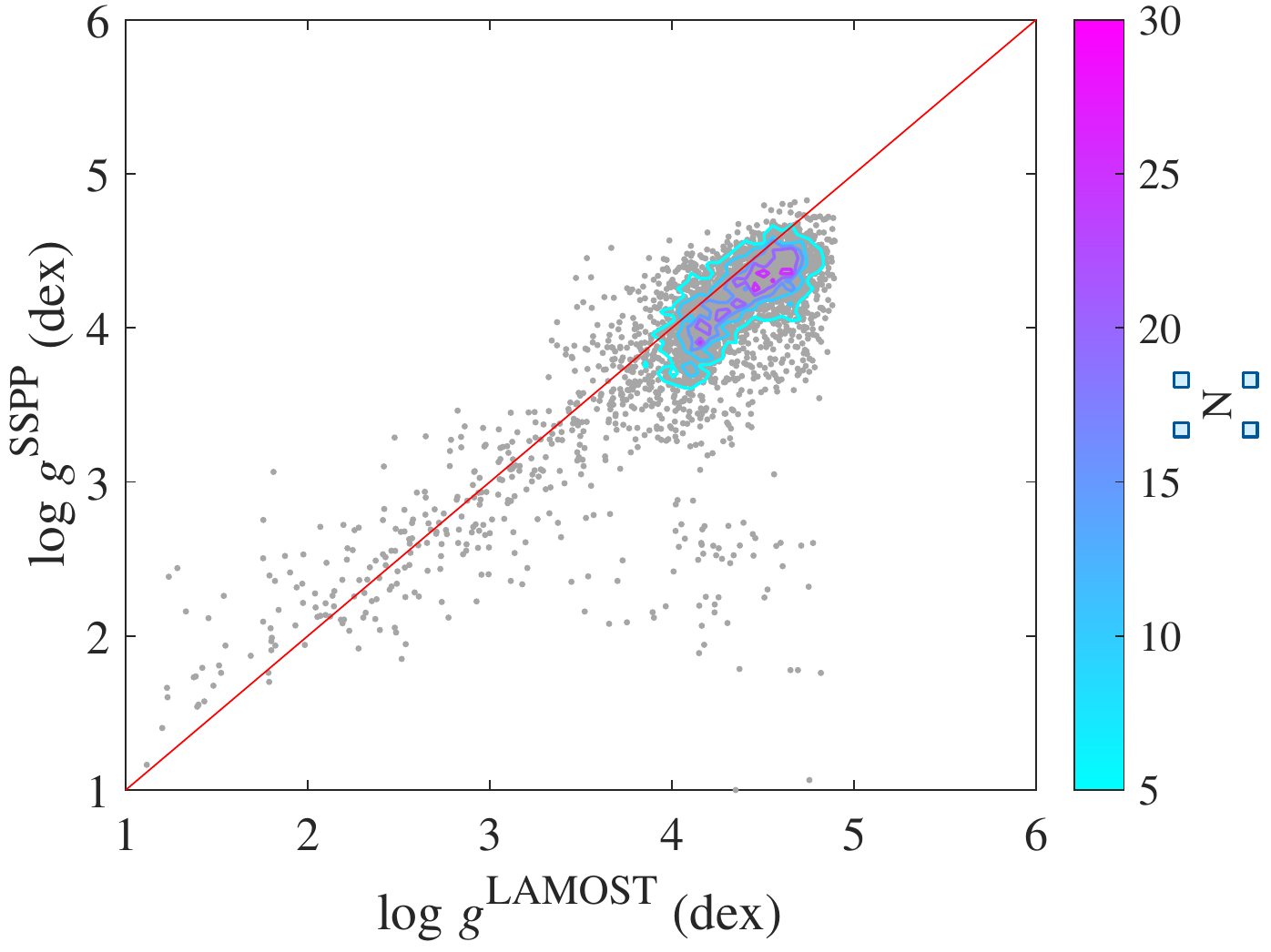}
    \includegraphics[width=0.45\textwidth]{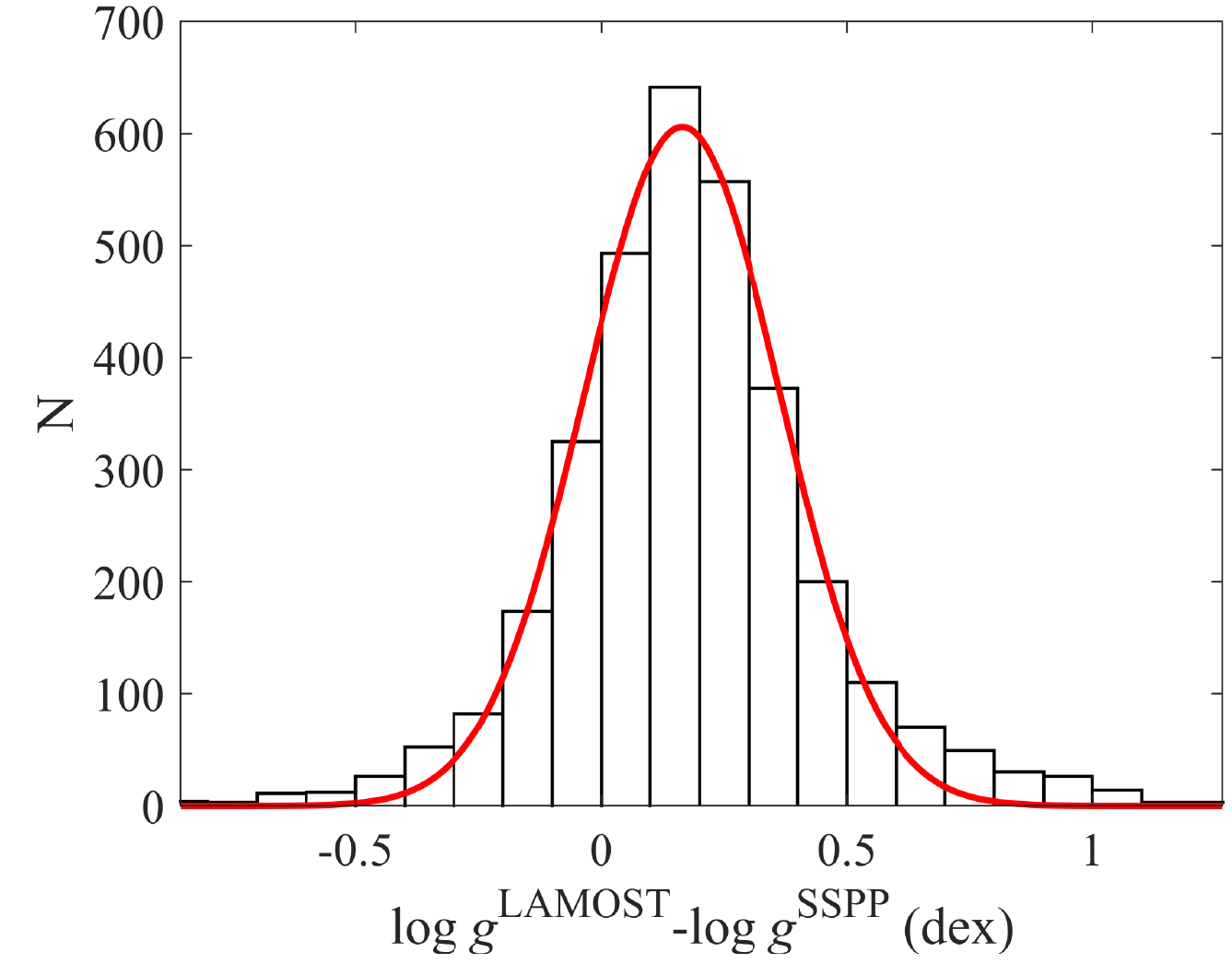}
    \includegraphics[width=0.45\textwidth]{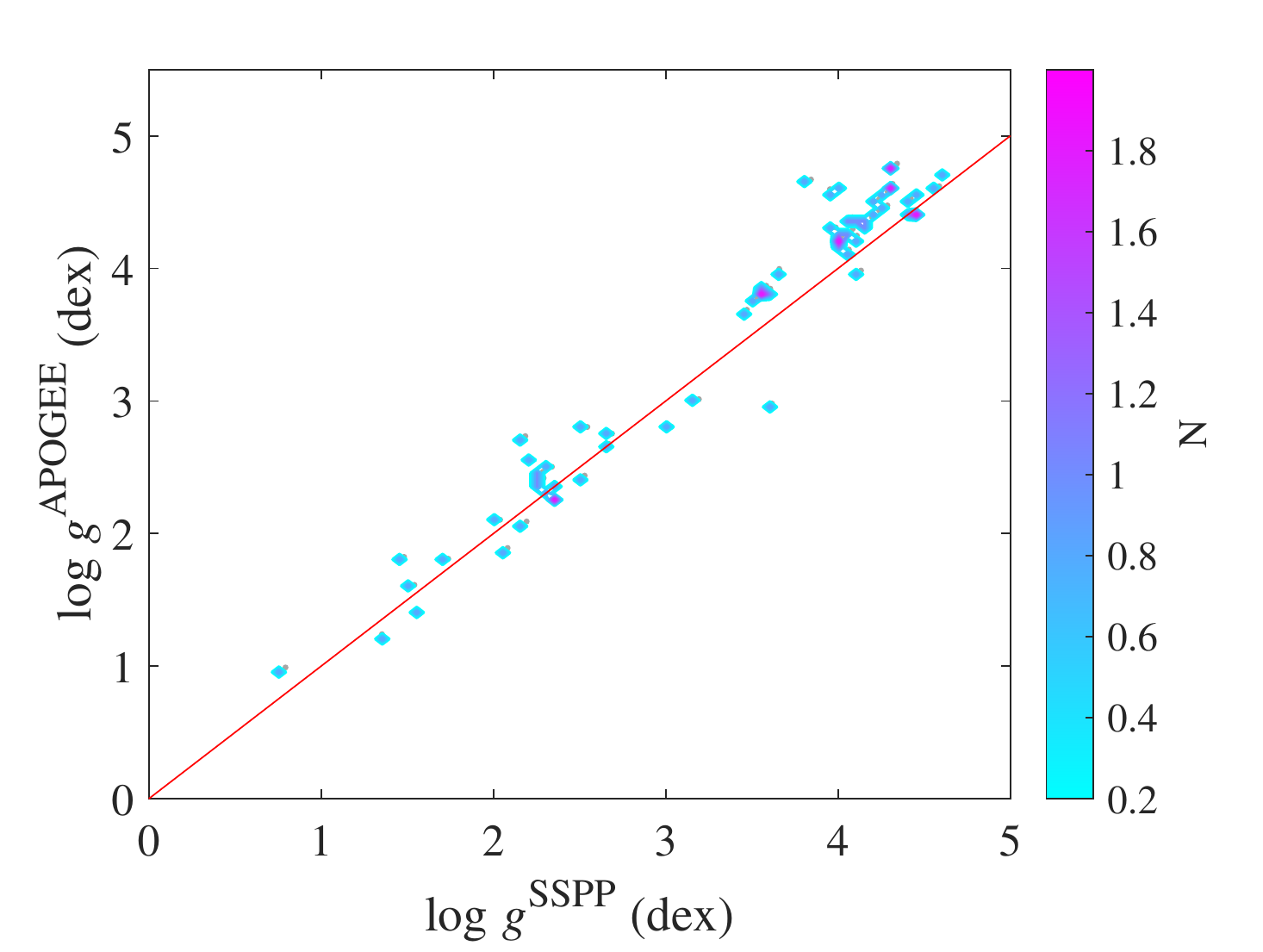}
    \includegraphics[width=0.45\textwidth]{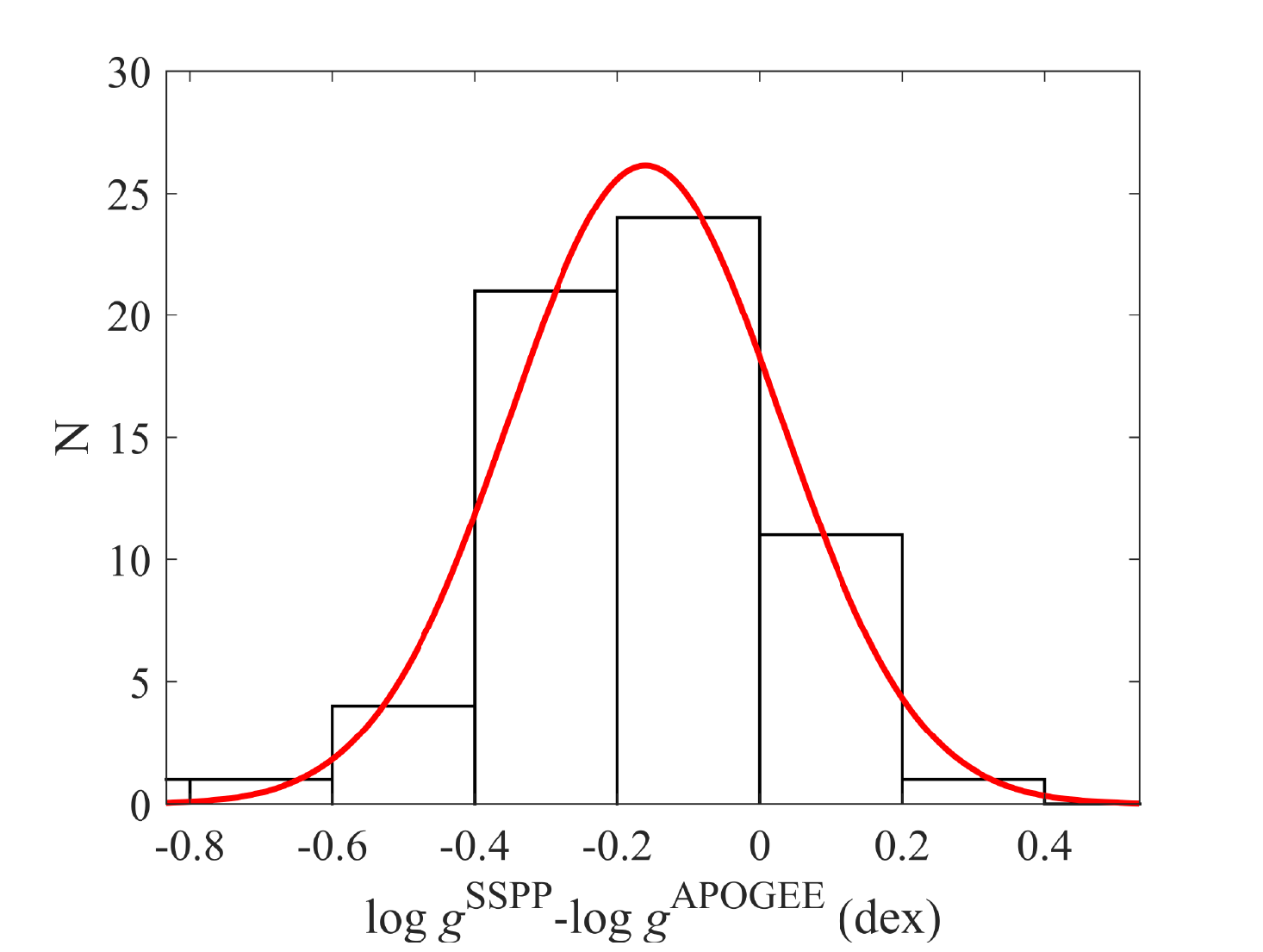}
    \caption{Pipeline difference of log $g$.}
\end{figure*}

\begin{figure*}
    \centering
    \includegraphics[width=0.45\textwidth]{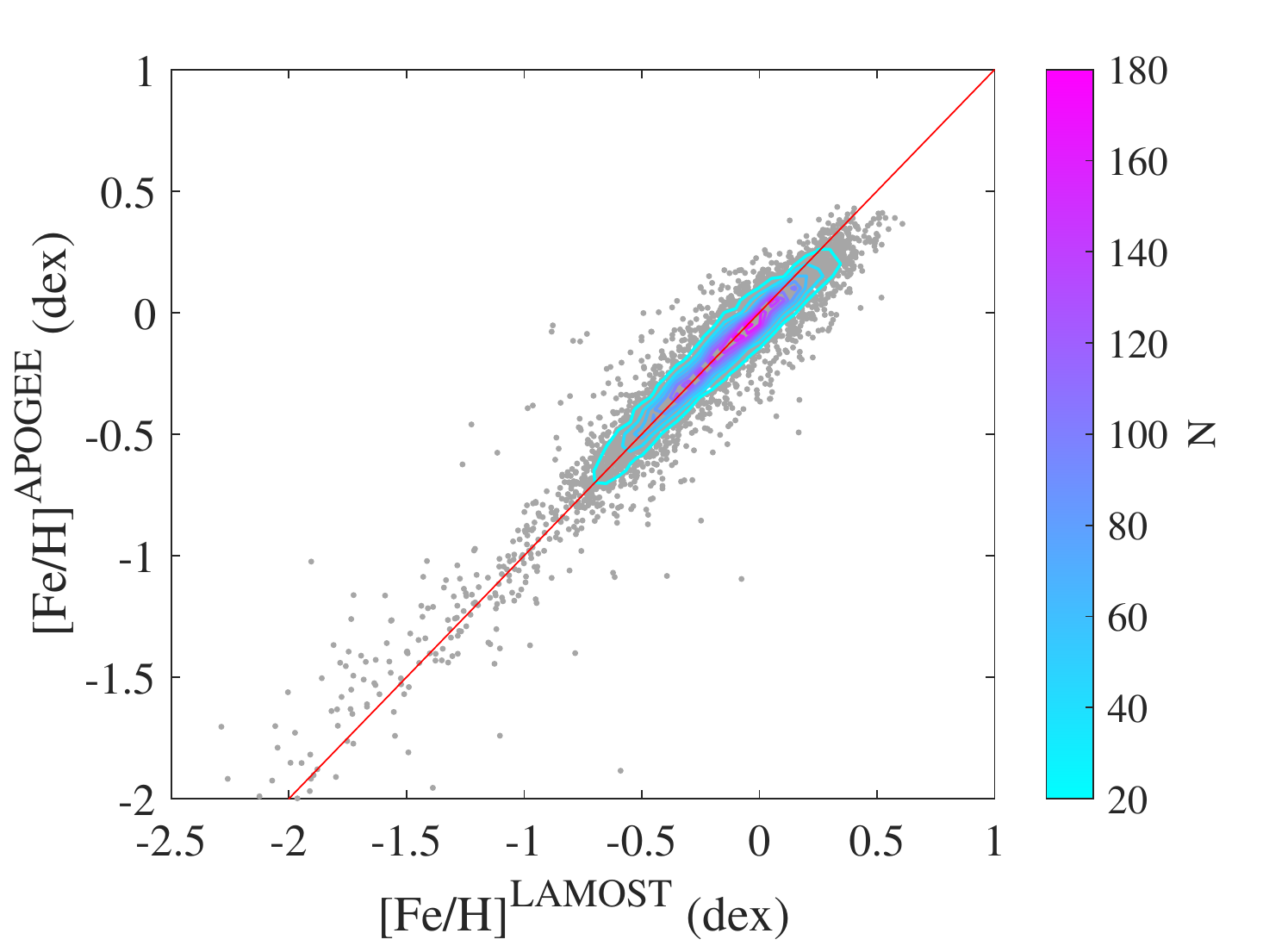}
    \includegraphics[width=0.45\textwidth]{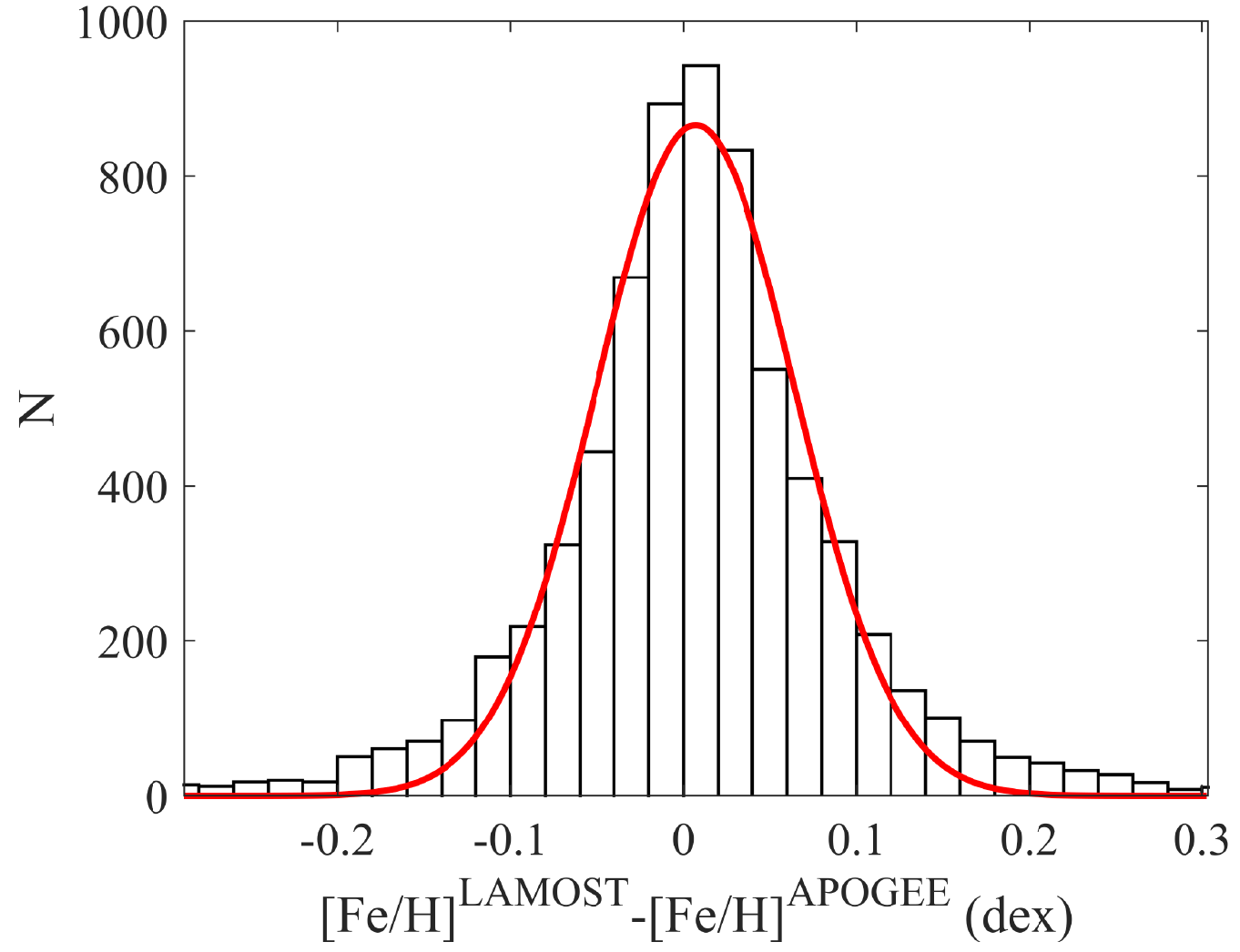}
    \includegraphics[width=0.45\textwidth]{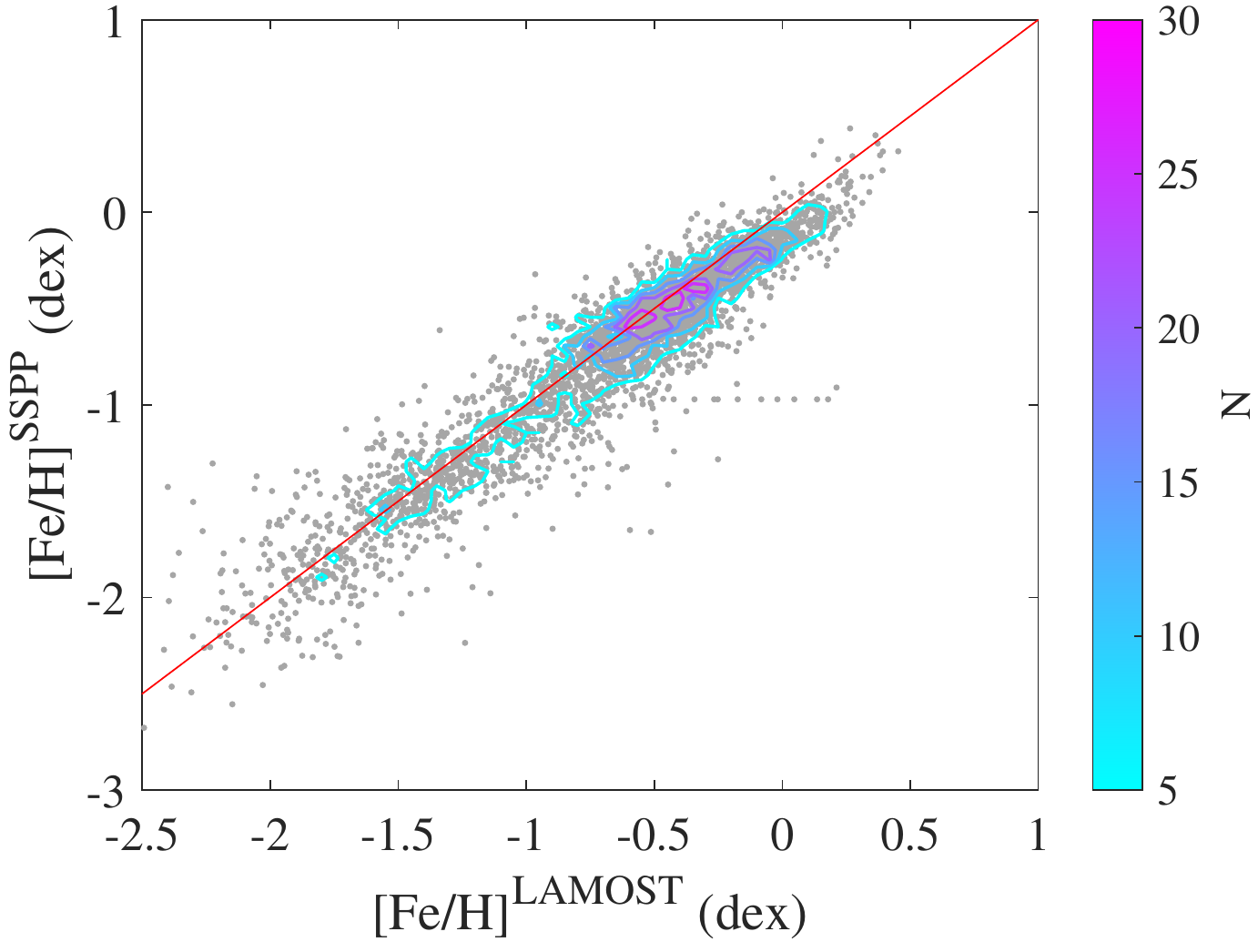}
    \includegraphics[width=0.45\textwidth]{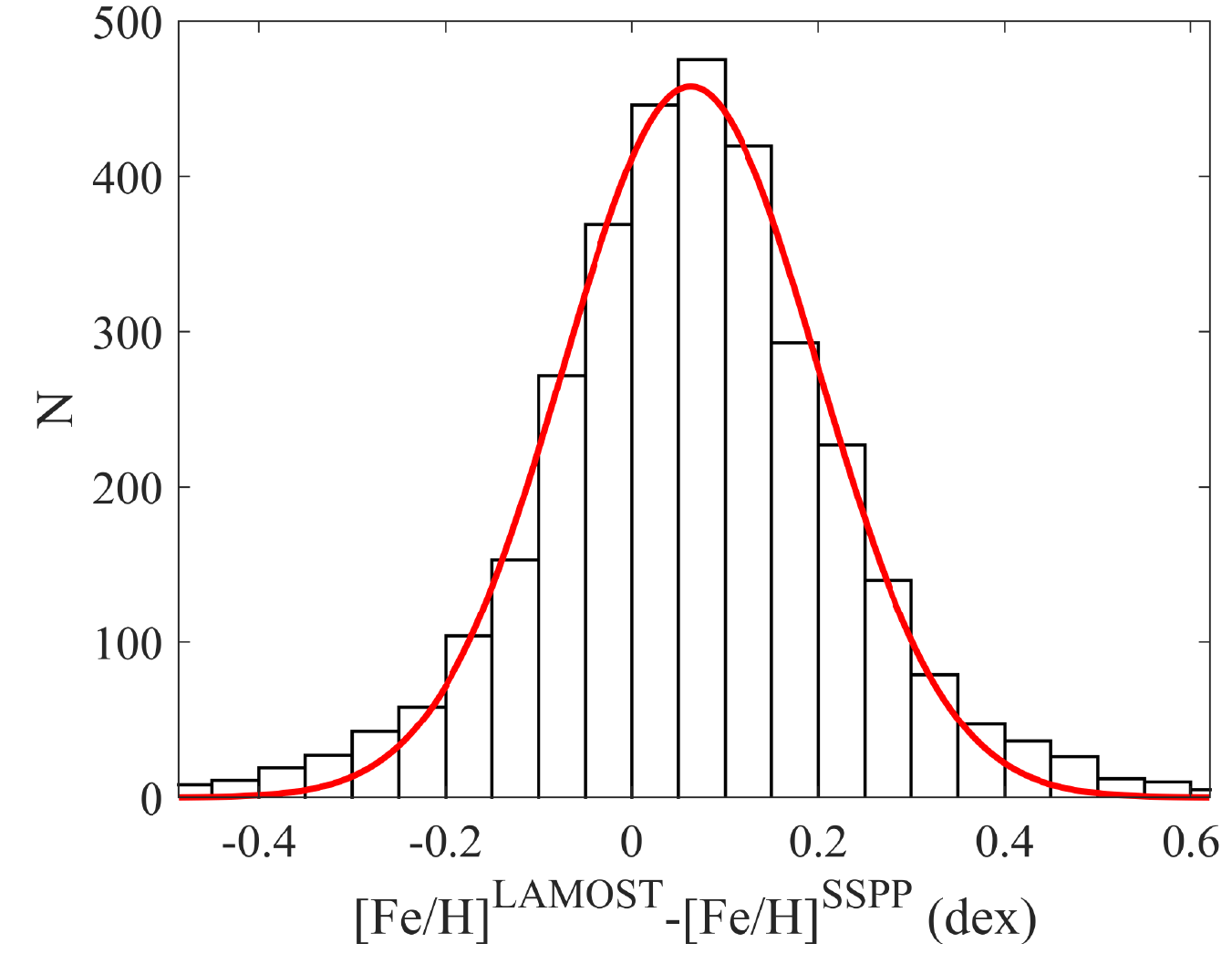}
    \includegraphics[width=0.5\textwidth]{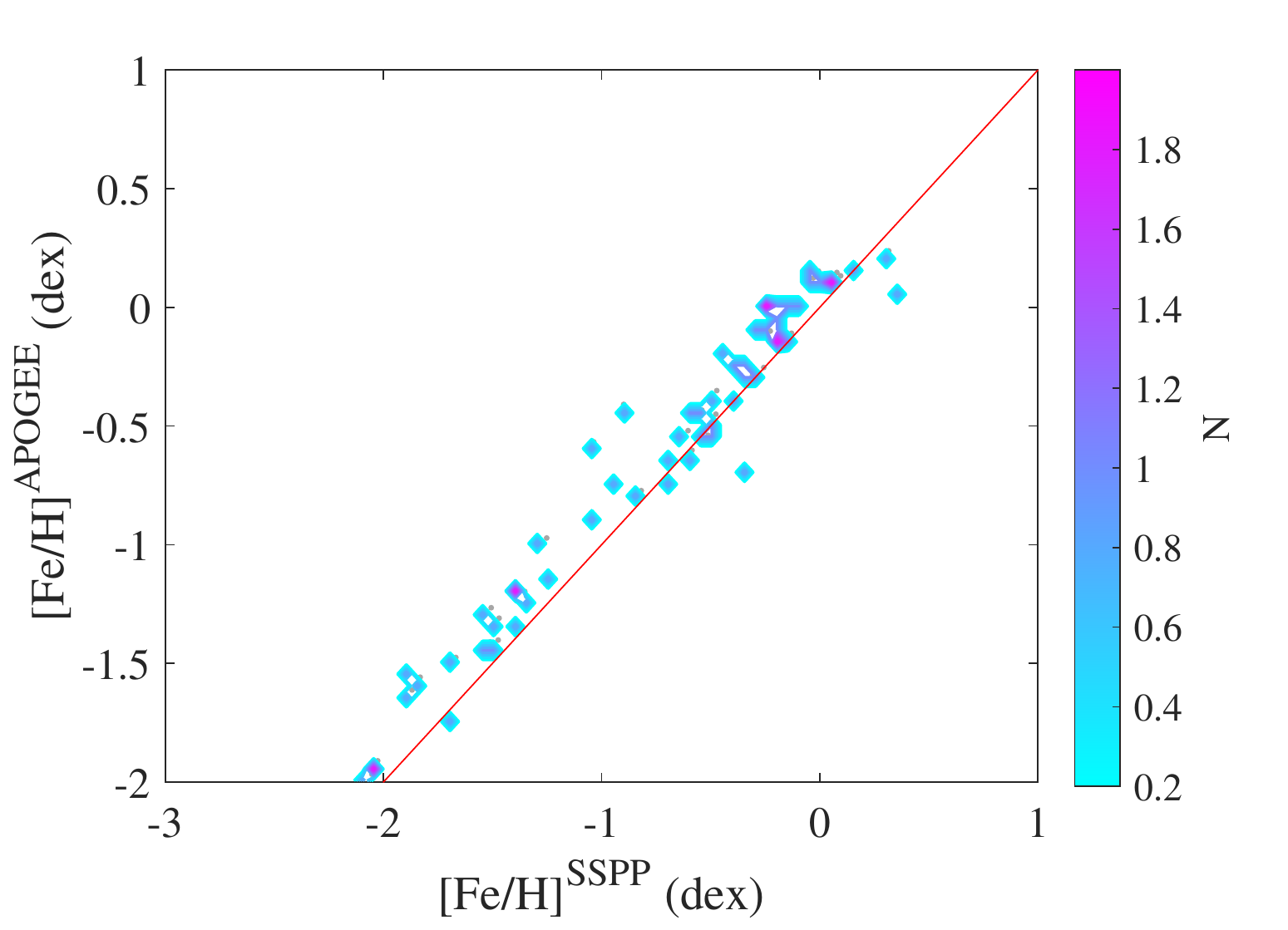}
    \includegraphics[width=0.45\textwidth]{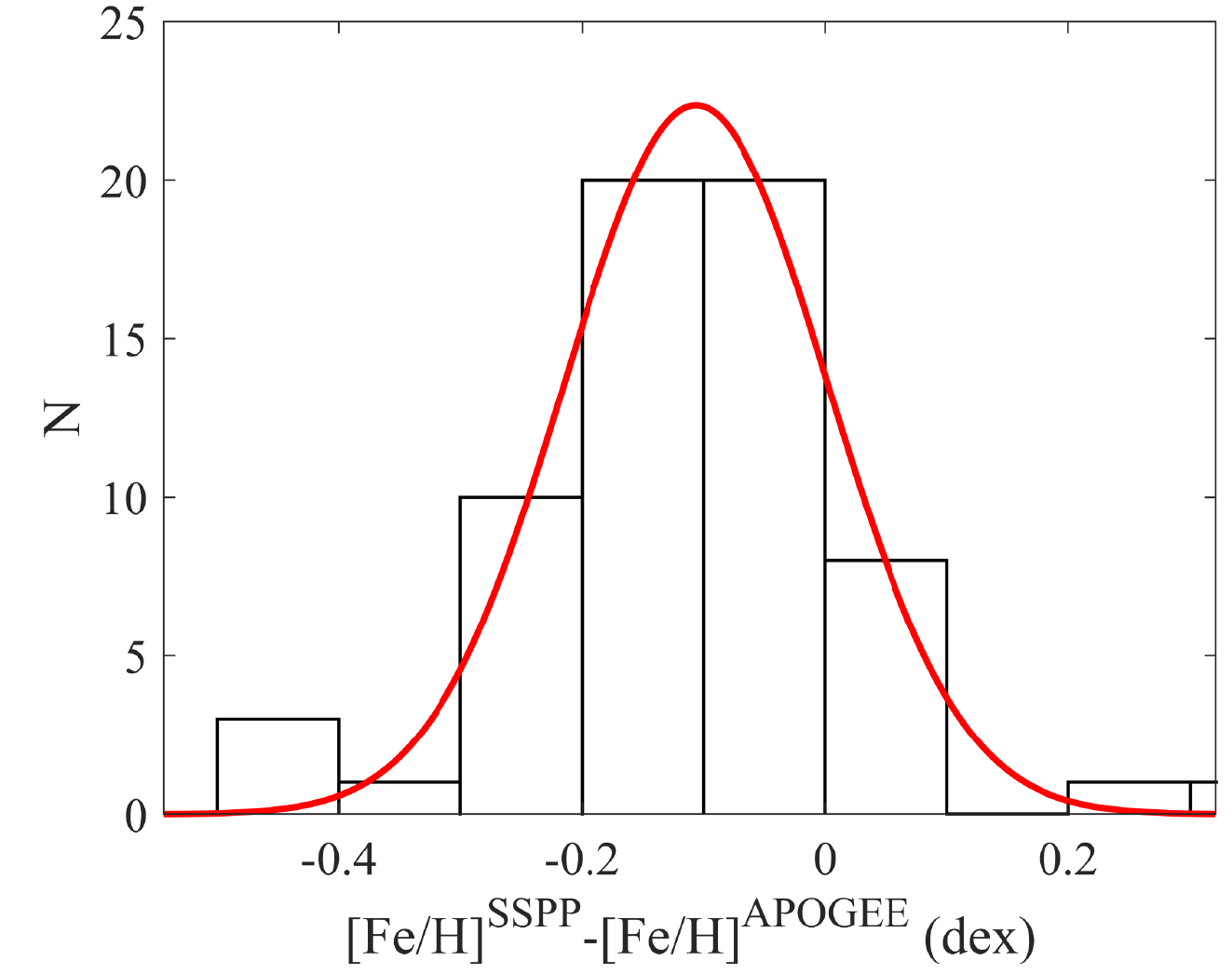}
    \caption{Pipeline difference of [Fe/H].}
\end{figure*}
\FloatBarrier

\end{appendix}

\end{document}